\input amstex.tex \input amsppt.sty 

%%%%%%%%%%%%%%%%%%%%%%%%%%%%%%%%%%%%%%%%%%%%%%%%%%%%%%%%%%%%%%%%%%%%%%%%%%
%%                                                                      %%
%%                         THE ACRES MACRO PACKAGE                      %%
%%                                                                      %%
%%%%%%%%%%%%%%%%%%%%%%%%%%%%%%%%%%%%%%%%%%%%%%%%%%%%%%%%%%%%%%%%%%%%%%%%%%
%
%
%   First we check whether the package has already been read, and
%   change the catcodes of @ and : in order to avoid clashes with
%   other packages.
%
\expandafter\ifx\csname ACRES chardef at\endcsname\relax\else\endinput\fi
\expandafter\chardef\csname ACRES chardef at\endcsname=\the\catcode`\@
\expandafter\chardef\csname ACRES chardef colon\endcsname=\the\catcode`\:
\catcode`\:=11\catcode`\@=11
%
%%%%%%%%%%%%%%%%%%%%%%%%%%%%%%%%%%%%%%%%%%%%%%%%%%%%%%%%%%%%%%%%%%%%%%%%%%
%
%
%   The version number of ACRES, the font in which the labels are printed
%   if asked for by \tracinglabels, the warning message of ACRES (see the 
%   TeX book page 227), and \@cres:newcount, which is the local version of
%   \newcount, but without the \outer part (see the TeX book page 347)
%   because we want to use it as a parameter.
%
\def\@cres:version{ACRES version 0.998 dd 20060403 by G.M. Tuynman }
\font\@cres:tracingfont=cmr5
\def\@cres:warning#1{\immediate\write16{ACRES: #1}}
\def\@cres:newcount{\alloc@0\count\countdef\insc@unt}
%
%%%%%%%%%%%%%%%%%%%%%%%%%%%%%%%%%%%%%%%%%%%%%%%%%%%%%%%%%%%%%%%%%%%%%%%%%%
%
%
\newread\@cres:myin
\newwrite\@cres:myout
\newif\if@cres:tracing\@cres:tracingfalse
\newif\if@cres:noteof
\def\@cres:chapsuppress{1}
\newif\if@cres:pagenos\@cres:pagenostrue
\@cres:newcount\@cres:chapdepth
\@cres:newcount\@cres:cntrdepth
\@cres:newcount\@cres:i
\@cres:newcount\@cres:rmi
\@cres:newcount\@cres:rmx
\@cres:newcount\@cres:rmc
\@cres:newcount\@cres:rmm
%
%%%%%%%%%%%%%%%%%%%%%%%%%%%%%%%%%%%%%%%%%%%%%%%%%%%%%%%%%%%%%%%%%%%%%%%%%%
%
%
%    \@cres:oponchap has two arguments : an "operator" #1, and a number
%    #2. It lets #1 operate on the counter \@cres:chap#2. Similarly for
%    \@cres:oponcntr : it lets #1 operate on \@cres:cntr#2.
%
\def\@cres:oponchap#1#2{\expandafter#1\csname @cres:chap#2\endcsname}
\def\@cres:oponcntr#1#2{\expandafter#1\csname @cres:cntr#2\endcsname}
%
%%%%%%%%%%%%%%%%%%%%%%%%%%%%%%%%%%%%%%%%%%%%%%%%%%%%%%%%%%%%%%%%%%%%%%%%%%
%
%
%    \@cres:preprm, \@cres:prtrm, \@cres:prtRM, \@cres:prta, \@cres:prtA,
%    \@cres:prepprint, and \@cres:PrintInMode prepare and print an
%    actual counter in the desired mode : lowercase roman, uppercase
%    roman, lowercase letter (only a-z), uppercase letter (only A-Z),
%    or just a number.
%
\def\@cres:preprm#1{\ifnum#1<1\@cres:rmi=10\@cres:rmx=0\@cres:rmc=0
  \@cres:rmm=0\else\@cres:rmi=#1\ifnum#1<4000\@cres:rmx=\@cres:rmi
  \divide\@cres:rmx by10 \multiply\@cres:rmx by10\advance\@cres:rmi
  by-\@cres:rmx \divide\@cres:rmx by10\@cres:rmc=\@cres:rmx\divide
  \@cres:rmc by10 \multiply\@cres:rmc by10\advance\@cres:rmx
  by-\@cres:rmc \divide\@cres:rmc by10\@cres:rmm=\@cres:rmc\divide
  \@cres:rmm by10 \multiply\@cres:rmm by10\advance\@cres:rmc
  by-\@cres:rmm \divide\@cres:rmm by10\else\@cres:rmx=0\@cres:rmc=0
  \@cres:rmm=#1\fi\fi\relax}
\def\@cres:prtrm{\ifcase\@cres:rmm\or m\or mm\or mmm\else\the\@cres:rmi\fi
  \ifcase \@cres:rmc \or c\or cc\or ccc\or cd\or d\or dc\or dcc\or
  dccc\or cm\fi \ifcase\@cres:rmx\or x\or xx\or xxx\or xl\or l\or lx\or
  lxx\or lxxx\or xc\fi\ifcase\@cres:rmi\or i\or ii\or iii\or iv\or v\or
  vi\or vii\or viii\or ix\or\fi}
\def\@cres:prtRM{\ifcase\@cres:rmm\or M\or MM\or MMM\else\the\@cres:rmi\fi
  \ifcase \@cres:rmc \or C\or CC\or CCC\or CD\or D\or DC\or DCC\or
  DCCC\or CM\fi \ifcase\@cres:rmx\or X\or XX\or XXX\or XL\or L\or LX\or
  LXX\or LXXX\or XC\fi\ifcase\@cres:rmi\or I\or II\or III\or IV\or V\or
  VI\or VII\or VIII\or IX\or\fi}
\def\@cres:prta#1{\ifcase#1 0\or a\or b\or c\or d\or e\or f\or g\or h\or
i\or j\or k\or l\or m\or n\or o\or p\or q\or r\or s\or t\or u\or v\or w\or
x\or y\or z\else#1\fi}
\def\@cres:prtA#1{\ifcase#1 0\or A\or B\or C\or D\or E\or F\or G\or H\or
I\or J\or K\or L\or M\or N\or O\or P\or Q\or R\or S\or T\or U\or V\or W\or
X\or Y\or Z\else#1\fi}
\def\@cres:prepprint{\ifcase\@cres:mode/\or\@cres:preprm{\@cres:number}\or
  \@cres:preprm{\@cres:number}\fi}
\def\@cres:PrintInMode{\ifcase\@cres:mode/ \@cres:number\or\@cres:prtrm\or
\@cres:prtRM\or\@cres:prta{\@cres:number}\or\@cres:prtA{\@cres:number}%
\fi}
%
%%%%%%%%%%%%%%%%%%%%%%%%%%%%%%%%%%%%%%%%%%%%%%%%%%%%%%%%%%%%%%%%%%%%%%%%%%
%
%
%    A reference has five parts : toplevel, separator, the rest, the
%    page number, and the title text. \@cres:printref only prints the 
%    first three.
%
\def\@cres:printref#1#2#3#4#5{%
\def\@cres:empty{}\def\@cres:tail{#3}%
\ifx\@cres:tail\@cres:empty 
   \def\@cres:noprint{?-.!;,}\def\@cres:sep{#2}%
   \ifx\@cres:sep\@cres:noprint\else#1\fi
\else\def\@cres:head{#1}%
    \ifx\@cres:head\@cres:empty #3%
    \else\edef\@cres:mode/{\csname @cres:chapmode1\endcsname}%
        \edef\@cres:number{\@cres:oponchap\the1}%
        \@cres:prepprint 
        \edef\@cres:chapno{\@cres:PrintInMode}%
        \ifcase\@cres:chapsuppress\relax #3%
        \or\ifx\@cres:head\@cres:chapno#3\else#1#2#3\fi
        \else#1#2#3%
        \fi 
    \fi
\fi}
%
%%%%%%%%%%%%%%%%%%%%%%%%%%%%%%%%%%%%%%%%%%%%%%%%%%%%%%%%%%%%%%%%%%%%%%%%%%
%
%
%    \@cres:printpageref prints (only) the page number of a reference.
%
\def\@cres:printpageref#1#2#3#4#5{\ifnum#4<0 
\romannumeral-#4\else#4\fi}
%
%%%%%%%%%%%%%%%%%%%%%%%%%%%%%%%%%%%%%%%%%%%%%%%%%%%%%%%%%%%%%%%%%%%%%%%%%%
%
%
%    \@cres:printtitletext prints (only) the title text of a reference.
%
\def\@cres:printtitletext#1#2#3#4#5{#5}
%
%%%%%%%%%%%%%%%%%%%%%%%%%%%%%%%%%%%%%%%%%%%%%%%%%%%%%%%%%%%%%%%%%%%%%%%%%%
%
%
%    \@cres:cntr does all the work : all counter macros are actually
%    defined as \@cres:cntr{level of the counter}
%
\def\@cres:cntr#1{%
%  We start to increase the counter
\global\@cres:oponcntr\advance#1 by1
%  We then retrieve the correct counter level, counter mode, and the
%  actual value of the counter
\edef\@cres:cntrlevel{\csname @cres:cntrlev#1\endcsname}%
\edef\@cres:mode/{\csname @cres:cntrmode#1\endcsname}%
\edef\@cres:number{\@cres:oponcntr\the#1}%
  \@cres:prepprint
  \edef\@cres:counter{\@cres:PrintInMode}%
\ifnum\@cres:cntrlevel<1\xdef\@cres:newref{{}{}{\@cres:counter}{\the
                                                            \pageno}{}}%
\else\ifnum\@cres:cntrlevel<2 \xdef\@cres:newref{\@cres:counter}%
        \xdef\@cres:cntrseptr{\csname @cres:cntrsep#1\endcsname}%
     \else\edef\@cres:mode/{\csname @cres:chapmode2\endcsname}%
         \edef\@cres:number{\@cres:oponchap\the2}%
         \@cres:prepprint\xdef\@cres:newref{\@cres:PrintInMode}\@cres:i=2
         \xdef\@cres:cntrseptr{\csname @cres:chapsep2\endcsname}%
         \loop\ifnum\@cres:i<\@cres:cntrlevel\advance\@cres:i by1
             \edef\@cres:mode/{%
              \csname @cres:chapmode\the\@cres:i\endcsname}%
             \edef\@cres:number{\@cres:oponchap\the{\the\@cres:i}}%
             \@cres:prepprint\xdef\@cres:newref{\@cres:newref
             \csname @cres:chapsep\the\@cres:i\endcsname\@cres:PrintInMode}%
         \repeat
         \xdef\@cres:newref{\@cres:newref
          \csname @cres:cntrsep#1\endcsname\@cres:counter}%
     \fi
     \edef\@cres:mode/{\csname @cres:chapmode1\endcsname}%
     \edef\@cres:number{\@cres:oponchap\the1}%
     \@cres:prepprint
     \xdef\@cres:newref{{\@cres:PrintInMode}{\@cres:cntrseptr
                                        }{\@cres:newref}{\the\pageno}{}}%
\fi
%  Finally we print the complete result
\expandafter\@cres:printref\@cres:newref}
%
%%%%%%%%%%%%%%%%%%%%%%%%%%%%%%%%%%%%%%%%%%%%%%%%%%%%%%%%%%%%%%%%%%%%%%%%%%
%
%
%    \newcounter{name}{mode}{separator}{level}
%
\def\newcounter#1#2#3#4{%
%  First we test if we have already initialized, if not we do it.
\expandafter
\ifx\csname ACRES initialize\endcsname\relax
   \@cres:initialize\expandafter\gdef\csname ACRES initialize\endcsname{}%
\fi
\advance\@cres:cntrdepth by1 
%  We create a new counter with the name "\@cres:cntr\@cres:cntrdepth"
\@cres:oponcntr\@cres:newcount{\the\@cres:cntrdepth}%
%  We define the actual new counter as \@cres:cntr{\@cres:cntrdepth}
\expandafter\xdef\csname#1\endcsname
  {\noexpand\@cres:cntr\the\@cres:cntrdepth}%
%  We define the separator for this counter as the third argument
\expandafter\gdef\csname @cres:cntrsep\the\@cres:cntrdepth\endcsname{#3}%
%  We define the mode of this counter
\expandafter\xdef\csname @cres:cntrmode\the\@cres:cntrdepth\endcsname
  {\ifx#2i1\else\ifx#2I2\else\ifx#2a3\else\ifx#2A4\else0\fi\fi\fi\fi}%
%  If we wish to go on with the counter beyond a chapterlevel,
%  we specify this in #4; it is stored in \@cres:cntrlev\@cres:cntrdepth
\ifnum#4<\@cres:chapdepth\@cres:i=#4\else\@cres:i=\@cres:chapdepth\fi
\expandafter\xdef\csname @cres:cntrlev\the\@cres:cntrdepth\endcsname
  {\the\@cres:i}}
%
%%%%%%%%%%%%%%%%%%%%%%%%%%%%%%%%%%%%%%%%%%%%%%%%%%%%%%%%%%%%%%%%%%%%%%%%%%
%
%
%    \@cres:chap is the equivalent of \@cres:cntr for the nested counters.
%
\def\@cres:chap#1{%
\@cres:i=#1\global\@cres:oponchap\advance#1 by1
\loop\ifnum\@cres:i<\@cres:chapdepth\advance\@cres:i by1
    \global\csname @cres:chap\the\@cres:i\endcsname=0%
\repeat
\@cres:i=0
\loop\ifnum\@cres:i<\@cres:cntrdepth\advance\@cres:i by1
    \expandafter
    \ifnum\csname @cres:cntrlev\the\@cres:i\endcsname<#1
    \else\global\csname @cres:cntr\the\@cres:i\endcsname=0
    \fi
\repeat
\ifnum#1<2 \xdef\@cres:newref{}\xdef\@cres:septr{}%
\else\edef\@cres:mode/{\csname @cres:chapmode2\endcsname}%
    \edef\@cres:number{\@cres:oponchap\the2}%
    \@cres:prepprint 
    \xdef\@cres:newref{\@cres:PrintInMode}%
    \@cres:i=2
    \xdef\@cres:septr{\csname @cres:chapsep2\endcsname}%
    \loop\ifnum\@cres:i<#1\advance\@cres:i by1
        \edef\@cres:mode/{\csname @cres:chapmode\the\@cres:i\endcsname}%
        \edef\@cres:number{\@cres:oponchap\the{\the\@cres:i}}%
        \@cres:prepprint
        \xdef\@cres:newref{\@cres:newref
            \csname @cres:chapsep\the\@cres:i\endcsname\@cres:PrintInMode}%
    \repeat
\fi
\edef\@cres:mode/{\csname @cres:chapmode1\endcsname}%
\edef\@cres:number{\@cres:oponchap\the1}%
\@cres:prepprint
\xdef\@cres:newref{{\@cres:PrintInMode}{\@cres:septr
                                         }{\@cres:newref}{\the\pageno}{}}%
\expandafter\@cres:printref\@cres:newref}
%
%%%%%%%%%%%%%%%%%%%%%%%%%%%%%%%%%%%%%%%%%%%%%%%%%%%%%%%%%%%%%%%%%%%%%%%%%%
%
%
%   \newlevel{name}{mode}{separator}
%
\def\newlevel#1#2#3{%
\expandafter
\ifx\csname ACRES initialize\endcsname\relax 
   \@cres:initialize\expandafter\gdef\csname ACRES initialize\endcsname{}%
\fi
\advance\@cres:chapdepth by1 
\@cres:oponchap\@cres:newcount{\the\@cres:chapdepth}%
\expandafter\xdef\csname#1\endcsname
 {\noexpand\@cres:chap\the\@cres:chapdepth}%
\expandafter\gdef\csname @cres:chapsep\the\@cres:chapdepth\endcsname{#3}%
\expandafter\xdef\csname @cres:chapmode\the\@cres:chapdepth\endcsname
    {\ifx#2i1\else\ifx#2I2\else\ifx#2a3\else\ifx#2A4\else0\fi\fi\fi\fi}}
%
%%%%%%%%%%%%%%%%%%%%%%%%%%%%%%%%%%%%%%%%%%%%%%%%%%%%%%%%%%%%%%%%%%%%%%%%%%
%
%
% \withoutnumber is a special counter with a special separator to 
% prevent it from printing, but otherwise it acts just as any other
% counter. It thus can be used with \memorize and \memorizetitle. Its
% main purpose is to be able to create labels without any visible
% sign/numbering.
%
\@cres:newcount\@cres:withoutnumber\@cres:withoutnumber=0
\def\withoutnumber{\global\advance\@cres:withoutnumber by1
   \xdef\@cres:newref{{\the\@cres:withoutnumber}{,,,,,}{}{\the\pageno}{}}}
%
%%%%%%%%%%%%%%%%%%%%%%%%%%%%%%%%%%%%%%%%%%%%%%%%%%%%%%%%%%%%%%%%%%%%%%%%%%
%
%
%    \@cres:undefined is the conditional macro that tests whether a
%    label has been defined before or not.
%
\def\@cres:undefined#1#2#3{%
\expandafter\ifx\csname @cres::#1\endcsname\relax#2\else#3\fi}
%
%%%%%%%%%%%%%%%%%%%%%%%%%%%%%%%%%%%%%%%%%%%%%%%%%%%%%%%%%%%%%%%%%%%%%%%%%%
%
%
%  \@cres:shipout : see the TeX book page 228/332 exercise 21.10
%
\def\@cres:shipout#1{{\let\the=0\edef\next{\write\@cres:myout{#1}}\next}}
%
%%%%%%%%%%%%%%%%%%%%%%%%%%%%%%%%%%%%%%%%%%%%%%%%%%%%%%%%%%%%%%%%%%%%%%%%%%
%
%
%  \@cres:memorize{label}{reference} : creates a macro \@cres:"label" with 
%  contents "reference".
%
\def\@cres:memorize#1#2{\expandafter\xdef\csname @cres::#1\endcsname{#2}}
%
%%%%%%%%%%%%%%%%%%%%%%%%%%%%%%%%%%%%%%%%%%%%%%%%%%%%%%%%%%%%%%%%%%%%%%%%%%
%
%
%  \memorize{name}={label} : memorizes the value of the counter \"name"
%  under the name \@cres:"label" (see also \@cres:memorize)
%
\def\memorize#1=#2{%
\if@cres:tracing{\@cres:tracingfont#2=}\fi
\def\@cres:tricky{#2}%
#1%
\expandafter\@cres:strippageandtext\@cres:newref
\edef\@cres:newrefwithoutpageandtext{\@cres:refwithoutpageandtext}%
\@cres:undefined{#2}{\@cres:memorize{#2}\@cres:newref}%
  {{\edef\@cres:oldref{\csname @cres::#2\endcsname}%
    \expandafter\@cres:strippageandtext\@cres:oldref
%EEold?\@cres:refwithoutpageandtext?EEnew?\@cres:newrefwithoutpageandtext?
    \ifx\@cres:refwithoutpageandtext\@cres:newrefwithoutpageandtext
    \else\@cres:warning{changed or duplicate "#2"}%
        \@cres:memorize{#2}\@cres:newref
    \fi}}%
% here ends the call to \@cres:undefined
\@cres:shipout{{#2}{\@cres:newrefwithoutpageandtext{\the\pageno}{}}}}
%
%%%%%%%%%%%%%%%%%%%%%%%%%%%%%%%%%%%%%%%%%%%%%%%%%%%%%%%%%%%%%%%%%%%%%%%%%%
%
%
%  \memorizetitle{name}={label}{intertext}{titletext} : memorizes the 
%  value of the counter \"name" and the title text under the name 
%  \@cres:"label" (see also \@cres:memorize). It prints the counter, 
%  the intertext and the title text.
%
\def\memorizetitle#1=#2#3#4{%
\if@cres:tracing{\@cres:tracingfont#2=}\fi
\def\@cres:tricky{#2}%
#1#3#4%
\expandafter\@cres:strippageandtext\@cres:newref
\edef\@cres:newrefwithoutpageandtext{\@cres:refwithoutpageandtext}%
\@cres:undefined{#2}{\@cres:memorize{#2}\@cres:newref}%
  {{\edef\@cres:oldref{\csname @cres::#2\endcsname}%
    \expandafter\@cres:strippageandtext\@cres:oldref
    \ifx\@cres:refwithoutpageandtext\@cres:newrefwithoutpageandtext
    \else\@cres:warning{changed or duplicate "#2"}%
        \@cres:memorize{#2}\@cres:newref
    \fi}}%
% here ends the call to \@cres:undefined
\@cres:shipout{{#2}{\@cres:newrefwithoutpageandtext{\the\pageno}{#4}}}}
%
%%%%%%%%%%%%%%%%%%%%%%%%%%%%%%%%%%%%%%%%%%%%%%%%%%%%%%%%%%%%%%%%%%%%%%%%%%
%
%
%  \@cres:strippageandtext{\@cres:newref} : strips the page number and the 
%  title text from the reference in order to add the actual page number in
%  shipout.
%
\def\@cres:strippageandtext#1#2#3#4#5{%
\xdef\@cres:refwithoutpageandtext{{#1}{#2}{#3}}}
%%%%%%%%%%%%%%%%%%%%%%%%%%%%%%%%%%%%%%%%%%%%%%%%%%%%%%%%%%%%%%%%%%%%%%%%%%
%
%
%    \recall{label} prints the counter corresponding to label.
%
\def\recall#1{{%
%  call to \@cres:undefined with its first argument
\@cres:undefined{#1}
%  the second argument
{\@cres:warning{undefined "#1"}%
    {{\bf???}\if@cres:tracing\kern.5em {\@cres:tracingfont#1}\fi}}
%  and finally the third argument
{{\edef\@cres:inter{\csname @cres::#1\endcsname}%
    \expandafter\@cres:printref\@cres:inter}}}}
%
%%%%%%%%%%%%%%%%%%%%%%%%%%%%%%%%%%%%%%%%%%%%%%%%%%%%%%%%%%%%%%%%%%%%%%%%%%
%
%
%    \recallpage{label} prints the page number on which the counter 
%    corresponding to label appears.
%
\def\recallpage#1{{%
\if@cres:pagenos\global\@cres:pagenosfalse
   \@cres:warning{page numbers : do not forget to typeset twice}\fi
%  call to \@cres:undefined with its first argument
\@cres:undefined{#1}
%  the second argument
{\@cres:warning{undefined "#1"}%
    {{\bf???}\if@cres:tracing\kern.5em {\@cres:tracingfont#1}\fi}}
%  and finally the third argument
{{\edef\@cres:inter{\csname @cres::#1\endcsname}%
    \expandafter\@cres:printpageref\@cres:inter}}}}
%
%%%%%%%%%%%%%%%%%%%%%%%%%%%%%%%%%%%%%%%%%%%%%%%%%%%%%%%%%%%%%%%%%%%%%%%%%%
%
%
%    \recalltitletext{label} prints the title text corresponding to 
%    the label. 
%
\def\recalltitletext#1{{%
%  call to \@cres:undefined with its first argument
\@cres:undefined{#1}
%  the second argument
{\@cres:warning{undefined "#1"}%
    {{\bf???}\if@cres:tracing{\kern.5em \@cres:tracingfont#1}\fi}}
%  and finally the third argument
{{\edef\@cres:inter{\csname @cres::#1\endcsname}%
    \expandafter\@cres:printtitletext\@cres:inter}}}}
%
%%%%%%%%%%%%%%%%%%%%%%%%%%%%%%%%%%%%%%%%%%%%%%%%%%%%%%%%%%%%%%%%%%%%%%%%%%
%
%
%    \tracinglabels is the externally accessible macro to activate
%    the tracing of labels in the typeset.
%
\def\tracinglabels{\@cres:tracingtrue}
%
%%%%%%%%%%%%%%%%%%%%%%%%%%%%%%%%%%%%%%%%%%%%%%%%%%%%%%%%%%%%%%%%%%%%%%%%%%
%
%
%    \previousreferences{filename} reads filename.acres
%
\def\previousreferences#1{{%
\openin\@cres:myin=#1.acres %
\ifeof\@cres:myin\@cres:warning{#1.acres does not exist}%
\else\read\@cres:myin to\@cres:reference
    \ifx\@cres:reference\@cres:version
        \loop\read\@cres:myin to\@cres:reference
            \ifeof\@cres:myin\@cres:noteoffalse\else\@cres:noteoftrue\fi
        \if@cres:noteof\expandafter\@cres:memorize\@cres:reference
% N.B. "\read file to ref" creert een extra spatie aan het einde van 
% "ref" die niet te onderdrukken schijnt te zijn.
        \repeat
    \else\@cres:warning{wrong version of ACRES}%
    \fi
\fi
\closein\@cres:myin}}
%
%  Check whether this ^ creates blanks in output at unwanted places
%
%%%%%%%%%%%%%%%%%%%%%%%%%%%%%%%%%%%%%%%%%%%%%%%%%%%%%%%%%%%%%%%%%%%%%%%%%%
%
%
%    \outerlevelinclude and \outerlevelsuppress are the externally
%    accessible macros to do what their names indicate.
%
\def\outerlevelinclude{\def\@cres:chapsuppress{2}}
\def\outerlevelsuppress{\def\@cres:chapsuppress{1}}
\def\Outerlevelsuppress{\def\@cres:chapsuppress{0}}
%
%%%%%%%%%%%%%%%%%%%%%%%%%%%%%%%%%%%%%%%%%%%%%%%%%%%%%%%%%%%%%%%%%%%%%%%%%%
%
%
%    \@cres:initialize memorizes the previously created labels and
%    starts a new file of labels.
%
\def\@cres:initialize{%
\previousreferences{\jobname}%
\openout\@cres:myout=\jobname.acres %
\@cres:shipout{\@cres:version}}
%
%%%%%%%%%%%%%%%%%%%%%%%%%%%%%%%%%%%%%%%%%%%%%%%%%%%%%%%%%%%%%%%%%%%%%%%%%%
%
%
%    \IncreaseByOne{\counter} and \DecreaseByOne{\counter} do exactly
%    what they say. They should be used with extreme caution.
%
\def\IncreaseByOne#1{\expandafter\@cres:stripperplus#1}
\def\DecreaseByOne#1{\expandafter\@cres:stripperminus#1}
\def\@cres:stripperminus#1#2{%
\ifx\@cres:cntr#1\global\@cres:oponcntr\advance#2 by-1 
\else\ifx\@cres:chap#1\global\@cres:oponchap\advance#2 by-1 
    \else\@cres:warning{DecreaseByOne does not apply}\fi
\fi}
\def\@cres:stripperplus#1#2{%
\ifx\@cres:cntr#1\global\@cres:oponcntr\advance#2 by1 
\else\ifx\@cres:chap#1\global\@cres:oponchap\advance#2 by1 
    \else\@cres:warning{IncreaseByOne does not apply}\fi
\fi}
%
%%%%%%%%%%%%%%%%%%%%%%%%%%%%%%%%%%%%%%%%%%%%%%%%%%%%%%%%%%%%%%%%%%%%%%%%%%
%%%%%%%%%%%%%%%%%%%%%%%%%%%%%%%%%%%%%%%%%%%%%%%%%%%%%%%%%%%%%%%%%%%%%%%%%%
%%%%%%%%%%%%%%%%%%%%%%%%%%%%%%%%%%%%%%%%%%%%%%%%%%%%%%%%%%%%%%%%%%%%%%%%%%
%
%
%    Here we give the macros that create index files for the program
%    makeindex. They create lines of the form 
%
%         \indexentry{text}{page number}
%
%    The "text" is copied texte from the argument, and "page number"
%    is the page number on which the reference is situated.
%    The macros are slightly modified copies of the corresponding macros
%    of LaTeX.
%
%%%%%%%%%%%%%%%%%%%%%%%%%%%%%%%%%%%%%%%%%%%%%%%%%%%%%%%%%%%%%%%%%%%%%%%%%%
%
%
%    \@cres:makeother changes the catcode of its argument to "other"
%    and \@cres:sanitize uses this to change the catcodes of most of 
%    the special characters to "other" so we can write the arguments 
%    literally on "index" files.
%
\def\@cres:makeother#1{\catcode`#112\relax}
\def\@cres:sanitize{\@cres:makeother\ \@cres:makeother\\%
\@cres:makeother\$\@cres:makeother\&\@cres:makeother\#\@cres:makeother\^%
\@cres:makeother\^^K\@cres:makeother\_\@cres:makeother\^^A%
\@cres:makeother\%\@cres:makeother\~\@cres:makeother\@}
%
%%%%%%%%%%%%%%%%%%%%%%%%%%%%%%%%%%%%%%%%%%%%%%%%%%%%%%%%%%%%%%%%%%%%%%%%%%
%
%
%    \@cres:bsphack and \@cres:esphack are used to create macros that
%    do not create any blanks, horizontal or vertical spacing at all,
%    independent of where these macros are called.
%
\newdimen\@cres:savsk
\newcount\@cres:savsf
\def\@cres:bsphack{\@cres:savsk\lastskip 
    \ifhmode\@cres:savsf\spacefactor\fi}
\def\@cres:esphack{\relax\ifhmode\spacefactor\@cres:savsf
     {}\ifdim \@cres:savsk >\z@ \ignorespaces 
  \fi \fi}
%
%%%%%%%%%%%%%%%%%%%%%%%%%%%%%%%%%%%%%%%%%%%%%%%%%%%%%%%%%%%%%%%%%%%%%%%%%%
% 
%
%    \@cres:ignorethis is a macro with a single argument. When used
%    it produces just nothing, not even additional spacing. Using this
%    we initialize various "index" macros as this macro. That way these
%    can be ignored until they are needed. Then a simple call to a
%    "makeindex" command creates the actual "index" macros.
%
\def\@cres:ignorethis{\@cres:bsphack\begingroup 
   \@cres:sanitize\@cres:ignore}
\def\@cres:ignore#1{\endgroup\@cres:esphack}
%
%%%%%%%%%%%%%%%%%%%%%%%%%%%%%%%%%%%%%%%%%%%%%%%%%%%%%%%%%%%%%%%%%%%%%%%%%%
%
%
%    \@cres:marginstar writes a star in the margin in case the option
%    \tracingindexentries is active. It does so whenever the index macro
%    is called. For the moment no distinction is made for the four
%    different index files.
%    The marginstar macro is copied from the TeX book p316, exercice 14.28.
%
\newif\if@cres:tracingindex\@cres:tracingindexfalse
\def\tracingindexentries{\@cres:tracingindextrue}
\def\strutdepth{\dp\strutbox}
\def\marginalstar#1{\strut\vadjust{\kern-\strutdepth\specialstar{#1}}}
\def\specialstar#1{\vtop to \strutdepth{\baselineskip\strutdepth
     \vss\llap{\rm#1}\null}}
\def\@cres:marginstar#1{\if@cres:tracingindex\marginalstar{#1}\fi}
%
%%%%%%%%%%%%%%%%%%%%%%%%%%%%%%%%%%%%%%%%%%%%%%%%%%%%%%%%%%%%%%%%%%%%%%%%%%
%
%
%    \makeindexfile#1#2#3#4#5 creates a macro \#3 with a single argument
%    and it opens the index file number #1 under the name
%
%        \jobname#2
%
%    only if #4 equals 1. If #4 is different from 1, the macro \#3 does 
%    strictly nothing, it suppresses the argument and does not create
%    blanks.  If #4=1, the macro \#3 writes its single argument verbatim
%    on the file \jobname#2.
%    For the moment only four different index files can be handled.
%    If \tracingindexentries is active, it writes #5 in the margin.
%
\newwrite\@cres:indexfileone    \def\@cres:one{1}
\newwrite\@cres:indexfiletwo    \def\@cres:two{2}
\newwrite\@cres:indexfilethree  \def\@cres:three{3}
\newwrite\@cres:indexfilefour   \def\@cres:four{4}
\def\makeindexfile#1#2#3#4#5{%
\def\@cres:testone{#1}\def\@cres:testfour{#4}%
\ifx\@cres:testfour\@cres:one  % should it be "active"?
\ifx\@cres:testone\@cres:one
% a nested if construction instead of an ifcase to avoid TeX error
% messages when not a number is provided
      \openout\@cres:indexfileone=\jobname#2
      \expandafter\def\csname#3\endcsname
         {\@cres:bsphack\begingroup
           \@cres:sanitize\@cres:marginstar{#5}%
% in case one wants to trace the index entries in the text
           \@cres:wrindex\@cres:indexfileone}%
   \else\ifx\@cres:testone\@cres:two
      \openout\@cres:indexfiletwo=\jobname#2
      \expandafter\def\csname#3\endcsname
         {\@cres:bsphack\begingroup
           \@cres:sanitize\@cres:marginstar{#5}%
           \@cres:wrindex\@cres:indexfiletwo}%
   \else\ifx\@cres:testone\@cres:three
      \openout\@cres:indexfilethree=\jobname#2
      \expandafter\def\csname#3\endcsname
         {\@cres:bsphack\begingroup
           \@cres:sanitize\@cres:marginstar{#5}%
           \@cres:wrindex\@cres:indexfilethree}%
   \else\ifx\@cres:testone\@cres:four
      \openout\@cres:indexfilefour=\jobname#2
      \expandafter\def\csname#3\endcsname
         {\@cres:bsphack\begingroup
           \@cres:sanitize\@cres:marginstar{#5}%
           \@cres:wrindex\@cres:indexfilefour}%
   \else\@cres:warning{#1 is an illegal index file number}%
        \expandafter\let\csname#3\endcsname\@cres:ignorethis
% if illegal index file number, then it is reset to "inactive"
   \fi\fi\fi\fi
\else
\expandafter\let\csname#3\endcsname\@cres:ignorethis
\fi
}
\def\@cres:wrindex#1#2{\let\the=0%
   \xdef\@cres:gtempa{\write#1{\string
      \indexentry{#2}{\the\pageno}}}\endgroup\@cres:gtempa
\@cres:esphack}
%
%%%%%%%%%%%%%%%%%%%%%%%%%%%%%%%%%%%%%%%%%%%%%%%%%%%%%%%%%%%%%%%%%%%%%%%%%%
%%%%%%%%%%%%%%%%%%%%%%%%%%%%%%%%%%%%%%%%%%%%%%%%%%%%%%%%%%%%%%%%%%%%%%%%%%
%%%%%%%%%%%%%%%%%%%%%%%%%%%%%%%%%%%%%%%%%%%%%%%%%%%%%%%%%%%%%%%%%%%%%%%%%%
%
%
%    Here we provide the macros to print the index file in double
%    column format. It is based on the output file of the program
%    makeindex. Since this creates a LaTeX type file, we have to
%    take care of \begin{theindex} and \end{theindex} commands.
%    We use these to switch to double column format (the \begin{theindex})
%    and back to single column format (the \end{theindex}). The user
%    has to provide himself the headers of the index.
%    The macro for double column format is copied from the book by
%    Paul W. Abrahams : TeX for the Impatient.
%
%%%%%%%%%%%%%%%%%%%%%%%%%%%%%%%%%%%%%%%%%%%%%%%%%%%%%%%%%%%%%%%%%%%%%%%%%%
%
%
%    We start with the double column format.
%
\newskip\@cres:abovedoublecolumnskip
\newskip\@cres:belowdoublecolumnskip
\@cres:abovedoublecolumnskip = \bigskipamount
\@cres:belowdoublecolumnskip = \bigskipamount
\newdimen\@cres:gutter   \@cres:gutter = 2pc
\newdimen\@cres:doublecolumnhsize \@cres:doublecolumnhsize = \hsize
\newbox\@cres:partialpage   \newdimen\@cres:singlecolumnhsize
\newdimen\@cres:singlecolumnvsize  \newtoks\@cres:previousoutput
\def\doublecolumns{%
   \@cres:doublecolumnhsize = \hsize
   \@cres:previousoutput = \expandafter{\the\output}%
   \advance\@cres:doublecolumnhsize by -\@cres:gutter
   \divide\@cres:doublecolumnhsize by 2
   \output = {\global\setbox\@cres:partialpage = 
      \vbox{\unvbox255\vskip\@cres:abovedoublecolumnskip}}%
   \par \pagegoal = \pagetotal \break
   \output = {\@cres:doublecolumnoutput}%
   \@cres:singlecolumnhsize = \hsize \@cres:singlecolumnvsize = \vsize
   \hsize = \@cres:doublecolumnhsize  \vsize = 2\vsize
}%
\def\@cres:doublecolumnsplit{%
   \splittopskip = \topskip  \splitmaxdepth = \maxdepth
   \dimen0 = \@cres:singlecolumnvsize
      \advance\dimen0 by -\ht\@cres:partialpage
      \advance\dimen0 by -\ht\footins
      \ifvoid\footins\else
   \advance\dimen0 by -\skip\footins \fi
      \advance\dimen0 by -\ht\topins
      \ifvoid\topins\else
   \advance\dimen0 by -\skip\topins \fi
   \begingroup
      \vbadness = 10000
      \global\setbox1 = \vsplit255 to \dimen0
      \global\setbox3 = \vsplit255 to \dimen0
      \wd1 = \hsize  \wd3 = \hsize
   \endgroup
   \global\setbox4 = 
      \vbox{\unvbox255 \penalty\outputpenalty}%
   \global\setbox255 = \vbox{\unvbox\@cres:partialpage
      \hbox to \@cres:singlecolumnhsize{\box1\hfil\box3}}}%
\def\@cres:doublecolumnoutput{\@cres:doublecolumnsplit
   \hsize = \@cres:singlecolumnhsize \vsize = \@cres:singlecolumnvsize
   \the\@cres:previousoutput \unvbox4}%
\def\singlecolumn{%
   \par
   \output = {\global\setbox1 = \box255}%
   \pagegoal = \pagetotal \break \setbox255 = \box1
   {  \@cres:singlecolumnvsize = \ht\@cres:partialpage
      \advance\@cres:singlecolumnvsize by \ht\footins
      \ifvoid\footins\else
   \advance\@cres:singlecolumnvsize by \skip\footins\fi
      \advance\@cres:singlecolumnvsize by \ht\topins
      \ifvoid\topins\else
   \advance\@cres:singlecolumnvsize by \skip\topins\fi
      \dimen0 = \ht255  \divide\dimen0 by 2
      \advance\@cres:singlecolumnvsize by \dimen0
      \advance\@cres:singlecolumnvsize by .5\baselineskip
      \@cres:doublecolumnsplit  }
   \hsize = \@cres:singlecolumnhsize  \vsize = \@cres:singlecolumnvsize
   \output = \expandafter{\the\@cres:previousoutput}%
   \unvbox255  \vskip\@cres:belowdoublecolumnskip
   \nointerlineskip
}%
%
%%%%%%%%%%%%%%%%%%%%%%%%%%%%%%%%%%%%%%%%%%%%%%%%%%%%%%%%%%%%%%%%%%%%%%%%%%
%
%
%    \printindexfile{suffix} reads the index file \jobname suffix
%    (note: no period between the jobname and the suffix!)
%    and prints it in double columns format. It supposes that this
%    index file is made by the program makeindex. This implies that
%    it supposes that this file begins with the LaTeX command
%    \begin{theindex} and ends with the LaTeX command \end{theindex}.
%    In between it supposes entries of the form
%        \item text
%        \subitem text
%        \subsubitem text
%        \indexspace
%    It correctly interprets these commands and reverts to the original
%    definitions afterward.
%    If the index file does not start with \begin{theindex}, the above
%    described macros either do not exist or are not correctly defined.
%
%
\def\printindexfile#1{%
  \def\begin##1{\begingroup
    \def\@cres:compareb{##1}\def\@cres:comparetheindex{theindex}%
    \ifx\@cres:compareb\@cres:comparetheindex
    \else\@cres:warning{Illegal LaTeX begin command: \@cres:compareb}\fi
    \parindent=0pt
    \def\item{\par\hangindent=40pt}%
    \def\subitem{\par\hangindent=40pt 
                     \vrule height0pt depth0pt width20pt}%
    \def\subsubitem{\par\hangindent=40pt 
                     \vrule height0pt depth0pt width30pt}%
    \def\end####1{\def\@cres:comparee{####1}%
      \ifx\@cres:comparee\@cres:comparetheindex
      \else\@cres:warning{Illegal LaTeX end command: \@cres:comparee}\fi
      \singlecolumn
      \endgroup
    }%
    \doublecolumns
  }%
  \input \jobname#1}%
%
%%%%%%%%%%%%%%%%%%%%%%%%%%%%%%%%%%%%%%%%%%%%%%%%%%%%%%%%%%%%%%%%%%%%%%%%%%
%%%%%%%%%%%%%%%%%%%%%%%%%%%%%%%%%%%%%%%%%%%%%%%%%%%%%%%%%%%%%%%%%%%%%%%%%%
%%%%%%%%%%%%%%%%%%%%%%%%%%%%%%%%%%%%%%%%%%%%%%%%%%%%%%%%%%%%%%%%%%%%%%%%%%
%%%%%%%%%%%%%%%%%%%%%%%%%%%%%%%%%%%%%%%%%%%%%%%%%%%%%%%%%%%%%%%%%%%%%%%%%%
%
%
%    And then we restore the catcodes of @ and : before returning control
%    to the calling program.
%
\catcode`\@=\csname ACRES chardef at\endcsname
\catcode`\:=\csname ACRES chardef colon\endcsname
%%%%%%%%%%%%%%%%%%%%%%%%%%%%%%%%%%%%%%%%%%%%%%%%%%%%%%%%%%%%%%%%%%%%%%%%%%
%
%  Here ends the ACRES macro package. The following macros are
%  defined for general use outside this package :
%
%  \newlevel{name}{mode}{separator}
%  \newcounter{name}{mode}{separator}{level}
%  \memorize{\name}={label}
%  \memorizetitle{\name}={label}{intertext}{titletext}
%  \recall{label}
%  \recallpage{label}
%  \recalltitletext{label}
%  \tracinglabels
%  \previousreferences{filename}
%  \outerlevelinclude
%  \outerlevelsuppress
%  \IncreaseByOne{\name}
%  \DecreaseByOne{\name}
%  \withoutnumber
%
%  \makeindexfile{file number}{file suffix}{index macro name}{active=1}
%
%  \doublecolumns
%  \singlecolumn
%
%%%%%%%%%%%%%%%%%%%%%%%%%%%%%%%%%%%%%%%%%%%%%%%%%%%%%%%%%%%%%%%%%%%%%%%%%%
%
%  Check what happens if there are more than 9 levels of chapters/counters
%  \noexpand\@cres:cntr   {   \the\@cres:cntrdepth    }   ???

\newlevel{sectionnum}{1}{.}
{1}{.}{1}

\def\thmm#1#2{{\memorize\theoremnummer={#1} #2}}
\def\formula#1{\tag{\memorize\theoremnummer={#1}}}

\def\recalt#1{[\recall{#1}]}
\def\recalf#1{(\recall{#1})}

\def\recals#1{\S\recall{#1}}

\outerlevelinclude

\def\oversetalign#1\to#2{\overset{\text{\rlap{\hss#1\vrule depth3pt width0pt}}}\to#2}

\def\Ad{{\operatorname{Ad}}}
\def\aglgrp/{an \glgrp/}
\def\agmfd/{an \gmfd/}
\def\agvs/{an \gvs/}
\def\ah{{\hat a}}
\def\alphah{{\hat \alpha}}
\def\Aut{\operatorname{Aut}}
\def\betah{{\hat\beta}}
\def\bh{{\hat b}}
\def\body{{\bold B}}
\def\CA{{\Cal A}}
\def\CC{{\bold C}}
\def\Coad{{\operatorname{Coad}}}
\def\contrf#1#2{\contrfoper(#1)#2}
\def\contrfoper{{\iota}}
\def\contrs#1#2{{\langle\mskip1.5mu}
    #1\mskip1.5mu |\mskip1.5mu #2
    {\mskip1.5mu\rangle}}
\def\datum{\the\day/\the\month/\the\year}
\def\eexp{{\operatorname{e}}}
\def\End{\operatorname{End}}
\def\extder{{\operatorname{d}}}
\def\etab{{\bar\eta}}

\def\fracp#1#2{\frac{\partial #1}{\partial #2}}
\def\gh{{\hat g}}
\def\glalg/{$\CA$-Lie algebra}
\def\glgrp/{$\CA$-Lie group}
\def\gmfd/{$\CA$-mani\-fold}
\def\gvs/{$\CA$-vector space}
\def\ie{i.e.}
\def\itemize{$\bullet$ }
\def\Liealg#1{{\frak#1}}
\def\Lieder{{\Cal L}}
\def\mapob{\kern.3em} 
\def\mo{^{-1}}
\def\NN{{\bold N}}
\def\Orbit{{\Cal O}}
\def\QEDbox{\hbox{\lower2.3pt\vbox{\hrule\hbox
   {\vrule\kern1pt\vbox{\kern1.7pt\hbox{$\scriptstyle
   QED$}\kern.6pt}\kern1pt\vrule}\hrule}}}
\def\QED{\hskip0.01em plus 40pt\null{} \null\nobreak\hfill
   \kern3pt\QEDbox} 
\def\RAd{{\overleftarrow{\operatorname{Ad}}}}
\def\repcomm#1,#2{\bigl[[#1,(#2)]\bigr]}
\def\repprod#1{(\!(#1)\!)}
\def\restricted{|}
\def\RR{{\bold R}}
\def\scirc{\,{\raise 0.8pt\hbox{$\scriptstyle\circ$}}\,}
\def\SS{{\bold S}}
\def\stresd#1{{\it#1\/}}
\def\stress#1{{\it #1\/}}
\def\xb{{\bar x}}
\def\xh{{\hat x}}
\def\xib{{\bar\xi}}
\def\xih{{\hat \xi}}
\def\xnul{x^o}
\def\xnulb{\xb^o}
\def\yb{{\bar y}}

\def\zb{{\bar z}}

\def\ZZ{{\bold Z}}
\def\[{{\mskip2mu\mathop[\mskip2mu}}
\def\]{{\mskip3mu\mathop]\mskip1.5mu}}

\topmatter
\title Geometric Quantization of Superorbits: a Case Study
\endtitle
\author G.M. Tuynman \endauthor
\address Laboratoire Paul Painlev{\'e},
U.M.R. CNRS 8524 et UFR de Math{\'e}matiques;
Universit{\'e} de Lille I; F-59655 Villeneuve d'Ascq Cedex; France\endaddress
\email Gijs.Tuynman\@univ-lille1.fr \endemail
\abstract  
By decomposing the regular representation of a particular (Heisenberg-like) Lie supergroup into irreducible subspaces, we show that not all of them can be obtained by applying geometric quantization to coadjoint orbits with an even symplectic form. However, all of them can be obtained by introducing coadjoint orbits through non-homogeneous points and with non-homogeneous symplectic forms as described in \cite{Tu10}. In this approach it turns out that the choice of a polarization can change (dramatically) the representation associated to an orbit. On the other hand, the procedure is not completely mechanical (meaning that some parts have to be done ``by hand''), hence work remains to be done in order to understand all details of what is happening.
\endabstract
\endtopmatter

\document

\head{\sectionnum. Introduction}\endhead

In \cite{Tu10} I introduced the notion of a non-homogeneous symplectic form on a supermanifold and I constructed a possible prequantization of such a symplectic supermanifold. The question remained whether non-homogenous symplectic forms are interesting to study and whether the proposed prequantization is the correct one to use. In this paper I will show by an explicit example that non-homogeneous symplectic forms play a role in representation theory. The example I will study is a Heisenberg-like group, meaning that we have two graded vector spaces $E$ and $C$ and an even graded skew-symmetric bilinear form $\Omega: E\times E \to C$ and that we look at the group $G$ which is as a manifold the even part of $E\times C$: $G=(E\times C)_0$ equipped with the group structure
$$
(a,b) \cdot (\hat a, \hat b) = (a+\hat a, b+ \hat b + \tfrac12 \contrf{a,\hat
a}\Omega)
\mapob.
$$
On this group I look at the (left-) regular representation consisting of square-integrable functions on $G$ with the group action 
$$
(\Phi_g f)(\gh) = f(g\mo\cdot \gh)
\mapob.
$$
Using ordinary Fourier direct integrals and Berezin-Fourier direct integrals, it is fairly easy to decompose this regular representation into invariant subspaces. Depending upon the Fourier parameters, some of these subspaces can be decomposed further in direct (Berezin-) Fourier integrals and others can be seen to be a direct sum of two invariant subspaces. A summary of the obtained decomposition is given at the end of \recals{naiveregularrep}.

Since we use direct integrals, these subspaces are not really subspaces in the strict sense of the word: the functions in question are no longer square-integrable over the group. But this is a standard technical detail. More important is the fact that these subspaces for an odd Berezin-Fourier parameter are not graded subspaces in any reasonable sense. We thus provide in \recals{invariantfamilies} a framework for dealing with this kind of subspaces by defining what we mean by an odd family decomposition of a graded vector space and what it should mean when we say that such a decomposition is irreducible. In the next section we apply these definitions to show that the decomposition of our regular representation obtained in \recals{naiveregularrep} is irreducible. We thus have completely decomposed the regular representation into irreducible parts. 

The next task is to see how the irreducible representations obtained in the decomposition of the regular representation correspond to the representations obtained by quantizing coadjoint orbits. It turns out that there are four types of coadjoint orbits: $0$-dimensional ones, $2\vert 2$-dimensional ones with an even symplectic form, $2\vert2$-dimensional ones with an odd symplectic form and $3\vert3$-dimensional ones with a non-homogenous symplectic form. In order to compute the associated representations, it turns out that we have to adjust the prequantization procedure given in \cite{Tu10} slightly by introducing an odd parameter for the odd part of the symplectic form. This parameter is the odd counterpart of the parameter $\hbar$ that is used for the even part. The difference is that the actual value of $\hbar$ can be taken to be $1$ by adjusting the physical units, whereas we cannot ``rescale'' the odd parameter. 

In the ungraded case it is (at least morally) true that the representation is independent of the choice of the invariant polarization. In the graded case this is true as long as the graded dimension of the polarization does not change; (invariant) polarizations with different graded dimensions will give rise to different representations. For the orbits of dimension $0$ and those of dimension $2\vert2$ with an even symplectic form there is only one graded dimension possible for an invariant polarization, but for $2\vert2$-dimensional orbits with an odd symplectic form there are three different possibilities and for $3\vert3$-dimensional orbits there are two possibilities. Each of these possibilities gives rise to a representation of our supergroup.

When we compare these representations with those obtained in the decomposition of the regular representation, we find the following result: all representations associated to $0$-dimensional orbits appear; some but not all representations associated to $2\vert2$-dimensional orbits with an even symplectic form appear; some but not all representations associated to $2\vert2$-dimensional orbits with an odd symplectic form and a $3\vert3$-dimensional polarization appear; and finally, some but not all representations associated to $3\vert3$-dimensional orbits (with a non-homogeneous symplectic form) with a $3\vert2$-dimensional polarization appear. Representations associated to orbits and polarizations of a different graded dimension do not appear in the regular representation. We thus see that representations associated to orbits with a non-homogeneous symplectic form intervene in the regular representation, showing that this kind of symplectic supermanifold is an interesting object to study. 

However, several problems still have to be solved. In the first place the question how to interpret the odd parameter introduced in the prequantization for the odd part of the symplectic form. And then how to make the correspondence between the irreducible representations appearing in the regular representation and those obtained by geometric quantization correctly. Because even though there is an obvious correspondence, there remains a problem how to identify real parameters with odd parameters (see the end of this paper). And then a rather long list of technical problems have to be solved, among others how to define densities for the quantization of symplectic supermanifolds with a non-homogeneous form (said differently, how to define in an intrinsic way the scalar product on the function spaces that appear in quantization) and how to define in a natural way the notion of equivalence between odd families of representations.

\head{\sectionnum. Preliminaries}\endhead

I will work with the geometric $H^\infty$ version of DeWitt supermanifolds,
which is equivalent to the theory of graded manifolds of Leites and Kostant
(see \cite{DW}, \cite{Ko}, \cite{Le}, \cite{Ro}, \cite{Tu04}). Any reader using a
(slightly) different version of supermanifolds should be able to translate the
results to her\slash his version of supermanifolds.

\bigskip

\itemize
The basic graded ring will
be denoted as $\CA$ and we will think of it as the exterior algebra $\CA
= \Lambda V$ of an infinite dimensional real vector space $V$.

\itemize
Any element $x$ in a graded space splits into an even and an odd part $x=x_0 + x_1$.

\itemize
All (graded) objects over the basic ring $\CA$ have an underlying real
structure, called their body, in which all nilpotent elements in $\CA$ are
ignored\slash killed. This forgetful map is called the body map, denoted by the
symbol $\body$. For the ring $\CA$ this is the  map\slash
projection $\body : \CA =\Lambda V \to \Lambda^0 V = \RR$.

\itemize
If $\omega$ is a $k$-form and $X$ a vector field, we denote the contraction of
the vector field $X$ with the $k$-form $\omega$ by $\contrf{X}{\omega}$, which
yields a $k-1$-form. If $X_1, \dots, X_\ell$ are $\ell\le k$ vector fields, we
denote the repeated contraction of $\omega$ by $\contrf{X_1, \cdots,
X_\ell}{\omega}$. More precisely:
$$
\contrf{X_1, \cdots, X_\ell}{\omega} = 
\Bigl( \contrf{X_1}{} \scirc \cdots \scirc \contrf{X_\ell}{}
\Bigr) \, \omega
\mapob.
$$
In the special case $\ell = k$ this definition differs by a factor
$(-1)^{k(k-1)/2}$ from the usual definition of the evaluation of a $k$-form on
$k$ vector fields. This difference is due to the fact that in ordinary
differential geometry repeated contraction with $k$ vector fields corresponds
to the direct evaluation in the reverse order. And indeed, $(-1)^{k(k-1)/2}$ is
the signature of the permutation changing $1,2,\dots,k$ in $k,k-1,\dots,2,1$.
However, in graded differential geometry this permutation not only introduces
this signature, but also signs depending upon the parities of the vector
fields. These additional signs are avoided by our definition.

\itemize
The evaluation of a \stress{left} linear map $\mu$ on a vector $v$ is denoted as
$\contrs{v}{\mu}$. For the contraction of a multi-linear form with a vector we
will use the same notation as for the contraction of a differential form with a
vector field. In particular, we denote the evaluation of a left bilinear map
$\Omega$ on a vector $v$ by $\contrf{v}{\Omega}$, which yields a left linear map
$w\mapsto \contrs{w}{\contrf{v}{\Omega}} \equiv \contrf{w,v}{\Omega}$.

\itemize
If $E$ is \agvs/, $E^*$ will denote the \stress{left} dual of $E$, \ie, the
space of all \stress{left} linear maps from $E$ to $\CA$.

\itemize
If $G$ is \aglgrp/, then its \glalg/ $\Liealg g$ is $\Liealg g = T_eG$, whose
Lie algebra structure is given by the commutator of left-invariant vector
fields (who are determined by their value at $e\in G$).

\itemize
If $\Phi : G \times M \to M$ denotes the (left) action of \aglgrp/ $G$ on
\agmfd/ $M$, then for all $v \in \Liealg g = T_eG$ the associated
\stress{fundamental vector field} $v^M$ on $M$ is defined as $v^M\restricted_m =
- T_{(e,m)}\Phi (v,0)$. The minus sign is conventional and ensures that the map
from $\Liealg g$ to vector fields on $M$ is a homomorphism of \glalg/s. 

Similarly, if $\Phi : M \times G \to M$ is a right action of $G$ on $M$, then
the fundamental vector field $v^M$ associated to $v\in \Liealg g$ is defined as 
$v^M\restricted_m = T_{(m,e)}\Phi (0,v)$. And again the map from $\Liealg g$ to
vector fields on $M$ is a morphism of \glalg/s. In the special case when $M =
G$ with the natural right action on itself, the fundamental vector fields are
exactly the left-invariant vector fields on $G$.

\head{\memorize\sectionnum={naiveregularrep}. The group and its (left) regular representation }\endhead

Consider the $\CA$-vector space $E$ of (graded) dimension $4\vert4$ with basis $e_1, e_2, e_3, k_0$, $e_4, e_5, e_6, k_1$ of which the first four are even (and the last four are odd). The group $G$ is the even part $G=E_0$ with coordinates $(a^1, a^2, a^3, b, \alpha^4, \alpha^5, \alpha^6, \beta)$ with respect to the given basis. The group law is given as
$$
\pmatrix \ah^1\\ \ah^2\\ \ah^3\\ \bh\\ \alphah^4\\ \alphah^5\\ \alphah^6\\ \betah
\endpmatrix
\cdot
\pmatrix a^1\\ a^2\\ a^3\\ b\\ \alpha^4\\ \alpha^5\\ \alpha^6\\ \beta
\endpmatrix
=
\pmatrix \ah^1+a^1\\ \ah^2+a^2\\ \ah^3+a^3
\\ 
\bh + b + \tfrac12(\ah^2a^1 - \ah^1a^2 - \alphah^5\alpha^5 + \alphah^6 \alpha^6)
\\ 
\alphah^4+\alpha^4\\ \alphah^5+\alpha^5\\ \alphah^6+\alpha^6
\\ 
\betah+\beta + \tfrac12 (\alphah^4a^1 - \ah^1 \alpha^4 + \alphah^5a^3 - \ah^3\alpha^5 )
\endpmatrix
\mapob.
$$
The neutral element has all coordinates zero and taking the inverse is reversing the sign of all coordinates. The action $\Phi_g$ of an element $g\in G$ on a function $f:G\to \CA^\CC$ is defined as
$$
(\Phi_{\gh} f)(g) = f(\gh\mo \cdot g)
\mapob.
$$
Any smooth function $f:G\to \CA^\CC$ can be written as
$$
\multline
f(a^i, \alpha^j, b,\beta)
=
f_{()}(a^i,b)
+
\beta\cdot f_{(0)}(a^i,b)
\\
\kern5em
+
\sum_{t=1}^3\ \sum_{4\le j_1<\cdots <j_t\le 6} 
\alpha^{j_1} \cdots \alpha^{j_t}\cdot f_{(j_1, \dots, j_t)}(a^i,b)
\\
+
\sum_{t=1}^3\ \sum_{4\le j_1<\cdots <j_t\le 6} 
\beta\cdot\alpha^{j_1} \cdots \alpha^{j_t}\cdot f_{(0,j_1, \dots, j_t)}(a^i,b)
\mapob,
\endmultline
\tag{\memorize\theoremnummer={LtwoformoffunctionsonG}}
$$
where the $16$ functions $f_{(\dots)}$ are (equivalent to) ordinary smooth functions of the $4$ real (even) coordinates $a^1, a^2, a^3,b$.

We now wish to study the (left) regular representation $V$ of $G$, which consists of functions on $G$ of the form \recalf{LtwoformoffunctionsonG} with $f_P$ an $L^2$ function on $\RR^4$, together with the action $\Phi$ defined above. Our purpose is to obtain a complete decomposition of this representation into irreducible subspaces. 
To do so, we start looking at the space $V_{(\ell_0,\ell_2,\ell_3,\lambda_0, \lambda_4)}$ of functions on $G$ of the form
$$
\multline
f(a^i, \alpha^j, b,\beta)
=
t_{(\ell_0,\ell_2,\ell_3,\lambda_0, \lambda_4)}(a^1, \alpha^5, \alpha^6) \cdot
\\
\eexp^{i\ell_2a^2}
\cdot\eexp^{i\ell_3a^3}
\cdot\eexp^{i\lambda_4\alpha^4}
\cdot\eexp^{i\ell_0(b+\frac12a^1a^2)}
\cdot\eexp^{i\lambda_0(\beta+ \frac12a^1\alpha^4 - \frac12a^3\alpha^5)}
\mapob,
\endmultline
\tag\memorize\theoremnummer={firstfouriermode}
$$
with $t_{(\ell_0,\ell_2,\ell_3,\lambda_0, \lambda_4)}$ a function of one even and two odd coordinates. This is essentially a Fourier mode with respect to the coordinates $a^2, a^3, \alpha^4, b,\beta$. More precisely, the functions $t_{(\ell_0,\ell_2,\ell_3,\lambda_0, \lambda_4)}$ are the Fourier transform with respect to the coordinates  $a^2, a^3, \alpha^4, c, \gamma$ of a function $f$  with 
$$
c= b + \tfrac12  a^1a^2
\qquad,\qquad
\gamma = \beta+ \tfrac12 a^1\alpha^4 -\tfrac12  a^3\alpha^5
$$
given explicitly by
$$
\multline
t_{(\ell_0,\ell_2,\ell_3,\lambda_0, \lambda_4)}(a^1, \alpha^5, \alpha^6)
= (2\pi)^{-3} \cdot
\int f\bigl(a^i, \alpha^j, c-\tfrac12 a^1a^2,\gamma- \tfrac12 a^1\alpha^4 +\tfrac12  a^3\alpha^5 \bigr) 
\cdot 
\\
\eexp^{-i\ell_2a^2 }
\cdot \eexp^{-i \ell_3a^3}
\cdot \eexp^{-i\lambda_4 \alpha^4}
\cdot \eexp^{-i\ell_0c}
\cdot \eexp^{-i\lambda_0\gamma}
\ \extder  c \,\extder  \gamma\,\extder a^2 \,\extder  a^3 \,\extder \alpha^4 
\mapob,
\endmultline
$$
where we use Berezin integration for the odd variables $\alpha^4$ and $\gamma$ (see also \cite{GS,\S7.1}). With our choice of the normalization constant, the inverse operation is given by
$$
\multline
f(a^i, \alpha^j, b,\beta) = 
\int t_{(\ell_0,\ell_2,\ell_3,\lambda_0, \lambda_4)}(a^1, \alpha^5, \alpha^6)
\cdot \eexp^{i\ell_2a^2 }
\cdot \eexp^{i \ell_3a^3}
\cdot \eexp^{i\lambda_4 \alpha^4}
\\
\cdot \eexp^{i\ell_0(b + \frac12  a^1x a^2)}
\cdot \eexp^{i\lambda_0(\beta+ \frac12 a^1\alpha^4 -\frac12  a^3\alpha^5)}
\ \extder  \ell_0 \, \extder \ell_2 \,\extder  \ell_3 \,\extder  \lambda_0\,\extder   \lambda_4 
\mapob.
\endmultline
$$
This shows that an arbitrary function on $G$ can be written as a direct integral of functions of the form \recalf{firstfouriermode}, or more precisely that we have a direct integral decomposition of $V$ as
$$
V = \int V_{(\ell_0,\ell_2,\ell_3,\lambda_0, \lambda_4)} \ \extder  \ell_0 \, \extder \ell_2 \,\extder  \ell_3 \,\extder  \lambda_0\,\extder   \lambda_4 
\mapob.
\formula{directintegraldecomp1}
$$

The action of $\Phi_\gh$ on a function $f$ of the form \recalf{firstfouriermode} is given by the formula
$$
\multline
(\Phi_\gh f)(g)
=
t_{(\ell_0,\ell_2,\ell_3,\lambda_0, \lambda_4)}(a^1 - \ah^1, \alpha^5 - \alphah^5, \alpha^6 - \alphah^6) \cdot
\\
\eexp^{-i\ell_0( \ah^2 a^1 - \frac12\alphah^5\alpha^5 + \frac12 \alphah^6\alpha^6)}\cdot
\eexp^{-i\lambda_0( \alphah^4 a^1 - \ah^3\alpha^5 )}\cdot
\\
\eexp^{-i\ell_2\ah^2}\cdot
\eexp^{-i\ell_3\ah^3}\cdot
\eexp^{-i\lambda_4\alphah^4}\cdot
\eexp^{-i\ell_0(\bh - \frac12 \ah^1\ah^2 )}\cdot
\eexp^{-i\lambda_0(\betah  - \frac12 \ah^1 \alphah^4 + \frac12 \ah^3 \alphah^5)}\cdot
\\
\eexp^{i\ell_2a^2}\cdot
\eexp^{i\ell_3a^3}\cdot
\eexp^{i\lambda_4\alpha^4}\cdot
\eexp^{i\ell_0(b+\frac12a^1a^2)}\cdot
\eexp^{i\lambda_0(\beta+ \frac12a^1\alpha^4 - \frac12a^3\alpha^5)}
\mapob,
\endmultline
$$
which shows that the spaces $V_{(\ell_0,\ell_2,\ell_3,\lambda_0, \lambda_4)}$ are invariant under the action of $G$. The direct integral decomposition \recalf{directintegraldecomp1} thus is a decomposition into invariant ``subspaces.'' 

In order to simplify notation, we will identify a function $f\in V_{(\ell_0,\ell_2,\ell_3,\lambda_0, \lambda_4)}$, i.e. of the form \recalf{firstfouriermode}, with the function $t_{(\ell_0,\ell_2,\ell_3,\lambda_0, \lambda_4)}$ of the three variables $a^1, \alpha^5, \alpha^6$. In terms of such a function $t$ of these three variables, the action $\Psi_\gh$ of $\gh\in G$ is given by
$$
\multline
(\Psi_\gh t)(a^1, \alpha^5, \alpha^6)
=
t(a^1 - \ah^1, \alpha^5 - \alphah^5, \alpha^6 - \alphah^6) \cdot
\\
\eexp^{-i\ell_0( \ah^2 a^1 - \frac12 \alphah^5\alpha^5 + \frac12 \alphah^6\alpha^6)}\cdot
\eexp^{-i\lambda_0( \alphah^4 a^1 - \ah^3\alpha^5 )}\cdot
\\
\eexp^{-i\ell_2\ah^2}\cdot
\eexp^{-i\ell_3\ah^3}\cdot
\eexp^{-i\lambda_4\alphah^4}\cdot
\eexp^{-i\ell_0(\bh - \frac12 \ah^1\ah^2 )}\cdot
\eexp^{-i\lambda_0(\betah  - \frac12 \ah^1 \alphah^4 + \frac12 \ah^3 \alphah^5)}
\mapob.
\endmultline
\tag\memorize\theoremnummer={overalltransformation}
$$

\proclaim{\thmm{firstdecomppm}{Proposition}}
If $\lambda_0$ is zero but $\ell_0$ is non-zero, then the space $V_{(\ell_0,\ell_2,\ell_3,0, \lambda_4)}$ splits as a direct sum of two $G$-invariant spaces $V_{(\ell_0,\ell_2,\ell_3,0, \lambda_4)}^\pm$

\endproclaim

\demo{Proof}
If $\lambda_0=0$, the $G$ action in terms of the functions $t$ \recalf{overalltransformation} is given by
$$
\multline
(\Psi_\gh t)(a^1, \alpha^5, \alpha^6)
=
t(a^1 - \ah^1, \alpha^5 - \alphah^5, \alpha^6 - \alphah^6) \cdot
\eexp^{-i\ell_0( \ah^2 a^1 - \frac12\alphah^5\alpha^5 + \frac12 \alphah^6\alpha^6)}\cdot
\\
\eexp^{-i\ell_2\ah^2}
\cdot\eexp^{-i\ell_3\ah^3}
\cdot\eexp^{-i\lambda_4\alphah^4}
\cdot\eexp^{-i\ell_0(\bh - \frac12 \ah^1\ah^2 )}
\mapob.
\endmultline
$$
It is not hard to see that functions of the form
$$
t(a^1, \alpha^5, \alpha^6)
=
h_\epsilon(a^1, \alpha^5+\epsilon\alpha^6) 
\cdot \eexp^{-\frac i2\ell_0\epsilon\alpha^5\alpha^6}
\formula{onlyonealpha}
$$
with $\epsilon=\pm1$ and $h_\epsilon$ a function of one even and one odd variable, transform (in terms of these functions $h_\epsilon$) under the action of $\Psi_\gh$ as
$$
\multline
(\Psi_\gh h_\epsilon)(a^1, \xi)
=
h_\epsilon(a^1 - \ah^1, \xi - (\alphah^5+\epsilon\alphah^6)) 
\cdot \eexp^{-i\ell_0( \ah^2 a^1 - \frac12 (\alphah^5 - \epsilon\alphah^6)\xi 
)}\cdot
\\
\eexp^{-i\ell_2\ah^2}\cdot
\eexp^{-i\ell_3\ah^3}\cdot
\eexp^{-i\lambda_4\alphah^4}\cdot
\eexp^{-i\ell_0(\bh - \frac12 \ah^1\ah^2 +\frac12\epsilon \alphah^5\alphah^6)}
\mapob.
\endmultline
\tag{\memorize\theoremnummer={ell0not0lambda0is0action}}
$$
This shows that these spaces are invariant under the $G$-action. Moreover, any smooth function $t$ of the variables $a^1, \alpha^5, \alpha^6$ can be written as
$$
t(a^1, \alpha^5, \alpha^6)
=
t_0(a^1) + \alpha^5\cdot t_5(a^1) + \alpha^6 \cdot t_6(a^1) + \alpha^5\alpha^6 \cdot t_{56}(a^1)
\formula{fourdecomposition}
$$
with four smooth functions $t_0, t_5, t_6, t_{56}$ of the single (even) variable $a^1$. In the same way, the functions $h_\epsilon(a^1, \xi)$ can be written as
$$
h_ \epsilon(a^1, \xi)
=
h_{\epsilon,0}(a^1) + \xi \cdot h_{\epsilon,1}(a^1)
\mapob.
$$
Taking the sum and writing $\eexp^{-\frac i2\epsilon\ell_0\alpha^5\alpha^6} = 1 -\frac i2\epsilon\ell_0 \alpha^5\alpha^6$, we obtain
$$
\multline
h_+(a^1, \alpha^5+\alpha^6)\cdot \eexp^{-\frac i2\ell_0\alpha^5\alpha^6}
+
h_-(a^1, \alpha^5 - \alpha^6)\cdot \eexp^{\frac i2\ell_0\alpha^5\alpha^6}
=
\\
\bigl(h_{+,0}(a^1) + h_{-,0}(a^1)\bigr)
+
\alpha^5\cdot \bigl(h_{+,1}(a^1) + h_{-,1}(a^1)\bigr)
\qquad\qquad
\\
+
\alpha^6\cdot \bigl(h_{+,1}(a^1) - h_{-,1}(a^1)\bigr)
+ 
\tfrac i2 \ell_0\alpha^5\alpha^6 \cdot \bigl(h_{-,0}(a^1) - h_{+,0}(a^1)\bigr)
\mapob.
\endmultline
$$
Comparing this with \recalf{fourdecomposition} shows that any (smooth) function of $(a^1, \alpha^5, \alpha^6)$ can be decomposed in a unique way as the sum of two functions of the form \recalf{onlyonealpha} (one of each kind).
\QED\enddemo

To analyse the case $\ell_0=0$, we start looking at the explicit transformation property in this case, which is given in terms of the function $t$ \recalf{overalltransformation} as
$$
\multline
(\Psi_\gh t)(a^1, \alpha^5, \alpha^6)
=
t(a^1 - \ah^1, \alpha^5 - \alphah^5, \alpha^6 - \alphah^6) \cdot
\eexp^{-i\lambda_0( \alphah^4 a^1 - \ah^3\alpha^5 )}\cdot
\\
\eexp^{-i\ell_2\ah^2}\cdot
\eexp^{-i\ell_3\ah^3}\cdot
\eexp^{-i\lambda_4\alphah^4}\cdot
\eexp^{-i\lambda_0(\betah  - \frac12 \ah^1 \alphah^4 + \frac12 \ah^3 \alphah^5)}
\mapob.
\endmultline
$$
Since $\alpha^6$ only appears as an argument in $t$, it is easy to see that we can perform another Berezin-Fourier transform. More precisely, we introduce the spaces $V_{(0,\ell_2,\ell_3,\lambda_0,\lambda_4),(\lambda_6)}$ of functions $t$ of the form
$$
t(a^1, \alpha^5, \alpha^6)
=
h(a^1, \alpha^5)\cdot \eexp^{i\lambda_6\alpha^6}
\mapob.
$$
In terms of the function $h$ the action of $\gh\in G$ is given by
$$
\multline
(\Psi_\gh h)(a^1, \alpha^5)
=
h(a^1 - \ah^1, \alpha^5 - \alphah^5) \cdot
\eexp^{-i\lambda_0( \alphah^4 a^1 - \ah^3\alpha^5 )}\cdot
\\
\eexp^{-i(\ell_2\ah^2+\ell_3\ah^3)}\cdot
\eexp^{-i(\lambda_4\alphah^4+ \lambda_6\alpha^6)}\cdot
\eexp^{-i\lambda_0(\betah  - \frac12 \ah^1 \alphah^4 + \frac12 \ah^3 \alphah^5)}
\mapob.
\endmultline
\tag{\memorize\theoremnummer={ell0zerobutnotlambda0action}}
$$ 
This shows that the spaces $V_{(0,\ell_2,\ell_3,\lambda_0,\lambda_4),(\lambda_6)} $  are invariant under the action of $G$, meaning that we have an invariant direct integral decomposition 
$$
V_{(0,\ell_2,\ell_3,\lambda_0,\lambda_4)} = \int V_{(0,\ell_2,\ell_3,\lambda_0,\lambda_4),(\lambda_6)} \ \extder \lambda_6
\mapob.
$$
If $\lambda_0=0$ the other coordinates $a^1$ and $\alpha^5$ also appear only in $t$ and we can perform a triple Berezin-Fourrier transform by looking at functions $t$ of the form
$$
t(a^1, \alpha^5, \alpha^6)
=
c\cdot \eexp^{i\ell_1 a^1} \cdot \eexp^{i\lambda_5 \alpha^5}\cdot \eexp^{i\lambda_6\alpha^6}
\mapob.
$$
These functions transform under the action of $\gh\in G$ as
$$
(\Psi_\gh t)(a^1, \alpha^5, \alpha^6)
=
c 
\cdot \eexp^{-i(\ell_1\ah^1+\ell_2\ah^2+\ell_3\ah^3)}
\cdot \eexp^{-i(\lambda_4\alphah^4+\lambda_5\alphah^5+\lambda_6\alphah^6)}
\mapob.
\formula{ell0zeroandlambda0zeroaction}
$$
This shows that the spaces $V_{(0,\ell_2,\ell_3,0,\lambda_4),(\ell_1, \lambda_5,\lambda_6)}$ consisting of this kind of functions are also invariant under the action of $G$ giving us an invariant direct integral decomposition 
$$
V_{(0,\ell_2,\ell_3,0,\lambda_4)} = \int V_{(0,\ell_2,\ell_3,0,\lambda_4),(\ell_1, \lambda_5,\lambda_6)} \ \extder \ell_1\,\extder\lambda_5\,\extder \lambda_6
\mapob.
$$

So far we thus have been able to decompose the regular representation in the following invariant ``subspaces'':
$$
\matrix
&&\rlap{\hss$V$}
\\
\noalign{\vskip2\jot}
&& @VV{\tsize \int} \extder \ell_0 \,\extder \ell_2\,\extder \ell_3 \,\extder \lambda_0 \,\extder \lambda_4 V
\\
\noalign{\vskip2\jot}
&&{\rlap{\hss$V_{(\ell_0,\ell_2,\ell_3,\lambda_0,\lambda_4)}$}}
\\
\noalign{\vskip3\jot}
\ell_0\neq0= \lambda_0
&& 
\ell_0=0\neq \lambda_0
&& 
\ell_0=0= \lambda_0
\\
\noalign{\vskip2\jot}
V_{(\ell_0,\ell_2,\ell_3,0,\lambda_4)}
&& 
V_{(0,\ell_2,\ell_3,\lambda_0,\lambda_4)}
&& 
V_{(0,\ell_2,\ell_3,0,\lambda_4)}
\\
\noalign{\vskip2\jot}
 @VV\oplus V  @VV{\tsize \int} \extder \lambda_6 V  @VV{\tsize \int} \extder \ell_1 \,\extder \lambda_5 \,\extder\lambda_6 V
\\
\noalign{\vskip2\jot}
 V_{(\ell_0,\ell_2,\ell_3,0,\lambda_4)}^\pm
&&V_{(0,\ell_2,\ell_3,\lambda_0,\lambda_4),(\lambda_6)}
&&V_{(0,\ell_2,\ell_3,0,\lambda_4),(\ell_1, \lambda_5,\lambda_6)}
\endmatrix
$$
The main question now is whether this decomposition is irreducible and, for the Berezin direct integrals, in what sense.

\head{\memorize\sectionnum={invariantfamilies}. Invariant odd families of subspaces}\endhead

Throughout this section, $V$ will denote a graded vector space, $G$ a graded Lie group and $\Phi$ a representation of $G$ on $V$. This means that $\Phi$ is a smooth homomorphism from $G$ to $\Aut(V)$, the group of automorphisms of $V$. As is customary, we will note the (even, bijective) linear map $\Phi(g)$ (for $g\in G$) often as $\Phi_g$. The purpose of this section is to extend the notion of an irreducible subspace and\slash or irreducible decomposition to incorporate in a convenient way subspaces that are indexed by an odd parameter. Most of the discussion will concentrate on motivations for the definitions \recalt{defoddfamdecomposition} and \recalt{definvardecomposition} and we will finish this section with two results concerning irreducibility that will be needed afterwards. Our arguments are based upon the notion of a graded vector space $V$ and smooth linear maps as defined in \cite{Tu04}, meaning in particular that we identify the body $\body V$ as a part of $V$, that a basis always consists of elements of $\body V$, that a graded subspace always admits a basis in $\body V$ (we thus make a distinction between (algebraic, linear) subspaces and graded subspaces) and that a smooth linear map always maps $\body V$ into the body of the target space. In this way there is a complete correspondence between graded vector spaces and smooth linear maps on the one hand and $\ZZ/2\ZZ$-graded vector spaces over $\RR$ with (ordinary) linear maps on the other (just tensor with $\CA$). The only difference being that in the latter case the distinction between left-linear and right-linear disappears.

\bigskip

If one starts to think about a family $W_\lambda\subset V$ of subspaces indexed by an odd parameter $\lambda \in \CA_1$, one soon arrives at the following description:
$$
W_\lambda = \{w+\lambda I(w)\mid w\in W\}
$$
for some subspace $W\subset V$ and a linear map $I:W\to V$. Even though it is not the most general possibility, it also seems reasonable to require that $W$ is a graded subspace and that $I$ is a smooth odd (right-)linear map. This gives us the following definition.

\definition{Definition}
An \stresd{odd family of subspaces} (of $V$) is a couple $(W,I)$ where $W\subset V$ is a graded subspace and $I:W\to V$ a smooth odd right-linear map.

\enddefinition

\definition{Remarks}
Right-linearity of the map $I$ implies that we have $I(v\cdot \lambda) = I(v) \cdot \lambda$ for all $v\in V$ and 	all $\lambda\in \CA$. However, the fact that $I$ is odd implies that we don't have the equality $I(\lambda\cdot v) = \lambda\cdot I(v)$ for all $v$ and $\lambda$. What we do have is the equality
$I(\lambda\cdot v) = (-1)^{\alpha}\lambda\cdot I(v)$ for all $\lambda\in \CA_\alpha$, which means in particular that we obtain a minus sign if $\lambda$ is odd.

\itemize
In general the subspaces $W_\lambda$ will not be graded subspaces when $\lambda\neq0$. The easiest example is the graded vector space $V$ with one even generator $e$ and one odd generator $f$, the graded subspace $W$ generated by $e$ and the smooth odd right-linear map $I:W\to V$ defined by $I(e) = f$. Then $W_\lambda$ consists of the elements $\alpha\cdot (e+\lambda f)$, which does not allow a basis of the form $xe+yf$ with $x,y\in \RR$.

\enddefinition

\proclaim{\thmm{isomorphismWandWlambda}{Lemma}}
Let $(W,I)$ be an odd family of subspaces and let $\lambda_1\in \CA_1$ be a fixed odd parameter. Then the map $W\to W_\lambda$ given by $w\mapsto w+\lambda I(w)$ is an even linear bijection.

\endproclaim

\demo{Proof}
It is obvious that the given map is linear and even and (by definition of $W_\lambda$) surjective. So suppose we have $w+\lambda I(w) = 0$. If we multiply by $\lambda$ and use that $\lambda^2=0$, we get the equation $\lambda w= 0$. But then we have $$
0 = w+\lambda I(w) = w - I(\lambda w) = w - I(0) = w
\eqno\QEDbox
$$

\enddemo

\definition{Definition}
Let $(W,I)$ be an odd family of subspaces of $V$. We will say that $(W,I)$ is an \stresd{odd family decomposition of $V$} if it satisfies the following two conditions:
\roster\item"(i)"
the map $I:W\to V$ is injective and

\item"(ii)"
we have a direct sum decomposition $V = W\oplus I(W)$.

\endroster

\enddefinition

The motivation for this definition is given by the following observations (which should be compared with a Fourier transform, see also example \recalt{exampFourierCA1}). We first note that the space $C^\infty(\CA_1, W)$ of smooth functions $f:\CA_1\to W$ is isomorphic to $\body W \oplus \body W$, the correspondence given by
$$
f(\lambda) = w_0 + \lambda w_1
\qquad,\qquad w_o,w_1\in \body W
$$
where smoothness of $f$ forces us to take $w_i$ in the body of $W$. Now for any smooth odd map $I:W\to V$ we can define, using the Berezin integral, the map $\Cal F_I:C^\infty(\CA_1, W)\to \body V$ by
$$
\align
\Cal F_I(f) 
&= 
\int \eexp^{\lambda I} f(\lambda) \ \extder\lambda
=
\int (id + \lambda I)(w_0 + \lambda w_1) \ \extder \lambda
=
w_1 + I (w_0)
\mapob ,
\endalign
$$
where the exponential map\slash series $\eexp^A = \sum_{n=0}^\infty {A^n}/{n!}$ defined for any linear map $A$ is in this case a finite sum with only two terms because $\lambda^2=0$.
Obviously the map $\Cal F_I$ is a (linear) bijection if and only if $I$ is injective and  $\body V$ is the direct sum $\body V = \body W \oplus I(\body W)$, which is the case if and only if $I$ is injective and we have a direct sum $V = W\oplus I(W)$. Tensoring with $\CA$, it is not hard to show that $C^\infty(\CA_1,W)\otimes \CA$ is a graded vector space isomorphic to $W\oplus \prod W$ where $\prod$ denotes the parity reversal operator. Moreover, $\Cal F_I$ extends to a smooth odd left-linear map $\Cal F_I : C^\infty(\CA_1,W)\otimes \CA \to V$. This map is bijective (we may not call it an isomorphisms because it is not even) if and only if $I$ is injective and we have the direct sum decomposition $V = W\oplus I(W)$.

On the other hand, for a fixed $\lambda\in \CA_1$ the element $w=f(\lambda)$ belongs to $W$ and thus
$$
\eexp^{\lambda I}\bigl(f(\lambda)\bigr) = (id +\lambda I)(w) = w+\lambda I(w)
$$
belongs to $W_\lambda$. Our conditions for an odd family decomposition thus are the necessary and sufficient conditions to guarantee that each element of $\body V$ can be obtained as a Berezin integral of a smooth family of elements in $W_\lambda$, and thus, by tensoring with $\CA$, that each element of $V$ is the Berezin integral of a family of elements in $W_\lambda$.

\medskip

When we want to analyze more precisely the invariance properties of subspaces appearing in odd family decompositions, we will need a preliminary result concerning the notion of elements in $\CA$ that are independent over $\RR$ (or $\CC$ if needed).

\proclaim{\thmm{lincombinWindependenceresultwithmfddep}{Lemma}}
Let $M$ be a supermanifold, let $v_1, \dots, v_n : M \to V$ be $n\in\NN$ smooth maps, let $\lambda_1, \dots, \lambda_n\in\CA$ be $n$ fixed elements that are independent over $\RR$ and let $W\subset V$ be a graded subspace. 

If the linear combination $\sum_{i=1}^n \lambda_i \, v_i(m)$ belongs to $W$ for all $ m\in M$, then $v_i$ maps $M$ into $W$ for all $1\le i\le n$.

\endproclaim

\definition{Remark}
If all the maps $v_i$ are constant (which happens if the manifold $M$ is a single point), the smoothness conditions implies that all $v_i(m)$ belong to $\body V$. And in that case we can be more precise, because then all $v_i(m)$ must lie in $\body W$, not just to $W$.

\enddefinition

\demo{Proof}
If $(x_1, \dots, x_p, \xi_1, \dots, \xi_q)$ is a local system of even and odd coordinates on $M$ (the $x_i$ even and the $\xi_j$ odd), any smooth map $f:M\to V$ is determined (locally on the given chart) by $2^q$ ordinary smooth maps $f_P:\RR^p\to \body V$, $P\subset\{1, \dots, q\}$ by
$$
f(x,\xi) = f_{\emptyset}(x) + \sum_{k=1}^q \ \sum_{1\le i_1<\cdots < i_k\le q} \xi_{i_k}\cdots \xi_{i_2}\cdot \xi_{i_1} \cdot f_{\{i_1, \dots, i_k\}}(x)
\mapob.
\formula{localexpressionsmoothfunction}
$$
Actually, the notation $f_P(x)$ should be interpreted as the extension of the smooth function on points with real coordinates in $\RR^p$ to the smooth function on points with arbitrary even coordinates in $(\CA_0)^p$ by Taylor expansion.
Using the local coordinates $(x,\xi)$ thus tells us that $\sum_{i=1}^k \lambda_i v_i(x,\xi)$ belongs to $W$ for all $(x,\xi)$ in the given chart. By taking derivatives at the point $(\body x,0)$ with respect to the odd coordinates $\xi_{i_1}$, \dots, $\xi_{i_k}$ we deduce that for all $P\subset \{1, \dots, q\}$ the vector
$$
\sum_{i=1}^k \lambda_i {v_{i,P}}(\body x,0)
$$
belongs to $W$ for all $(x,0)$ in the given chart. 

Now let $(e_\ell)_{\ell\in L}$ be a basis for $V$ such that a subset forms a basis of $W$. By the smoothness assumption the $v_{i,P}(\body x,0)$ have real coefficients with respect to this basis. Let $\ell\in L$ be such that $e_\ell$ is not in the basis for $W$ and let $a_i\in \RR$ be the coefficients of $v_{i,P}(\body x,0)$ with respect to this basis element. Then the condition that $\sum_{i=1}^k \lambda_i v_{i,P}(\body x,0)$ belongs to $W$ implies that we must have $\sum_{i=1}^k \lambda_i a_i = 0$. But then independence of the $\lambda_i$ over $\RR$ implies that all $a_i$ must be zero. It follows that the coefficients of each vector $v_{i,P}(\body x,0)$ with respect to the basis element $e_\ell$ is zero, showing that each $v_{i,P}(\body x,0)$ separately belongs to $W$. But if these all separately belong to $W$, then so do the linear combinations 
$$
v_i(m) = \sum_{P\subset \{1, \dots, q\}} (\xi)_P \cdot v_{i,P}(x,0)
\mapob.
$$
This shows that all $v_i$ separately map the local chart into $W$. Since the local chart is arbitrary, the result follows.
\QED\enddemo

\proclaim{\thmm{lincombinWindependenceresultwithmfddepBIS}{Corollary}}
Let $M$ be a supermanifold, let $v_1, \dots, v_n : M \to V$ be $n\in\NN$ smooth maps, let $\lambda_1, \dots, \lambda_n\in\CA$ be $n$ fixed elements, not all zero and let $W\subset V$ be a graded subspace. 

If the linear combination $\sum_{i=1}^n \lambda_i \, v_i(m)$ belongs to $W$ for all $ m\in M$, 
then there exists $1\le k\le n$, a permutation $\sigma$ of $\{1, \dots, n\}$ and a real valued matrix $A_{ij}\in M\bigl(k\times (n-k),\RR\bigr)$ such that
\roster
\item"(i)"
$\dsize \forall j=k+1, \dots, n 
:
\lambda_{\sigma(j)} = \sum_{i=1}^k \lambda_{\sigma(i)} A_{ij}
$

\item"(ii)"
$\forall m\in M \ \dsize \forall i=1, \dots, k
:
v_{\sigma(i)}(m) + \sum_{j=k+1}^n A_{ij} v_{\sigma(j)}(m) \in W
$.

\endroster

\endproclaim

\demo{Proof}
The subspace over $\RR$ of $\CA$ spanned by the $\lambda_i$ has real dimension at most $n$ and at least $1$ because not all $\lambda_i$ are zero. It follows that there exists $1\le k\le n$, a permutation $\sigma$ of $\{1, \dots, n\}$ and a real valued matrix $A_{ij}$ such that $\lambda_{\sigma(1)}, \dots, \lambda_{\sigma(k)}$ are indepenndent and such that the remaining $\lambda_{\sigma(k+1)}, \dots, \lambda_{\sigma(n)}$ are linear combinations of them with real coefficients as in
$$
\forall j=k+1, \dots, n 
:
\lambda_{\sigma(j)} = \sum_{i=1}^k \lambda_{\sigma(i)} A_{ij}
\mapob.
$$
For $1\le i\le k$ we now introduce the elements $\mu_i=\lambda_{\sigma(i)}$ and the smooth maps $w_i:M\to V$ defined by
$$
w_i  = v_{\sigma(i)} + \sum_{j=k+1}^n A_{ij} v_{\sigma(j)}
\mapob.
$$
They are smooth because they are linear combinations with real coefficients of smooth maps.
We then have the equality (of smooth maps)
$$
\align
\sum_{i=1}^n \lambda_i v_i 
&= 
\sum_{i=1}^n \lambda_{\sigma(i)} v_{\sigma(i)}
=
\sum_{i=1}^k \mu_i  v_{\sigma(i)} 
+ 
\sum_{j=k+1}^n 
\sum_{i=1}^k \lambda_{\sigma(i)} A_{ij}
v_{\sigma(j)}
\\&
=
\sum_{i=1}^k \mu_i \Bigl( v_{\sigma(i)} + \sum_{j=k+1}^n 
\lambda_{\sigma(i)} A_{ij}
v_{\sigma(j)}
\Bigr)
=
\sum_{i=1}^k \mu_i w_i
\mapob.
\endalign
$$
We thus have $k$ independent (over $\RR$) elements $\mu_i$ and $k$ smooth maps $w_i:M\to V$ such that the linear combination $\sum_{i=1}^k \mu_i w_i$ maps $M$ into $W$. It then suffices to invoke \recalt{lincombinWindependenceresultwithmfddep} to conclude.
\QED\enddemo

\definition{\thmm{notabenenonindeplincomb}{Nota Bene}}
The way to interpret \recalt{lincombinWindependenceresultwithmfddepBIS} is as follows. If the $n$ coefficients $\lambda_i$ are not independent, some will be a linear combination with real coefficients of others. Substituting these linear combinations in the sum $\sum_i \lambda_i v_i(m)$ thus contains less than $n$ coefficients $\lambda_i$ because some are rewritten as a linear combination of others. Now reorder the terms in this sum in such a way that all terms with the same $\lambda_i$ are grouped together. The corollary then says that the applications that are the separate ``coefficients'' of each remaining $\lambda_i$ map $M$ to $W$. 

\enddefinition

\proclaim{\thmm{genericelementinvariantallinvariant}{Lemma}}
Let $(W,I)$ be an odd family of subspaces of $V$ and let $\lambda\neq0$ be a fixed odd parameter. If the subspace $W_\lambda$ is invariant under the action $\Phi$, \ie, for all $g\in G$ we have $\Phi_g(W_\lambda) \subset W_\lambda$, then the following two conditions are satisfied.
\roster\item"(i)"
$W$ itself is invariant under $\Phi$, \ie, $\forall g\in G$: $\Phi_g(W) \subset W$.

\item"(ii)" $W$ is invariant under all (odd) maps $\[\Phi_g,I\] = \Phi_g \scirc I - I\scirc \Phi_g$, \ie, $\forall g\in G$: $\[\Phi_g,I\](W) \subset W$.

\endroster
Conversely, if these two conditions are satisfied then all subspaces $W_\mu$, $\mu\in \CA_1$ are invariant under $\Phi$.

\endproclaim

\demo{Proof}
Let $g\in G$, $w\in W$ and $\mu\in \CA_1$ be arbitrary. Then, since $\Phi_g$ is even and thus right- and left-linear, we have
$$
\align
\Phi_g(w+\mu I(w))
&=
\Phi_g(w) + \mu\Phi_g(I(w))
=
\Phi_g(w) + \mu \[\Phi_g,I\](w) + \mu I(\Phi_g(w))
\\&
=
\bigl(\Phi_g(w) + \mu \[\Phi_g,I\](w) \bigr) + \mu I\bigl(\Phi_g(w) + \mu \[\Phi_g,I\](w) \bigr)
\mapob.
\endalign
$$
It follows immediately that if the conditions (i) and (ii) are satisfied, then $W_\mu$ is invariant under $\Phi$, simply because $\Phi_g(w) + \mu \[\Phi_g,I\](w)$ belongs to $W$.

For the converse we suppose that $W_\lambda$ is invariant under $\Phi$. This means that for all $w\in W$ and all $g\in G$ there exists and element $z\in W$ such that 
$$
\Phi_g(w) + \lambda \Phi_g(I(w))
=
\Phi_g(w+ \lambda  I(w)) = z+ \lambda  I(z)
\mapob.
\formula{imagePhiinW}
$$
Multiplying this equation on the left with $\lambda$ gives us the equality
$
\lambda \Phi_g(w) = \lambda z
$.
Since $z$ and thus $\lambda  z$ belongs to $W$, it follows that $\lambda\Phi_g(w)$ belongs to $W$. If we take $w\in \body W$, then the map $g\mapsto \Phi_g(w)$ is smooth. We thus can apply \recalt{lincombinWindependenceresultwithmfddep} to conclude that $\Phi_g(w)$ itself belongs to $W$. But $\Phi_g$ is linear and $\body W$ generates $W$, so $\Phi_g(w)$ belongs to $W$ for all $g\in G$ and all $w\in W$, \ie, $W$ is invariant under $\Phi$, proving (i).

Now using the linearity of $I$ and substituting the equality $\lambda\Phi_g(w) = \lambda  z$ in \recalf{imagePhiinW}, we get
$$
\Phi_g(w) + \lambda \Phi_g(I(w))
=
z+ \lambda  I(\Phi_g(w))
\mapob,
$$
which gives us the solution for $z$ as
$$
z = \Phi_g(w) + \lambda \[\Phi_g,I\](w) 
\mapob.
\formula{inducedactiononWcasen=1}
$$
We know already that $\Phi_g(w)$ belongs to $W$, so we deduce that $\lambda\[\Phi_g,I\](w)$ belongs to $W$. As before, we note that for $w\in \body W$ the map $g\mapsto [\Phi_g,I\](w)$ is smooth and thus by \recalt{lincombinWindependenceresultwithmfddep} $[\Phi_g,I\](w)$ belongs to $W$. And again as before, $\body W$ generates $W$ and the map $[\Phi_g,I\]$ is linear, so $[\Phi_g,I\](w)$ belongs to $W$ for all $g\in G$ and all $w\in W$, proving (ii).
\QED\enddemo

\definition{\thmm{definvariantoddfamdecomp}{Definition}}
We will say that an odd family of subspaces $(W,I)$ of $V$ is \stresd{invariant} (under the action $\Phi$) if the following two conditions are satisfied:
\roster\item"(i)"
$\forall g\in G$: $\Phi_g(W)\subset W$ and

\item"(ii)"
$\forall g\in G$: $\[\Phi_g,I\](W) \subset W$.

\endroster
An \stresd{invariant odd family decomposition of $V$} is an odd family decomposition $(W,I)$ of $V$ such that $(W,I)$ is invariant as an odd family of subspaces.

\enddefinition

\definition{Remark}
It should be noted that if $(W,I)$ is an invariant odd family of subspaces of $V$ under the action $\Phi$, then in particular $W=W_0$ is an invariant subspace. However, it is quite possible that $W=W_0$ is an invariant subspace without $(W,I)$ being an invariant odd family, \ie, without any $W_\lambda$ with $\lambda\neq0$ being invariant.

\enddefinition

\definition{\thmm{exampFourierCA1}{Example}}
Consider the real graded space $X$ of smooth functions on $\CA_1$ with values in the complexified ring $\CA^\CC$, $f:\CA_1\to \CA^\CC$. Since any such smooth function is of the form $f(\xi) = c_0 + \xi\cdot c_1$ with $c_i\in \CC$, $V$ has dimension $1\vert 1$. It follows that $V=X\otimes \CA$ is a graded vector space of graded dimension $1\vert1$. On $V$ we let act the group $G=\CA_1$ by 
$$
\bigl(\Phi_\tau(f\otimes \alpha)\bigr)(\xi) = f(\xi-\tau)\otimes \alpha
\mapob.
$$
A basis for $V$ is given by the functions $f_0, f_1$ defined by $f_0(\xi) = 1$ and $f_1(\xi) = \xi$. In terms of this basis the action of $G$ is given as
$$
\Phi_\tau f_0 = f_0
\qquad\text{and}\qquad
\Phi_\tau f_1 = \tau\cdot f_0 + f_1
\mapob.
$$
The graded subspace $W$ generated by $f_0$ thus is an invariant subspace, but it does not admit an invariant supplement. However, if we consider the odd right-linear map $I$ defined on $W$ with values in $V$ by $I(f_0) = i \cdot f_1$, then $(W,I)$ is an invariant odd family decomposition of $V$. In fact, the subspace $W_\lambda$ is given as
$$
W_\lambda = \{ c_0\cdot f_0 + \lambda I(c_0\cdot f_0) \mid c_0\in \CA^\CC\}
=
\{ \,(f_0 + i \lambda\cdot f_1)\cdot c_0\mid c_0\in \CA^\CC\}
\mapob.
$$
But the function $f_\lambda = f_0 + i \lambda\cdot f_1$ is defined as
$$
f_\lambda(\xi) = f_0(\xi) + i \lambda\cdot f_1(\xi)
=
1+i \lambda\xi = \eexp^{i \lambda\xi}
\mapob.
$$
And indeed $\Phi_\tau f_\lambda = \eexp^{-i \lambda\tau} \cdot f_\lambda$. Moreover, any function $f\in V$ can be obtained as a Berezin integral over a function of the form $c_0(\lambda)\cdot \eexp^{i\lambda\xi}$. This is exactly the odd Fourier transform described (for instance) in \cite{GS,\S7.1}.

\enddefinition

\definition{Definition}
Let $(W,I)$ be an odd family of subspaces of $V$ and let $\lambda\in \CA_1$ be fixed. We will say that a linear subpace $L\subset W_\lambda$ is a graded subspace of $W_\lambda$ if there exists a graded subspace $X\subset W$ such that we have the equality $L=X_\lambda$ where $X_\lambda$ is defined as
$$
X_\lambda
=
\{ x+\lambda I(x) \mid x\in X\}
\mapob,
\formula{inducedsubspaceonfamilyofsubspaces}
$$
\ie, if $L$ is the image of $X$ under the even linear bijection $W\to W_\lambda$ given in \recalt{isomorphismWandWlambda}. Note that a graded subspace of $W_\lambda$ is in general not a graded subspace of $V$.

\enddefinition

\proclaim{Proposition}
Let $(W,I)$ be an odd family of subspaces of $V$, let $\lambda\in\CA_1$ be a fixed non-zero odd element and let $X\subset W$ be a graded subspace. If $X_\lambda$ (see \recalf{inducedsubspaceonfamilyofsubspaces}) is invariant under $\Phi$ (\ie, $\Phi_g(X_\lambda) \subset X_\lambda$ for all $g\in G$), then $X_\mu\subset W_\mu$ is invariant under $\Phi$ for all $\mu\in \CA_1$. Moreover, this is the case if and only if the following two conditions are satisfied:
\roster\item"(i)"
$\forall g\in G$: $\Phi_g(X)\subset X$ and

\item"(ii)"
$\forall g\in G$: $\[\Phi_g,I\](X) \subset X$.

\endroster

\endproclaim

\demo{Proof}
$(X,I\restricted_X)$ is an odd family of subspaces and $X_\lambda$ is invariant. The result then follows from \recalt{genericelementinvariantallinvariant}.
\QED\enddemo

\definition{Definition}
Let $(W,I)$ be an odd family of subspaces of $V$. A graded subspace $X\subset W$ is called an \stresd{invariant subspace of the couple $(W,I)$} if $(X,I\restricted_X)$ is an invariant family of subspaces.

\enddefinition

\proclaim{Corollary}
If $X$ is an invariant subspace of an odd family decomposition $(W,I)$, then not only $X$, but also $X\oplus I(X)$ is an invariant subspace of $V$ under the action $\Phi$. Moreover, $(X,I\restricted_X)$ is an invariant odd family decomposition of $X\oplus I(X)$.

\endproclaim

\proclaim{Corollary}
Suppose $(W,I)$ is an invariant odd family decomposition of $V$. If $X$ and $Y$ are two invariant subspaces of the odd family $(W,I)$ and if $W=X\oplus Y$, then $(X\oplus I(X))\oplus (Y\oplus I(Y))$ is a decomposition into two invariant subspaces of $V$, each of which admits an invariant odd family decomposition.

\endproclaim

The two conditions given in \recalt{definvariantoddfamdecomp} for an odd family of subspaces $(W,I)$ to be invariant are not independent in the sense that the second condition $\[\Phi_g,I\](W) \subset W$ cannot be formulated when we do not know already the first condition. More precisely, the map $I$ is defined only on $W$, so for the map $\[\Phi_g,I\] = \Phi_g \scirc I - I \scirc \Phi_g$ to make sense on $W$, we need to know that $\Phi_g$ maps $W$ into $W$.

In the situation of an odd family decomposition $(W,I)$ of $V$ we can make the two conditions independent by changing\slash extending the definition of an odd family decomposition slightly. If $I$ is injectiive and $V=W\oplus I(W)$, we can extend the map $I:W\to V$ to the whole of $V$ by saying that on $I(W)\subset V$ it is the inverse of $I$. We then have an odd bijective smooth right-linear map $I:V\to V$ satisfying $I\scirc I = id_V$ such that $V=W\oplus I(W)$. (Another way to state these conditions is to say that we have an isomorphism of graded vector spaces between $V$ and $W\oplus \prod W$, where $\prod$ is the parity reversal operation.) Once we have defined $I$ on the whole of $V$, the two conditions for an odd family decomposition to be invariant are independent. Since this reformulation of an odd family decomposition will be usefull in other situations, we thus modify its definition.

\definition{Modified definition}
An \stresd{odd family decomposition of $V$} is a couple $(W,I)$ with $W\subset V$ a graded subspace and $I:V\to V$ a smooth odd right-linear map satisfying $I\scirc I = id_V$ such that $V$ is the direct sum of $W$ and $I(W)$. It thus is in particular an odd family of subspaces of $V$.

\enddefinition

Let us now consider an odd family of subspaces $(W,I)$ of $V$ and an odd family of subspaces $(X,J)$ of $W$. This means that $W\subset V$ is a graded subspace, that $I:W\to V$ is a smooth odd right-linear map, that $X\subset W$ is a graded subspace and that $J:X\to W$ is a smooth odd right-linear map. We then can use the subspaces $X_\mu\subset W$, $\mu\in \CA_1$ to define subspaces $X_{\mu,\lambda}\subset W_\lambda\subset V$ by
$$
X_{\mu,\lambda} = \{\,(x+\mu J(x))+\lambda I(x+\mu J(x)) \mid x\in X\,\}
\mapob.
\formula{twodimoddfamily}
$$
It seems reasonable to call the collection $(X_{\mu,\lambda})_{\mu,\lambda\in \CA_1}$ a 2-dimensional odd family of subspaces of $V$. Since we are more interested in odd family decompositions of $V$, we will not formalize this general definition but we will concentrate on the odd decomposition case with our modified definition.

If $(W,I)$ is an odd family decomposition of $V$ and if $(X,J)$ is an odd family decomposition of $W$, then we have the direct sum decomposition
$$
V=(X\oplus J(X)) \oplus I(X\oplus J(X))
=
X\oplus J(X) \oplus I(X) \oplus I(J(X))
\mapob.
$$
Moreover, $I$ maps $V$ to itself and $J$ maps $W$ to itself.
We now extend the map $J$ to a map $J':V\to V$ on the whole of $V$ by
$$
J'\restricted_{W} = J
\qquad\text{and}\qquad
J'\restricted_{I(W)} = - I\scirc J\scirc I
\mapob.
$$
The extended (odd right-linear) map $J'$ satisfies $J'\scirc J' = id_V$ and it commutes (in the graded sense) with $I$, \ie, $I\scirc J' = -J'\scirc I$ on the whole of $V$. Moreover, if we define $Z=X\oplus I(X)$, then $(Z, J')$ is an odd family decomposition of $V$ and $(X,I\restricted_{Z})$ is an odd family decomposition of $Z$. 

Going one step further, we can consider an odd family decomposition $(Y,K)$ of $X$. This gives us a direct sum decomposition of $V$ as
$$
V = Y\oplus K(Y) \oplus J(Y) \oplus J(K(Y)) \oplus I(Y) \oplus I(K(Y)) \oplus I(J(Y)) \oplus I(J(K(Y)))
\mapob.
$$
The map $K:X\to X$ can be extended to a smooth odd right-linear map $K':V\to V$ by
$$
\alignat2
K'\restricted_X 
&
= K
\qquad
,\qquad
&
K'\restricted_{J(X)} 
&
= -J\scirc K\scirc J
\\ 
K'\restricted_{I(X)} 
&
= -I\scirc K\scirc I
\qquad
,\qquad
&
K'\restricted_{I(J(X))} 
&
= I\scirc J\scirc K\scirc J\scirc I
\mapob.
\endalignat
$$
It is not hard to see that $K'\scirc K'=id_V$ and that $K'$ commutes (in the graded sense) with the maps $I$ and $J'$. Moreover, it is not hard to show (for instance) that $(Y,J)$ is an odd family decomposition of $Y\oplus J(Y)$, that $(Y\oplus J(Y), K')$ is an odd family decomposition of $Y\oplus J(Y) \oplus K(Y) \oplus K'(J(Y))$ and that $(Y\oplus J(Y) \oplus K(Y) \oplus K'(J(Y)),I)$ is an odd family decomposition of $V$, simply because we have the obvious equality of subspaces $K'(J(Y)) = J(K(Y))$. 

In order to prepare the general definition of an $n$-dimensional odd family decomposition of $V$, we introduce some notation. Let $I_1, \dots, I_n$ be $n$ linear maps, let $\lambda_1, \dots, \lambda_n\in \CA$ be $n$ elements of the ring $\CA$ and let $P\subset \{1, \dots, n\}$ be any subset written as $P=\{i_1, \dots, i_k\}$ with $i_1<\cdots<i_k$. We then define the element $\repprod\lambda_P\in \CA$ and the linear map $\repprod I_P$ by
$$
\repprod\lambda_P = \lambda_{i_k} \cdots \lambda_{i_2}\cdot \lambda_{i_1}
\qquad\text{and}\qquad
\repprod I_P = I_{i_k} \scirc \cdots \scirc I_{i_1}
\mapob,
$$
with the convention that $\repprod\lambda_\emptyset = 1$ and $\repprod I_\emptyset = id_V$. We obviously assume that this composition makes sense, a condition which will always be satisfied in our use of this notation. In the same vein we define the map $\repprod\lambda I_P$ as the map $\repprod J_P$ for the linear maps $J_i = \lambda_i \cdot I_i$, i.e.,
$$
\repprod\lambda I_P = (\lambda_{i_k} \cdot I_{i_k}) \scirc \dots \scirc  (\lambda_{i_1} \cdot I_{i_1})
\mapob.
$$
We also define the multi-commutator $\repcomm A,I_P$ for any linear map $A$ by
$$
\repcomm A,I_P = \[\[\cdots\[\[A,I_{i_1}\], I_{i_2}\]Ê\cdots\],I_{i_k}\]
\mapob,
$$
where again we suppose that these (graded!) commutators make sense. By convention we define $\repcomm A,I_\emptyset = A$.

\definition{\thmm{defoddfamdecomposition}{Definition}}
An \stresd{$n$-dimensional odd family decomposition of $V$} is a  graded subspace $X\subset V$ and $n$ smooth odd right-linear maps $I_1, \dots, I_n:V\to V$ satisfying the following three conditions:
\roster\item"(i)"
$\forall 1\le i\le n: I_i\scirc I_i = id_V$,

\item"(ii)"
$\forall 1\le i<j\le n : \[I_i,I_j\] = 0$ and

\item"(iii)"
we have a direct sum decomposition
$$
V = 
\bigoplus_{P\subset \{1, \dots, n\}} \repprod I_P(X)
=
X \oplus \bigoplus_{k=1}^n \quad \bigoplus_{1\le i_1<i_2<\cdots<i_k\le n} I_{i_k}(\cdots (I_{i_2}(I_{i_1}(X))) \cdots)
$$

\endroster

\enddefinition

\definition{Remark}
Using the fact that for an odd map $I$ we have $\[I,I\]=I\scirc I + I\scirc I$, we can write conditions (i) and (ii) in a single formula as $\[I_i,I_j\] = 2 \delta_{ij} \cdot id_V$ for all $1\le i,j\le n$ (where $\delta_{ij}$ is the Kronecker $\delta$).

\enddefinition

In terms of subspaces indexed by $n$ odd parameters $\lambda_1, \dots, \lambda_n\in \CA_1$ we obtain the subspaces $X_{\lambda_1, \dots, \lambda_n}$ as
$$
\multline
X_{\lambda_1, \dots, \lambda_n}
=
\Bigl\{\, \sum_{P\subset \{1, \dots, n\}} \repprod{\lambda I}_P(x)
\mid x\in X\,\Bigr\}
\\
=
\Bigl\{\, x+\sum_{k=1}^n \sum_{1\le i_1<i_2<\cdots<i_k\le n} \lambda_{i_k}I_{i_k} ( \cdots (\lambda_{i_2}I_{i_2}(\lambda_{i_1}I_{i_1}(x))))
\mid x\in X\,\Bigr\}
\mapob.
\endmultline
\tag\memorize\theoremnummer={ndimensionaloddfamily}
$$
Since the maps $\lambda_{i_k}I_{i_k}$ are even, commute and are of square zero ($\lambda_i^2=0$), it is not hard to show that we can write this in the following shortened form
$$
X_{\lambda_1, \dots, \lambda_n}
=
\bigl\{\, \eexp^{\lambda_1I_1 + \cdots + \lambda_nI_n}(x)
\mid x\in X\,\bigr\}
\mapob,
\formula{ndimensionaloddfamilyinexpform}
$$
where, as in the case $n=1$, the exponential map $\eexp^A = \sum_{k=0}^\infty A^k/k!$ is a finite sum because the map $A=\lambda_1I_1 + \cdots + \lambda_nI_n$ is nilpotent (of order less than $n+1$). Once we have seen this way to write $X_{\lambda_1, \dots, \lambda_n}$, we can give a more or less independent motivation for our definition along the same lines as given for $n=1$. Forgetting about tensoring with $\CA$ when needed (seee the discussion for $n=1$ for what to do), we start to recall that the set $C^\infty(\CA_1^n,X)$ of smooth functions of $n$ odd variables into $X$ is a graded vector space over $\RR$ isomorphic to $\bigwedge \RR^n \otimes \body X$: a smooth function $f$ of $n$ odd variables with values in the graded vector space $X$ is completely determined by $2^n$ elements $x_P\in \body X$ indexed by a subset $P\subset \{1, \dots, n\}$ as (see also \recalf{localexpressionsmoothfunction})
$$
f(\lambda_1, \dots, \lambda_n)
=
x_\emptyset + \sum_{k=1}^n\ \sum_{1\le i_1<\cdots<i_k\le n} \lambda_{i_k}\cdots\lambda_{i_2}\cdot\lambda_{i_1}\cdot x_{i_1\dotsi_k} 
=
\sum_{P\subset \{1, \dots, n\}}
\repprod{\lambda}_P\cdot x_P
\mapob.
$$

Now let $I_i:X_i\to V$ be maps defined on an increasing set of subspaces $X_i\subset V$ defined by $X_1=X$ and $X_{i+1}=X_i+I_i(X_i)$. We then can define the Fourier-Berezin transformation $\Cal F_I$ relative to the maps $I_i$ as the map $\Cal F_I:C^\infty(\CA_1^n, X)\to \body V$ given by
$$
\align
\Cal F_I(f)
&=
\int\cdots\int 
\bigl(id_V + \lambda_nI_n\bigr) \scirc \cdots \scirc \bigl(id_V + \lambda_2I_2\bigr)
\\&
\kern3cm
\scirc \bigl(id_V + \lambda_1I_1\bigr)
\bigl( f(\lambda_1, \dots, \lambda_n)\bigr) \ \extder\lambda_1 \cdots\extder \lambda_n
\\&
=
\sum_{P,Q\subset \{1, \dots, n\}}
\int\cdots\int \repprod{\lambda I}_Q (\repprod{\lambda}_P \cdot x_P)
\ \extder\lambda_1 \cdots\extder \lambda_n
\\&
=
\sum_{P\subset \{1, \dots, n\}} \varepsilon_P \cdot \repprod{I}_{P^c}(x_P)
\mapob,
\formula{generalFourierBerezinformula}
\endalign
$$
where $P^c = \{1, \dots, n\}\setminus P$ denotes the complement of $P$ and where the sign $\varepsilon_P=\pm1$ is determined by
$$
\repprod{\lambda I}_{P^c}\bigl(\repprod{\lambda}_P \cdot x_P\bigr) = \varepsilon_P \cdot \lambda_n \cdots \lambda_2\cdot \lambda_1\cdot \repprod I_{P^c}(x_P)
\mapob,
$$
which means that we have normalized the Berezin integration by the condition
$$
\int \lambda_n \cdots \lambda_2 \cdot \lambda_1 \ \extder\lambda_1 \cdots\extder \lambda_n
=
1
\mapob.
$$
As in the case $n=1$ this implies that $\Cal F_I$ is a bijection onto $\body V$ if and only if the maps $I_i$ are injective and we have a direct sum decomposition like in \recalt{defoddfamdecomposition}. If this is the case, then one can extend the maps $I_i$ to the whole of $V$ such that the conditions (i) and (ii) of \recalt{defoddfamdecomposition} are satisfied. And if we finally realize that we then have the equality
$$
\bigl(id_V + \lambda_nI_n\bigr) \scirc \cdots \scirc \bigl(id_V + \lambda_2I_2\bigr)
\scirc \bigl(id_V + \lambda_1I_1\bigr)
=
\eexp^{\sum_{i=1}^n \lambda_i I_i} 
\mapob,
$$
then the circle is closed and we see that our definition of an $n$-dimensional odd family decomposition is exactly what is needed to make a Fourier-Berezin decomposition of $V$ in terms of functions with values in $X$.

\proclaim{\thmm{invariantsubspaceoftwodimfamily}{Proposition}}
Let $(W,I)$ be an odd family of subspaces of $V$, let $(X,J)$ be an odd family of subspaces of $W$ and let $\lambda, \mu\in \CA_1$ be two fixed elements. If $\lambda\mu\neq 0$ and if the set $X_{\mu, \lambda}$ \recalf{twodimoddfamily} is invariant under the action $\Phi$, then the following four conditions are satisfied:

\roster\item"(i)"
$\forall g\in G$: $\Phi_g(X)\subset X$,

\item"(ii)"
$\forall g\in G$: $\[\Phi_g,I\](X) \subset X$,

\item"(iii)"
$\forall g\in G$: $\[\Phi_g,J\](X) \subset X$ and

\item"(iv)"
$\forall g\in G$: $\[\[\Phi_g,I\],J\](X) \subset X$ with $\[\[\Phi_g,I\],J\] = \[\Phi_g,I\] \scirc J + J\scirc \[\Phi_g,I\]$ the graded commutator.

\endroster
Conversely, if these four conditions are satisfied, then all subspaces $X_{\rho,\nu}$, $\nu,\rho\in\CA_1$ \recalf{twodimoddfamily} are invariant under the action $\Phi$.

\endproclaim

\demo{Proof}
The fact that $X_{\mu, \lambda}$ is invariant under $\Phi$ means that for all $x\in X$ there exists $y\in X$ such that
$$
\multline
\Phi_g\bigl(x+ \mu  J(x) + \lambda  I(x)+ \lambda  I(\mu J(x))\bigr) 
= 
y+ \mu J(y) + \lambda I(y)+ \lambda I(\mu J(y))
\mapob.
\endmultline
\tag \memorize\theoremnummer={Xlambdamuinvariance}
$$
If we multiply this equation on the left by $\mu\lambda$, we get the equation
$$
\mu\lambda\Phi_g(x) = \mu\lambda y
\mapob.
\formula{secondXlambdamuinvariance}
$$
Since $y$ belongs to $X$, it follows that $\mu\lambda\Phi_g(x)$ belongs to $X$. If we now take $x\in \body V$, then the map $g\mapsto \Phi_g(x)$ is smooth.  By \recalt{lincombinWindependenceresultwithmfddep} and the fact that $\lambda\mu\neq0$, it follows that $\Phi_g(x)$ lies in $X$. But then, since $X$ is generated by $\body X$ and since $\Phi_g$ is linear, $\Phi_g(x)$ lies in $X$ for all $x\in X$, which is condition (i).

If we multiply \recalf{Xlambdamuinvariance} on the left by $\mu$ and substitute the result \recalf{secondXlambdamuinvariance} (note that the linear maps $I$ and $J$ are odd, as are $\lambda$ and $\mu$, so $\lambda I(\mu J(y)) = \mu\lambda I(J(y)) = I(J(\mu\lambda y))$), we get the equation
$$
\mu\Phi_g(x) + \mu\lambda \Phi_g(I(x))
=
\mu y + \mu\lambda I(\Phi_g(x))
\mapob.
\formula{thirdXlambdamuinvariance}
$$
Since we alreay know that $\mu y$ and $\mu\Phi_g(x)$ belong to $X$, it follows that $\mu\lambda\[\Phi_g,I\](x) $ belongs to $X$. But then we can apply the same argument as for $\mu\lambda\Phi_g(x)$ to show that $\[\Phi_g,I\](x)$ itself belongs to $X$, which is condition (ii).

We now solve \recalf{thirdXlambdamuinvariance} for $\mu y$, which gives
$$
\mu y = \mu\Phi_g(x) + \mu\lambda\[\Phi_g,I\](x) 
\mapob,
\formula{fourthXlambdamuinvariance}
$$
and we substitute this together with \recalf{secondXlambdamuinvariance} in \recalf{Xlambdamuinvariance} and multiply the result by $\lambda$ on the left. This gives us the equation
$$
\lambda\Phi_g(x) + \lambda\mu \Phi_g(J(x))
= 
\lambda y + \lambda \mu J(\Phi_g(x)) 
\mapob.
\formula{fifthXlambdamuinvariance}
$$
Again because we already know that $\Phi_g(x)$ and $y$ belong to $X$, it follows that $\lambda\mu\[\Phi_g,J\](x)$ belongs to $X$. And as before, it follows that $\[\Phi_g,J\](x)$ belongs to $X$, which is condition (iii).

Solving \recalf{fifthXlambdamuinvariance} for $\lambda y$, which gives
$
\lambda y = \lambda\Phi_g(x) + \lambda\mu \[\Phi_g,J\](x)
$,
and substituting this result as well as \recalf{fourthXlambdamuinvariance} and \recalf{secondXlambdamuinvariance} in \recalf{Xlambdamuinvariance}, we get the equation
$$
\multline
\Phi_g(x)+ \mu\[\Phi_g,J\](x) + \lambda\[\Phi_g,I\](x)- \lambda\mu\Phi_g(I(J(x))) 
\\
= 
y+\mu\lambda J(\[\Phi_g,I\](x))  +\lambda\mu I( \[\Phi_g,J\](x))+ \lambda\mu I(J(\Phi_g(x)))
\mapob.
\endmultline
\tag{\memorize\theoremnummer={inducedactiononWcasen=2}}
$$
As before it follows that
$$
\lambda\mu
\Bigl( \Phi_g(I(J(x))) + J(\[\Phi_g,I\](x)) - I( \[\Phi_g,J\](x)) - I(J(\Phi_g(x))) \Bigr)
$$
belongs to $X$, and thus, still by the same argument,
$$
\Phi_g(I(J(x))) + J\bigl(\[\Phi_g,I\](x)\bigr) - I\bigl( \[\Phi_g,J\](x)\bigr) - I\bigl(J(\Phi_g(x))\bigr) 
=
\[\[\Phi_g,I\],J\](x)
$$
belongs to $X$, which proves condition (iv).

\medskip

On the other hand, if these four conditions are satisfied, we can define, for each $x\in X$, an $y\in X$ by
$$
y= 
\Phi_g(x) + \rho\cdot \[\Phi_g,J\](x)  + \nu\cdot\[\Phi_g,I\](x)  + \rho\nu\cdot \[\[\Phi_g,I\],J\](x) 
\mapob.
$$
An elementary computation then shows that we have the equality
$$
\multline
\Phi_g\bigl(x+\rho\cdot J(x) + \nu\cdot I(x) +\rho\nu\cdot I(J(x))\bigr) 
\\= 
y+\rho\cdot J(y) +\nu\cdot I(y) + \rho\nu\cdot I(J(y))
\mapob,
\endmultline
$$
proving that $X_{\rho,\nu}$ is invariant under the action $\Phi$.
\QED\enddemo

The generalization of this result to a sequence of $n$ nested odd families of subspaces gives the following result, whose proof follows the same ideas and is left to the reader.

\proclaim{\thmm{generalndimfamilyinvariant}{Proposition}}
Suppose we have a sequence of couples $(W_i,I_i)$, $1\le i\le n$ such that the couple $(W_{{i}}, I_{i})$ is an odd family of subspaces of $W_{{i+1}}$, $1\le i<n$ and $(W_n,I_n)$ an odd family of subspaces of $V$. Associated to $\lambda_1, \dots, \lambda_n\in \CA_1$ we define a subspace $W_{\lambda_1,\dots, \lambda_n}\subset V$ by
$$
W_{\lambda_1,\dots, \lambda_n}
=
\Bigl\{\,\sum_{P\subset \{1, \dots, n\}} \repprod{\lambda I}_P(w)
\mid w\in W_1\,\Bigr\}
\mapob.
$$
If the $n$-fold product $\lambda_1\lambda_2\cdots\lambda_n$ is non-zero, and if the subspace $W_{\lambda_1,\dots, \lambda_n}$ is invariant under the action $\Phi$, then the following conditions are satisfied:
$$
\forall P\subset \{1, \dots, n\}\quad \forall g\in G : \repcomm{\Phi_g},I_P(W_1) \subset W_1
\mapob.
$$
Conversely, if these conditions are satisfied, then all subspaces $W_{\mu_1,\dots, \mu_n}\subset V$ are invariant under the action $\Phi$.

\endproclaim

\definition{\thmm{definvardecomposition}{Definition}}
An \stresd{invariant $n$-dimensional odd family decomposition of $V$} is an $n$-dimensional odd family decomposition $(W,I_1, \dots, I_n)$ of $V$ satisfying the conditions:
$$
\forall P\subset \{1, \dots, n\} \quad \forall g\in G : \repcomm{\Phi_g},I_P(W) \subset W
\mapob.
$$
This is equivalent to requiring that all subspaces $W_{\lambda_1, \dots, \lambda_n}$ \recalf{ndimensionaloddfamily} or \recalf{ndimensionaloddfamilyinexpform} are invariant under the action $\Phi$.

\enddefinition

\definition{Remark}
Instead of looking at the action $\Phi$ on a subspace $W_{\lambda_1, \dots, \lambda_n}$, we can also look at the induced action $\Psi$ on $W$ via the map $w\mapsto \eexp^{\sum_{i=1}^n \lambda_iI_i}(w)$ from $W$ to $W_{\lambda_1, \dots, \lambda_n}$. This induced action $\Psi$ depends (obviously) upon the odd parameters $\lambda_i$ and satisfies by definition 
$$
\eexp^{\sum_{i=1}^n \lambda_iI_i}(\Psi_g (w))
=
\Phi_g\bigl( \eexp^{\sum_{i=1}^n \lambda_iI_i}(w)\bigr)
\mapob.
$$
According to \recalf{inducedactiononWcasen=1}, for $n=1$ it is given as 
$$
w\mapsto \Phi_g(w) + \lambda \[\Phi_g,I\](w)
$$
and according to \recalf{inducedactiononWcasen=2} for $n=2$ it is given by
$$
w\mapsto \Phi_g(w) + \lambda \[\Phi_g,I\](w) + \mu \[\Phi_g,J\](w) +\mu \lambda \[\[\Phi_g, I\], J\](w)
\mapob.
$$
If we introduce for $M\in \End(V)$ the right-adjoint map $\RAd(M): \End(V) \to \End(V)$ by 
$$
\RAd(M)(X) = \[X,M\]
\mapob,
$$
then we can write these two cases as
$$
\gather
w\mapsto 
\Bigl(\Phi_g + \RAd{(\lambda I)}(\Phi_g)\Bigr)(w)
\quad\text{and}\quad
\\
w\mapsto 
\Bigl(\Phi_g + \RAd{(\lambda I)}(\Phi_g) + \RAd{(\mu J)}(\Phi_g) + \RAd{(\mu J)}\bigl(\RAd{(\lambda I)}(\Phi_g)\bigr)
\Bigr)(w)
\mapob.
\endgather
$$
But the maps $\RAd{(\lambda_iI_i)}$ are even, of square zero (beacuse $\lambda_i^2=0$) and commute (because of the (graded) Jacobi identity and the fact that the $I_i$ (graded) commute). It then is not hard to show that the general case can be written as
$$
w\mapsto \Psi_g(w) =  \bigl(\eexp^{\RAd{(\sum_{i=1}^n \lambda_iI_i})}(\Phi_g)\bigr)(w)
\mapob,
$$
simply because one can easily show the equality (as maps on $V$)
$$
\Phi_g \scirc \eexp^{\sum_{i=1}^n \lambda_iI_i}
=
\eexp^{\sum_{i=1}^n \lambda_iI_i} \scirc \eexp^{\RAd{(\sum_{i=1}^n \lambda_iI_i})}(\Phi_g)
\mapob.
$$
The separate maps $\repcomm{\Phi_g},I_P$ then can easily be recovered from the action $\Psi_g$ by taking the derivative with respect to the odd variables in $P\subset \{1, \dots, n\}$ (at zero values for the odd variables).

\enddefinition

\definition{Definition}
Let $(W,I_1, \dots, I_n)$ be an invariant $n$-dimensional odd family decomposition of $V$ and let $\lambda_1, \dots,\lambda_n\in \CA_1$ be $n$ odd elements. We will say that the subspace $W_{\lambda_1, \dots, \lambda_n}$ is \stresd{irreducible} if there exists no non-trivial graded subspace $X\subset W$ such that the corresponding subspace $X_{\lambda_1, \dots, \lambda_n}\subset W_{\lambda_1, \dots, \lambda_n}$ is invariant under $\Phi$. Non-trivial meaning $X\neq\{0\}$ and $X\neq W$.

\enddefinition

When one tries to find criteria for a subspace $W_{\lambda_1, \dots, \lambda_n}$ to be irreducible in terms of the graded subspace $W$ and the maps $I_i$, then the description becomes very rapidly very complicated. Let us briefly outline why by looking at subspaces $X_{\lambda_1, \dots, \lambda_n} \subset W_{\lambda_1, \dots, \lambda_n}$ for $n=1$ and $n=2$.

For $n=1$ the situation is rather easy: if a single $X_\lambda$ with $\lambda\neq0$ is invariant, then all $X_\mu$ are invariant. And this is the case if and only if $X$ itself is invariant under both $\Phi_g$ and $\[\Phi_g,I_1\]$. And $X_0=X$ is invariant if and only if $X$ is invariant under $\Phi$ (a tautology). However, as can be seen in examples, it is quite possible that $X_0$ (which is $X$) is invariant but that no $X_\lambda$ with $\lambda\neq0$ is invariant. There thus are just three cases: (i) all $W_{\lambda_1}$ are irreducible, which is the case if $W$ does not admit a non-trivial invariant graded subspace; (ii) all $W_{\lambda_1}$ with $\lambda_1\neq0$ are irreducible but $W_0$ is not, which is the case if and only if $W$ admits a non-trivial invariant graded subspace and if no invariant graded subspace of $W$ is also invariant under the maps $\[\Phi_g,I\]$; (iii) none of the spaces $W_{\lambda_1}$ is irreducible.

But already for $n=2$ the situation becomes more complicated. If we replace the maps $I_1$, $I_2$ by the maps $J_1$, $J_2$ defined as $J_i = \sum_{j=1}^2 a_{ij} I_j$ with $a_{ij}\in \RR$, then the triple $(W,J_1,J_2)$ satisfies the conditions of \recalt{defoddfamdecomposition} if and only if the matrix $a_{ij}$ is orthogonal (belongs to $O(2)$). As $W_{\lambda_1,\lambda_2}$ is defined by the map $\lambda_1I_1+\lambda_2I_2$ \recalf{ndimensionaloddfamilyinexpform}, it follows that \stress{the same} subspace is defined by the map $\mu_1J_1+\mu_2J_2$ with $\mu_j = \sum_{i} \mu_i a_{ij}$, simply because it is the same map. Since there exist examples in which a subspace $X_{\lambda_1=0,\lambda_2}$ is invariant but $X_{\lambda_1,\lambda_2=0}$ is not (one of the $\lambda$-parameters being zero and the other not), one could be tempted to think that it is simply a case of looking at zero values for some of the parameters. But had we performed the same computations using the $\mu$ parameters, neither $X_{\mu_1=0,\mu_2}$ nor $X_{\mu_1,\mu_2=0}$ would have been invariant. As it turns out, the right approach is to distinguish the following four cases for an invariant subspace $X_{\lambda_1,\lambda_2} \subset W_{\lambda_1,\lambda_2}$: (i) the product $\lambda_1\lambda_2$ is non-zero, in which case all three elements $\lambda_1$, $\lambda_2$ and $\lambda_1\lambda_2$ are independent over $\RR$ (in the $\RR$-algebra $\CA$); (ii) $\lambda_1\lambda_2=0$ and $\lambda_1$ and $\lambda_2$ are independent over $\RR$; (iii) $\lambda_1$ and $\lambda_2$ are dependent over $\RR$ but not both zero; and (iv) $\lambda_1=\lambda_2=0$. And then the fact that it is an invariant subspace has to be translated into terms involving only the graded subspace $W$ and the maps $I_1$ and $I_2$. 

As one can imagine, the case $n=3$ is even more complicated. The right approach is to look at the number of independent elements \stress{over $\RR$} among the $2^n -1 = 7$ elements $\lambda_1$, $\lambda_2$, $\lambda_3$, $\lambda_1\lambda_2$, $\lambda_1\lambda_3$, $\lambda_2\lambda_3$, $\lambda_1\lambda_2\lambda_3$ (and which ones). We will not go into a detailed analysis, but will restrict our attention to the case that will concern us in our study of the decomposition of the regular representation of our group $G$ of section \recall{naiveregularrep}.

\proclaim{\thmm{twodiminvdecompirredfornonzero}{Proposition}}
Let $(W,I_1,I_2)$ be an invariant odd family decomposition of $V$. If there is no non-trivial graded subspace $X\subset W$ which is invariant under all maps $\Phi_g$ and $[\Phi_g,I_1+aI_2]$ ($g\in G$ and $a\in\RR$), then all spaces $W_{\lambda_1,\lambda_2}$ with $\lambda_1\neq0$ are irreducible.

\endproclaim

\demo{Proof}
To show that none of the spaces $W_{\lambda_1,\lambda_2}$ with $\lambda_1\neq0$ admits a non-trivial invariant subspace, we suppose that $X_{\lambda_1,\lambda_2}$ is such a non-trivial invariant subspace. 
Invariance of $X_{\lambda_1,\lambda_2}$ implies that for each $x\in X$ (and each $g\in G$) there exists an $y\in X$ such that
$$
\Phi_g\bigl(  \eexp^{\sum_{i=1}^2 \lambda_iI_i}(x) \bigr) = \eexp^{\sum_{i=1}^2 \lambda_iI_i}(y)
\mapob,
$$
which expands to
$$
\multline
\Phi_g(
x+ \lambda_1I_1x 
+ \lambda_2I_2x 
+ \lambda_1\lambda_2I_2I_1x 
) 
= 
y+ \lambda_1I_1y + \lambda_2I_2y +
\lambda_1\lambda_2I_2I_1y 
\endmultline
\tag{\memorize\theoremnummer={conditioninvariantsubspacetwodimproof}}
$$
Multiplying this by $\lambda_1\lambda_2$ gives us the equation $\lambda_1\lambda_2\Phi_g(x) = \lambda_1\lambda_2y$, which we can substitute back into \recalf{conditioninvariantsubspacetwodimproof}. If we then multiply by $\lambda_i$, we can solve for $\lambda_iy$ and substitute back. This is exactly what we have done in the proof of \recalt{invariantsubspaceoftwodimfamily}. As in that proof, we obtain the following expression for $y$ in terms of $x$:
$$
y
=
\Phi_gx + \lambda_1[\Phi_g,I_1]x + \lambda_2[\Phi_g,I_2]x +  \lambda_1\lambda_2 [[\Phi_g,I_2],I_1]x
\mapob.
\formula{defxinvariantsubspaceVtwodimproof}
$$
We now restrict attention to $x\in \body X$, which garantees that the maps $$
g\mapsto \Phi_g x
\quad,\quad
g\mapsto [\Phi_g,I_i]x
\quad\text{and}\quad
g\mapsto [[\Phi_g,I_2],I_1]x
\formula{smoothmapsintwodimproof}
$$
are smooth. The important observation is that these $x$ generate the whole of $X$ (over $\CA$).

We now reason with respect to the different possibilities of independent (over $\RR$!) coefficients among the $4$ coefficients appearing in \recalf{defxinvariantsubspaceVtwodimproof}: $1$ (the coefficient of $\Phi_gx$), $\lambda_1$, $\lambda_2$ and $\lambda_1\lambda_2$, just as suggested in the discussion above. Now if we have a relation with real coefficients among these four elements, we can take the body map to conclude that the coefficient of $1$ must be zero. And if we separate into homogeneous parts, we get a relation among $\lambda_1$ and $\lambda_2$ for the odd part and a relation ``among'' $\lambda_1\lambda_2$ for the even part. For $\lambda_1\lambda_2$ there thus are just two possibilities: zero (if the coefficient of $\lambda_1\lambda_2$ in the relation is non-zero) or not.

\itemize
It is not hard to show that if $\lambda_1\lambda_2\neq0$, then $\lambda_1$ and $\lambda_2$ are also independent over $\RR$. All four elements are thus independent over $\RR$ in $\CA$. Hence by \recalt{lincombinWindependenceresultwithmfddep} all four maps in \recalf{smoothmapsintwodimproof} lie in $X$. In particular $\Phi_gx$ and $[\Phi_g, I_1]x\in X$. Since $X$ is generated by our choice of $x$'s and since this map is linear, it follows that $X$ is invariant under $\Phi_g$ and $[\Phi_g,I_1+aI_2]$ for $a=0$.

\itemize
If $\lambda_1\lambda_2=0$, there are three possible cases to consider for the two elements $\lambda_1$ and $\lambda_2$: zero, one or two independent relations among them. 

--- If there is no relation, then $\lambda_1$ and $\lambda_2$ are independent over $\RR$. As $\lambda_1\lambda_2=0$, we only have three terms in \recalf{defxinvariantsubspaceVtwodimproof} with three independent elements over $\RR$ in $\CA$. Hence by \recalt{lincombinWindependenceresultwithmfddep} all three maps in \recalf{smoothmapsintwodimproof} lie in $X$. In particular $\Phi_gx$ and $[\Phi_g, I_1]x\in X$. Again since $X$ is generated by our choice of $x$'s and since this map is linear, it follows that $X$ is invariant under $\Phi_g$ and $[\Phi_g,I_1+aI_2]$ for $a=0$.

--- If there is a single independent relation of the form $b\lambda_1+c\lambda_2=0$ (not both $b$ and $c$ zero), we may assume that $c\neq0$. Because when $c=0$, we must have $b\neq0$ and then the relation would tell us that we have $\lambda_1=0$, contrary to the assumption $\lambda_1\neq0$. We thus can rewrite the relation as $\lambda_2=a\lambda_1$ with $a=-b/c$. And then we can apply \recalt{lincombinWindependenceresultwithmfddepBIS} and \recalt{notabenenonindeplincomb} to conclude that the two functions $\Phi_gx$ and the linear combination $[\Phi_g,I_1]x + a[\Phi_g,I_2]x$ belong to $X$. And as before, this implies that $X$ is invariant under the maps $\Phi_g$ and $[\Phi_g,I_1+aI_2]$.

--- The case of two independent relations among $\lambda_1$ and $\lambda_2$ is excluded because then both $\lambda_1$ and $\lambda_2$ must be zero, which is contrary to the assumption $\lambda_1\neq0$. 

\itemize
The conclusion is that in all cases $X$ is invariant under $\Phi_g$ and under a map of the form $[\Phi_g,I_1+aI_2]$ for some $a\in\RR$. Since this is excluded by hypothesis, there does not exist a non-trivial invariant subspace of $W_{\lambda_1,\lambda_2}$.
\QED\enddemo

\proclaim{\thmm{threediminvdecompirredfornonzero}{Proposition}}
Let $(W,I_1,I_2,I_3)$ be an invariant odd family decomposition of $V$. If there is no non-trivial graded subspace $X\subset W$ which is invariant under all maps $\Phi_g$ and $[\Phi_g,I_1+aI_2+bI_3]$ ($g\in G$ and $a,b\in\RR$), then all spaces $W_{\lambda_1,\lambda_2,\lambda_3}$ with $\lambda_1\neq0$ are irreducible.

\endproclaim

\demo{Proof}
To show that none of the spaces $W_{\lambda_1,\lambda_2,\lambda_3}$ with $\lambda_1\neq0$ admits a non-trivial invariant subspace, we proceed as in the proof of \recalt{twodiminvdecompirredfornonzero}. We thus suppose that $X_{\lambda_1,\lambda_2,\lambda_3}$ is such a non-trivial invariant subspace. 
Invariance of $X_{\lambda_1,\lambda_2,\lambda_3}$ implies that for each $x\in X$ (and each $g\in G$) there exists an $y\in X$ such that
$$
\Phi_g\bigl(  \eexp^{\sum_{i=1}^3 \lambda_iI_i}(x) \bigr) = \eexp^{\sum_{i=1}^3 \lambda_iI_i}(y)
\mapob,
$$
which expands to
$$
\multline
\Phi_g\Bigl(
x+\sum_{i=1}^3 \lambda_iI_ix 
+ \sum_{1\le i<j\le 3} \lambda_i\lambda_jI_jI_ix 
+ \lambda_1\lambda_2\lambda_3I_3I_2I_1x 
\Bigr) 
\\
= 
y+ \sum_{i=1}^3\lambda_iI_iy +
\sum_{1\le i<j\le 3} \lambda_i\lambda_jI_jI_iy
+ \lambda_1\lambda_2\lambda_3I_3I_2I_1y 
\endmultline
\tag{\memorize\theoremnummer={conditioninvariantsubspacesecondproof}}
$$
Multiplying this by $\lambda_1\lambda_2\lambda_3$ gives us the equation $\lambda_1\lambda_2\lambda_3\Phi_g(x) = \lambda_1\lambda_2\lambda_3y$, which we can substitute back into \recalf{conditioninvariantsubspacesecondproof}. If we then multiply by a double product $\lambda_i\lambda_j$, we can solve for $\lambda_i\lambda_jy$ and substitute back. Continuing this process, we finally end up with the following expression for $y$ in terms of $x$:
$$
y
=
\Phi_gx + \sum_i \lambda_i[\Phi_g,I_i]x + \sum_{i<j} \lambda_j\lambda_i [[\Phi_g,I_i],I_j]x
+
\lambda_3\lambda_2\lambda_1[[[\Phi_g,I_1],I_2],I_3]x
\mapob.
\formula{defxinvariantsubspaceVsecondproof}
$$
We now restrict attention to $x\in \body X$, which garantees that the maps $$
g\mapsto \Phi_g x
\quad,\quad
g\mapsto [\Phi_g,I_i]x
\quad,\quad
g\mapsto [[\Phi_g,I_i],I_j]x
\quad\text{and}\quad
g \mapsto [[[\Phi_g,I_1],I_2],I_3]x
$$
are smooth. As in the proof of \recalt{twodiminvdecompirredfornonzero}, the important observation is that these $x$ generate the whole of $X$ (over $\CA$).

We now reason with respect to the different possibilities of independent coefficients among the $8$ coefficients appearing in \recalf{defxinvariantsubspaceVsecondproof}: $1$ (the coefficient of $\Phi_gx$), $\lambda_1$, $\lambda_2$, $\lambda_3$, $\lambda_1\lambda_2$, $\lambda_1\lambda_3$, $\lambda_2\lambda_3$, $\lambda_1\lambda_2\lambda_3$. 

\itemize
It is not hard to show that if $\lambda_1\lambda_2\lambda_3\neq0$, then all these $8$ elements are independent over $\RR$ in $\CA$. Hence by \recalt{lincombinWindependenceresultwithmfddep} we must have (among others) $\Phi_gx\in X$ as ``coefficient'' of $1$ and $[\Phi_g, I_1]x\in X$ as ``coefficient'' of $\lambda_1$. Since $X$ is generated by our choice of $x$'s and since this map is linear, it follows that $X$ is invariant under $\Phi_g$ and $[\Phi_g,I_1+aI_2+bI_3]$ for $a=b=0$.

\itemize
If $\lambda_1\lambda_2\lambda_3=0$, we suppose that we have a relation (with real coefficients) among the remaining $7$ elements. By taking the body of this relation, it follows that the element $1$ does not appear (\ie, has coefficient zero in the relation). Next we use parity to split such a relation into two relations, one among the three odd elements $\lambda_i$ and one among the three even elements $\lambda_i\lambda_j$ (remember that $1$ does not appear). Solving these relations for independent ones (and $1$ is among the independent ones!) thus will tell us in particular that some of the $\lambda_i$ are linear combinations of the others. Using \recalt{lincombinWindependenceresultwithmfddepBIS}, \recalt{notabenenonindeplincomb} it then follows that the vectors appearing as ``coefficients'' of these independent elements belong to $X$.

--- The first case to consider is when all three $\lambda_i$ are independent (over $\RR$), in which case it follows (among others) that $\Phi_gx\in X$ as ``coefficient of $1$ and $[\Phi_g,I_1]x\in X$ as ``coefficient'' of $\lambda_1$ and thus (as always because $x\in \body X$ generate $X$ and because the map is linear) that $\Phi_g$ and $[\Phi_g,I_1]$ preserve $X$.

--- If there is a single independent relation among the $\lambda_i$, we can write it as $\sum_i a_i\lambda_i=0$ for some $a_i\in \RR$, not all zero. But the case $a_2=a_3=0$ is also excluded as we suppose by hypothesis that $\lambda_1\neq0$. Without loss of generality we thus may assume that $a_2\neq0$, which allows us to write $\lambda_2 = -(a_1\lambda_1+a_3\lambda_3)/a_2$ and to say that $\lambda_1$ and $\lambda_3$ are independent over $\RR$. It then follows in particular that $[\Phi_g, I_1-(a_1/a_2)I_2]x\in X$, this being the ``coefficient'' of $\lambda_1$. And hence $X$ is invariant under $[\Phi_g, I_1-(a_1/a_2)I_2]$. And if we look at the ``coefficient'' of $1$, we find as before that $X$ is invariant under $\Phi_g$.

--- If there are two independent relations among the $\lambda_i$, this means that we can express two of them as multiples of the third. Since $\lambda_1\neq0$, we may assume that it is $\lambda_2$ and $\lambda_3$ that are multiples of $\lambda_1$, say $\lambda_2=a\lambda_1$ and $\lambda_3=b\lambda_1$. But then the ``coefficient'' of $\lambda_1$ will be $[\Phi_g,I_1+xI_2+yI_3]x$, which thus belongs to $X$ and thus $X$ is invariant under $[\Phi_g,I_1+xI_2+yI_3]$. And as always, looking at the ``coefficient'' of $1$ tells us that $X$ is invariant under $\Phi_g$.

--- The case of three independent relations among the $\lambda_i$ would mean that we have $\lambda_i=0$ for all $i$, which is excluded by the condition $\lambda_1\neq0$.

\itemize
The conclusion is that in all cases $X$ is invariant under $\Phi_g$ and under a map of the form $[\Phi_g,I_1+aI_2+bI_3]$ for some $a,b\in\RR$. Since this is excluded by hypothesis, there does not exist a non-trivial invariant subspace of $W_{\lambda_1,\lambda_2,\lambda_3}$.
\QED\enddemo

\head{\sectionnum. The regular representation revisited }\endhead

In order to cast our decomposition of the (left) regular representation of $G$ into the setting of odd family decompositions, we introduce $4$ odd maps $I_0$, $I_4$, $I_5$ and $I_6$ on $V$ (the space of ($L^2$) functions on $G$) defined as
$$
\alignat2
I_0 
&
= i(\beta+ \tfrac12a^1\alpha^4 - \tfrac12a^3\alpha^5) -i \partial_{\beta}
&
\qquad
I_5
&
=
i\alpha^5 - i(\partial_{\alpha^5} + \tfrac12 a^3 \partial_\beta)
\\
I_6
&
=
i\alpha^6 - i\partial_{\alpha^6}
&
I_4 
&
= i\alpha^4 -i (\partial_{\alpha^4} - \tfrac12 a^1 \partial_{\beta})
\mapob.
\endalignat
$$
If we apply a change of coordinates $\beta\to\gamma = \beta+ \tfrac12 a^1\alpha^4 -\tfrac12  a^3\alpha^5$, then these maps\slash operators take the simpler form
$$
I_0 
= i\gamma -i \partial_{\gamma}
\quad,\quad
I_4 
= i\alpha^4 -i \partial_{\alpha^4} 
\quad,\quad
I_5
=
i\alpha^5 - i\partial_{\alpha^5}
\quad,\quad
I_6
=
i\alpha^6 - i\partial_{\alpha^6}
\mapob.
$$
It is not hard to show that these maps verify the conditions $I_i\scirc I_i= id_V$ and $i\neq j \Rightarrow \[I_i,I_j\] = 0$. We define the graded subsapce $E\subset V$ as consisting of functions independent of the odd variables $\alpha^4$, $\alpha^5$, $\alpha^6$ and $\beta$, \ie, $E$ is the space of ($L^2$) functions of the four even (real) coordinates $a^1$, $a^2$, $a^3$ and $b$. It thus has no odd dimensions. It is not hard to show that $(E,I_0, I_4, I_5, I_6)$ is a $4$-dimensional odd family decomposition of $V$. More precisely, $I_i$ with $i=4,5,6$ adds a factor $\alpha^i$ to a function in $E$, whereras $I_0$ adds a factor $\gamma$, which is essentially $\beta$. The various graded subspaces $(I)_P(E)$ thus contain the corresponding factors $\alpha$ (and\slash or a modified $\beta$).

To make the link with the subspaces $V_{(\ell_0,\ell_2,\ell_3,\lambda_0, \lambda_4)}$ given in \recals{naiveregularrep} we introduce the space $W_{(\ell_0, \ell_2, \ell_3)} $ consisting of functions (on $G$) of the form
$$
f(a^i, \alpha^j, b,\beta)
=
h(a^1, \alpha^4, \alpha^5, \alpha^6, \gamma) 
\cdot\eexp^{i\ell_2a^2}
\cdot\eexp^{i\ell_3a^3}
\cdot\eexp^{i\ell_0(b+\frac12a^1a^2)}
\mapob,
$$
with $\gamma= \beta+ \tfrac12 a^1\alpha^4 -\tfrac12  a^3\alpha^5$ as before and where $h$ is a function of one even and $4$ odd coordinates. In terms of the functions $h$ the action of $\gh\in G$ is given by
$$
\align
(\Psi_\gh h)(a^1, \alpha^i, \gamma)
&
=
h(a^1 - \ah^1, \alpha^i - \alphah^i, \gamma - \alphah^4 a^1 + \ah^3 \alpha^5 + \tfrac12 \ah^1 \alphah^4 - \tfrac12 \ah^3 \alphah^5 - \betah)
\\&
\qquad
\cdot \eexp^{-i(\ell_2 \ah^2 + \ell_3 \ah^3)}
\cdot \eexp^{i\ell_0(-\bh + \frac12 \ah^1\ah^2 - \ah^2 a^1 + \frac12 \alphah^5 \alpha^5 - \frac12 \alphah^6 \alpha^6)}
\mapob.
\endalign
$$
These spaces are invariant under the action $\Phi$ and they are preserved by the maps $I_i$ (note that there is an $\alpha^4$ and $\alpha^5$ dependence via $\gamma$). 
In terms of $L^2$ functions this means that we have written $V$ as the (usual) direct integral of the Fourier modes
$$
V = \int W_{(\ell_0, \ell_2, \ell_3)} \ \extder \ell_0\,\extder \ell_2 \,\extder \ell_3
\mapob,
$$
given by the map
$$
f(a^i, \alpha^j, b,\beta) = 
\int h_{(\ell_0,\ell_2,\ell_3)}(a^1, \alpha^j, \gamma)
\cdot \eexp^{i(\ell_2a^2+ \ell_3a^3+\ell_0(b + \frac12  a^1 a^2))}
\ \extder  \ell_0 \, \extder \ell_2 \,\extder  \ell_3 
\mapob.
$$

A direct computation shows that the graded subspace $X_{(\ell_0, \ell_2, \ell_3)} \subset W_{(\ell_0, \ell_2, \ell_3)}$ of functions independent of $\alpha^4$ and $\beta$\slash$\gamma$, \ie, of the form
$$
f(a^i, \alpha^j, b,\beta)
=
t(a^1, \alpha^5, \alpha^6) 
\cdot\eexp^{i\ell_2a^2}
\cdot\eexp^{i\ell_3a^3}
\cdot\eexp^{i\ell_0(b+\frac12a^1a^2)}
\mapob,
\formula{fouriermodeswithoutlambdas}
$$
is invariant and that $(X_{(\ell_0, \ell_2, \ell_3)} , I_0, I_4)$ is an invariant $2$-dimensional odd family decomposition of $W_{(\ell_0, \ell_2, \ell_3)} $. 
The link with the subspaces $V_{(\ell_0,\ell_2,\ell_3,\lambda_0, \lambda_4)}$ described in \recals{naiveregularrep} is given by \recalf{ndimensionaloddfamilyinexpform}:
$$
V_{(\ell_0,\ell_2,\ell_3,\lambda_0, \lambda_4)}
=
\bigl\{\,\eexp^{\lambda_0I_0 + \lambda_4I_4} f \mid f\in X_{(\ell_0, \ell_2, \ell_3)} \,\bigr\}
\mapob.
\formula{bothnonzerointermsofIs}
$$

\proclaim{\thmm{ellnulnonzeropmisirreducible}{Proposition}}
The subspace $X_{(\ell_0, \ell_2, \ell_3)}$ can be written as the direct sum of two invariant graded subspaces  $X_{(\ell_0, \ell_2, \ell_3)}^\pm \subset X_{(\ell_0, \ell_2, \ell_3)}$:
$$
X_{(\ell_0, \ell_2, \ell_3)} = X_{(\ell_0, \ell_2, \ell_3)}^+\oplus X_{(\ell_0, \ell_2, \ell_3)}^-
\mapob.
$$
These subspaces are also invariant under all maps $\[\Phi_g,I_4\]$. If $\ell_0$ is non-zero, these two subspaces are the only non-trivial invariant subspaces; they are thus in particular irreducible.
Moreover, the subspaces $V^\pm_{(\ell_0,\ell_2,\ell_3,0, \lambda_4)}$ defined in \recalt{firstdecomppm} are related to $X_{(\ell_0, \ell_2, \ell_3)}^\pm$ by
$$
V^\pm_{(\ell_0,\ell_2,\ell_3,0, \lambda_4)}
=
\bigl\{\,\eexp^{\lambda_4I_4} f \mid f\in X^\pm_{(\ell_0, \ell_2, \ell_3)} \,\bigr\}
\mapob.
$$

\endproclaim

\demo{Proof}
Mimicking the proof of \recalt{firstdecomppm} we first note that the action of $G$ on the elements of $X_{(\ell_0, \ell_2, \ell_3)}$ in terms of the functions $t$ \recalf{fouriermodeswithoutlambdas} (see also \recalf{firstfouriermode}, \recalf{overalltransformation}) is given by the formula
$$
\multline
(\Psi_\gh t)(a^1, \alpha^5, \alpha^6)
=
t(a^1 - \ah^1, \alpha^5 - \alphah^5, \alpha^6 - \alphah^6) \cdot
\eexp^{-i\ell_0( \ah^2 a^1 - \frac12\alphah^5\alpha^5 + \frac12 \alphah^6\alpha^6)}\cdot
\\
\eexp^{-i\ell_2\ah^2}
\cdot\eexp^{-i\ell_3\ah^3}
\cdot\eexp^{-i\ell_0(\bh - \frac12 \ah^1\ah^2 )}
\mapob.
\endmultline
$$
As in the proof of \recalt{firstdecomppm} it is not hard to see that the graded subspaces $X_{(\ell_0, \ell_2, \ell_3)}^\pm \subset X_{(\ell_0, \ell_2, \ell_3)}$ of functions of the form
$$
t(a^1, \alpha^5, \alpha^6)
=
h_\epsilon(a^1, \alpha^5+\epsilon\alpha^6) 
\cdot \eexp^{-\frac i2\ell_0\epsilon\alpha^5\alpha^6}
\formula{onlyonealphainoddfamdecomp}
$$
(with $\epsilon=\pm1$ and $h_\epsilon$ a function of one even and one odd variable) transform under the action of $\Psi_\gh$ as
$$
\multline
(\Psi_\gh t)(a^1, \alpha^5, \alpha^6)
=
h_\epsilon(a^1 - \ah^1, (\alpha^5+\epsilon\alpha^6) - (\alphah^5+\epsilon\alphah^6)) \cdot 
\eexp^{- \frac i2\ell_0\epsilon\alpha^5\alpha^6} \cdot
\\
\eexp^{-i\ell_0( \ah^2 a^1 - \frac12 (\alphah^5 - \epsilon\alphah^6)(\alpha^5 + \epsilon\alpha^6) 
)}\cdot
\\
\eexp^{-i\ell_2\ah^2}\cdot
\eexp^{-i\ell_3\ah^3}\cdot
\eexp^{-i\ell_0(\bh - \frac12 \ah^1\ah^2 +\frac12\epsilon \alphah^5\alphah^6)}
\mapob.
\endmultline
$$
This shows that these spaces are invariant under the $G$-action. And again as in the proof of \recalt{firstdecomppm} one can show that this gives a direct sum decomposition into two invariant subspaces
$$
X_{(\ell_0, \ell_2, \ell_3)}
=
X_{(\ell_0, \ell_2, \ell_3)}^+
\oplus
X_{(\ell_0, \ell_2, \ell_3)}^-
\mapob.
$$

An elementary computation shows that the action of $\[\Phi_g,I_4\]$ in terms of the functions $t$  \recalf{fouriermodeswithoutlambdas} is given by
$$
([\Psi_\gh,I_4]\,t)(a^1, \alpha^5, \alpha^6)
=
-\alphah^4\cdot (\Psi_\gh t)(a^1, \alpha^5, \alpha^6)
\mapob.
$$
Since we already know that we have $\Psi_\gh(X_{(\ell_0, \ell_2, \ell_3)}^\pm) \subset X_{(\ell_0, \ell_2, \ell_3)}^\pm$, it follows that the $X_{(\ell_0, \ell_2, \ell_3)}^\pm$ are also invariant under the maps $\[\Phi_g,I_4\]$.

Next we have to show that these two spaces are irreducible when $\ell_0\neq0$. If we don't write the (obligatory) factors $\eexp^{-\frac i2\ell_0\epsilon\alpha^5\alpha^6}$, $\eexp^{i\ell_2a^2}$, $\eexp^{i\ell_3a^3}$  and $\eexp^{i\ell_0(b+\frac12a^1a^2)}$, we have to deal with functions $h_\epsilon$ of the two variables $a^1, \xi$ which transform under the $G$ action as
$$
\multline
(\Psi_\gh h_\epsilon)(a^1, \xi)
=
h_\epsilon(a^1 - \ah^1, \xi - (\alphah^5+\epsilon\alphah^6)) \cdot 
\eexp^{-i\ell_0( \ah^2 a^1 - \frac12 (\alphah^5 - \epsilon\alphah^6)\xi 
)}\cdot
\\
\eexp^{-i\ell_2\ah^2}\cdot
\eexp^{-i\ell_3\ah^3}\cdot
\eexp^{-i\ell_0(\bh - \frac12 \ah^1\ah^2 +\frac12\epsilon \alphah^5\alphah^6)}\mapob.
\endmultline
$$
If we restrict for the moment our attention to the subgroup in which only $\ah^1, \ah^2, \bh$ are non-zero, the group law is given as
$$
(\ah^1, \ah^2,\bh) \cdot (a^1, a^2, b) = (\ah^1+a^1, \ah^2+a^2, \bh+b +\tfrac12 (\ah^2a^1 - \ah^1a^2))
\mapob,
$$
which we recognize as the Heisenberg group.
On this subgroup we get the transformation property
$$
(\Psi_\gh h_\epsilon)(a^1, \xi)
=
h_\epsilon(a^1 - \ah^1, \xi) \cdot 
\eexp^{-i\ell_2\ah^2}\cdot
\eexp^{-i\ell_0(\bh + \ah^2 a^1  - \frac12 \ah^1\ah^2 
)}
\mapob,
$$
which we recognize as the irreducible representation of the Heisenberg group, provided we disregard the dependence of the odd coordinate $\xi$. Writing as before $h_\epsilon(a^1, \xi)
=
h_{\epsilon,0}(a^1) + \xi \cdot h_{\epsilon,1}(a^1)$, each of the two components $h_{\epsilon,0}$ and $h_{\epsilon,1}$ transforms as the irreducible representation of the Heisenberg group. Since the function $h_\epsilon$ is even when $h_{\epsilon,1}$ is zero and odd when $h_{\epsilon,0}$ is zero, it follows that the subspaces $V_{\epsilon,i} \subset X_{(\ell_0, \ell_2, \ell_3)}^\epsilon$ ($i=0,1$), defined as the set of those functions $h_\epsilon$ for which $h_{\epsilon,1-i}$ is zero, give us the decomposition into homogeneous parts of $X_{(\ell_0, \ell_2, \ell_3)}^\epsilon$:
$$
X_{(\ell_0, \ell_2, \ell_3)}^\epsilon = V_{\epsilon,0} \oplus V_{\epsilon,1}
\qquad,\qquad
\bigl( X_{(\ell_0, \ell_2, \ell_3)}^\epsilon\bigr)_i = V_{\epsilon,i}
\mapob.
$$
Moreover, each $V_{\epsilon,i}$ is invariant and irreducible under the Heisenberg subgroup.

If we now look at the generators of the $\alphah^5$ and $\alphah^6$ action (take the derivative with respect to these variables in the group action), we get
$$
h_\epsilon \mapsto -\fracp{h_\epsilon}{\xi} + \frac{i\ell_0}2\cdot \xi\cdot h_\epsilon
\qquad\text{and}\qquad
h_\epsilon \mapsto -\epsilon\fracp{h_\epsilon}{\xi} -\epsilon\cdot \frac{i\ell_0}2\cdot \xi\cdot h_\epsilon
\mapob.
\formula{infgenforalpha5and6}
$$
In terms of the decomposition $h_\epsilon = (h_{\epsilon,0},h_{\epsilon,1})$ this corresponds to the matrices
$$
\pmatrix 0 & -1 \\ \frac{i\ell_0}2 & 0 \endpmatrix
\qquad\text{and}\qquad
\pmatrix 0 & -\epsilon \\ -\epsilon\frac{i\ell_0}2 & 0 \endpmatrix
\mapob.
$$
If a subspace $H\subset X_{(\ell_0, \ell_2, \ell_3)}^\epsilon$ is $G$-invariant, then it must be invariant under these operations and also under linear combinations of these operations and thus in particular under the matrices ($\ell_0\neq 0$!)
$$
\pmatrix 0 & -1 \\ 0 & 0 \endpmatrix
\qquad\text{and}\qquad
\pmatrix 0 & 0 \\ 1 & 0 \endpmatrix
\mapob.
\formula{suitablematricescompltoHeisenberg}
$$
Now if $H$ is invariant, it is also invariant under the Heisenberg subgroup. Which implies that the subspaces $H\cap V_{\epsilon,i} \subset V_{\epsilon,i}$ are also invariant under the Heisenberg subgroup. But $V_{\epsilon,i}$ is irreducible under this subgroup, thus $H\cap V_{\epsilon,i}$ is either $V_{\epsilon,i}$ or $\{0\}$. Now suppose that $H\cap V_{\epsilon,0}=V_{\epsilon,0}$, then $V_{\epsilon,0}\subset H$. But if we then act with the second matrix of \recalf{suitablematricescompltoHeisenberg} on $V_{\epsilon,0}\subset H$ we get $V_{\epsilon,1}$, which by invariance of $H$ must also lie in $H$. But then $H= 
V_{\epsilon,0} \oplus V_{\epsilon,1} = X_{(\ell_0, \ell_2, \ell_3)}^\epsilon$. A similar argument applies when $H\cap V_{\epsilon,1}=V_{\epsilon,1}$. And if both intersections yield $\{0\}$, we recall that an invariant subspace is in particular a graded submodule, and thus the even and odd parts of $H$ are given by the intersections with the even and odd parts of $X_{(\ell_0, \ell_2, \ell_3)}^\epsilon$, i.e., by $H\cap V_{\epsilon,i}$. And thus if both are $\{0\}$, we must have $H=\{0\}$. This shows that $X_{(\ell_0, \ell_2, \ell_3)}^\epsilon$ is indeed irreducible.

To finish the proof, we have to show that there are no other non-trivial invariant subspaces when $\ell_0$ is non-zero. We thus suppose that $H$ is an invariant subspace (now of the whole space $X_{(\ell_0, \ell_2, \ell_3)}$, not of a subspace $X_{(\ell_0, \ell_2, \ell_3)}^\epsilon$) and we consider an element $h\in H$. Using the decomposition $X_{(\ell_0, \ell_2, \ell_3)}  = X_{(\ell_0, \ell_2, \ell_3)}^+ \oplus X_{(\ell_0, \ell_2, \ell_3)}^-$, there exists $h_\epsilon\in X_{(\ell_0, \ell_2, \ell_3)}^\epsilon$ such that we have $h=h_+\oplus h_-$. According to \recalf{infgenforalpha5and6}, if we add the infinitesimal generators associated to the $\alphah^5$ and $\alphah^6$ actions we get
$$
\operatorname{inf}(\alphah^5) + \operatorname{inf}(\alphah^6)
: h_+ \mapsto -2\fracp{h_+}{\xi} \text{ and } h_- \mapsto i\ell_0\xi h_-
$$
and for the difference of these two infinitesimal operators we get
$$
\operatorname{inf}(\alphah^5) - \operatorname{inf}(\alphah^6)
: h_+ \mapsto i\ell_0\xi h_+ \text{ and } h_- \mapsto -2\fracp{h_-}{\xi}
\mapob.
$$
We now act first with the sum and then with the difference of these two operators on our element $h\in H$. Writing as before $h_\epsilon(a^1,\xi) = h_{\epsilon,0}(a^1) + \xi h_{\epsilon,1}(a^1)$, we thus obtain
$$
\multline
\bigl(\,
\bigl(\,\operatorname{inf}(\alphah^5) - \operatorname{inf}(\alphah^6) \,\bigr)
\bigl(\, \operatorname{inf}(\alphah^5) - \operatorname{inf}(\alphah^6) \,\bigr) (h_+\oplus h_-)
\,\bigr)(a^1,\xi)
\\
=
(-2i\ell_0 \xi h_{+,1}(a^1)) \oplus (-2i\ell_0  h_{-,0}(a^1) )
\mapob.
\endmultline
$$
Since $H$ is invariant, this image belongs to $H$. But $-2i\ell_0 \xi h_{+,1}(a^1)$ is the odd part of this image and $-2i\ell_0  h_{-,0}(a^1)$ its even part. Hence $\xi h_{+,1}(a^1)$ and $h_{-,0}(a^1)$ separately belong to $H$ (being a graded subspace). 
On the other hand, if we apply first the difference and then the sum, the parities are reversed and we obtain that $\xi h_{-,1}(a^1)$ and $h_{+,0}(a^1)$ separately belong to $H$. But then $h_+(a^1,\xi) = h_{+,0}(a^1) + \xi h_{+,1}(a^1)$ and $h_-(a^1,\xi) = h_{-,0}(a^1) + \xi h_{-,1}(a^1)$ separately belong to $H$. In particular we thus have $h_\epsilon\in H\cap X_{(\ell_0, \ell_2, \ell_3)}^\epsilon$. But $X_{(\ell_0, \ell_2, \ell_3)}^\epsilon$ is irreducible, so if $h_\epsilon\neq0$, then $X_{(\ell_0, \ell_2, \ell_3)}^\epsilon \subset H$. This shows that there are only four possibilities for $H$: $\{0\}$, $X_{(\ell_0, \ell_2, \ell_3)}^\pm$ and $X_{(\ell_0, \ell_2, \ell_3)}$.
\QED\enddemo

\proclaim{Corollary}
For $\ell_0\neq0$ the representations $V_{(\ell_0,\ell_2,\ell_3,0, \lambda_4)}^\pm $ are irreducible.

\endproclaim

\proclaim{Proposition}
If $\ell_0$ is non-zero, then the representations $V_{(\ell_0,\ell_2,\ell_3,\lambda_0, \lambda_4)}$ \recalf{bothnonzerointermsofIs} with $\lambda_0\neq0$ 
are irreducible.

\endproclaim

\demo{Proof}
According to \recalt{twodiminvdecompirredfornonzero} it suffices to show that there is no non-trivial subspace of $X_{(\ell_0, \ell_2, \ell_3)}$ which is invariant under all maps $\Phi_g$ and $\[\Phi_g,I_0+xI_4\]$. In \recalt{ellnulnonzeropmisirreducible} we have shown that, when $\ell_0\neq0$, the only non-trivial invariant subspaces of $X_{(\ell_0, \ell_2, \ell_3)}$ are $X_{(\ell_0, \ell_2, \ell_3)}^\pm$. It thus suffices to show that these subspaces are \stress{not} invariant under the maps $\[\Phi_g,I_0+xI_4\]$. An elementary computation shows that we have, in terms of the functions $t$ \recalf{fouriermodeswithoutlambdas}
$$
\multline
([\Psi_\gh,I_0+xI_4]t)(a^1, \alpha^5, \alpha^6)
=
\\
i(-\betah + \tfrac12 \ah^1\alphah^4 - \tfrac12 \ah^3\alphah^5 - \alphah^4 a^1 + \ah^3 \alpha^5 - x\alphah^4)\cdot (\Psi_\gh t)(a^1, \alpha^5, \alpha^6)
\mapob.
\endmultline
$$
This computation is a (partial) confirmation that $(X_{(\ell_0, \ell_2, \ell_3)}, I_0, I_4)$ is indeed an invariant 2-dimensional odd family decomposition. But more important is that it shows that the space of functions of the form \recalf{onlyonealphainoddfamdecomp} is not invariant under the maps $[\psi_\gh, I_0+xI_4]$, simply beacuse of the single $\alpha^5$ factor in front of $(\Psi_\gh t)(a^1, \alpha^5, \alpha^6)$, which does not come in the combination $\alpha^5 + \epsilon \alpha^6$. It follows that $V_{(\ell_0,\ell_2,\ell_3,\lambda_0, \lambda_4)}$ is irreducible.
\QED\enddemo

To analyse the case $\ell_0=0$, we start looking at the explicit transformation property of elements in $X_{(0, \ell_2, \ell_3)}$ in terms of the functions $t$ \recalf{fouriermodeswithoutlambdas} (see also \recalf{firstfouriermode}, \recalf{overalltransformation}) when $\ell_0=0$. This gives us
$$
(\Psi_\gh t)(a^1, \alpha^5, \alpha^6)
=
t(a^1 - \ah^1, \alpha^5 - \alphah^5, \alpha^6 - \alphah^6) \cdot
\eexp^{-i\ell_2\ah^2}
\cdot\eexp^{-i\ell_3\ah^3}
\mapob.
$$
With the results of \recals{naiveregularrep} in mind, we introduce the subspaces $Y_{(\ell_2,\ell_3)}\subset X_{(0,\ell_2,\ell_3)}$ of functions independent of $\alpha^6$. It turns out that $(Y_{(\ell_2,\ell_3)},I_0, I_4, I_6)$ is an invariant $3$-dimensional odd family decomposition of $W_{(0, \ell_2, \ell_3)}$ for which the corresponding subspaces indexed by odd parameters are exactly the spaces $V_{(0,\ell_2,\ell_3,\lambda_0,\lambda_4),(\lambda_6)}$:
$$
V_{(0,\ell_2,\ell_3,\lambda_0,\lambda_4),(\lambda_6)}
=
\{\eexp^{\lambda_0I_0+\lambda_4I_4+\lambda_6I_6} f \mid f\in Y_{(\ell_2,\ell_3)} \}
\mapob.
$$

\proclaim{Proposition}
The representations $V_{(0,\ell_2,\ell_3,\lambda_0, \lambda_4),(\lambda_6)}$ with $\lambda_0\neq0$ 
are irreducible.

\endproclaim

\demo{Proof}
According to \recalt{threediminvdecompirredfornonzero}, $V_{(0,\ell_2,\ell_3,\lambda_0, \lambda_4),(\lambda_6)}$ with $\lambda_0\neq0$ is irreducible if there does not exists a graded subspace of $Y_{(\ell_2,\ell_3)}$ which is invariant under the action of $\Phi_g$ and $\[\Phi_g,I_0+xI_4+yI_6\]$. An element $f$ of $Y_{(\ell_2,\ell_3)}$ is of the form (compare with \recalf{fouriermodeswithoutlambdas})
$$
f(a^i, \alpha^j, b,\beta)
=
s(a^1, \alpha^5) 
\cdot\eexp^{i\ell_2a^2}
\cdot\eexp^{i\ell_3a^3}
$$
for some function $s$ of one even and one odd variable.
The action of $\Phi_\gh$ and $\[\Phi_\gh,I_0+xI_4+yI_6\]$ on such a function is given (in terms of the function $s$) by
$$
\align
(\Phi_\gh s)(a^1, \alpha^5)
&
=
s(a^1-\ah^1, \alpha^5-\alphah^5)
\cdot\eexp^{-i\ell_2 \ah^2}
\cdot \eexp^{-i\ell_3\ah^3}
\\
(\[\Phi_\gh,I_0+xI_4+yI_6\] s)(a^1,  \alpha^5)
&=
-i\bigl(\betah + \alphah^4 a^1 - \ah^3\alpha^5 -\tfrac12 \ah^1\ah^4 
+ \tfrac12 \ah^3\alphah^5 
\\&
\qquad\qquad\qquad
+ x\alphah^4 
+ y\alphah^6\bigr)
\cdot 
(\Phi_\gh s)(a^1, \alpha^5)
\mapob.
\endalign
$$

If we compute the generators of the  $\ah^1$ and $\alphah^4$ action (taking the derivative with respect to these variables in the group action), we find the operators
$$
s\mapsto -\fracp s{a^1}
\qquad,\qquad
s\mapsto -i ( a^1 + x)\cdot s
\mapob.
$$
If we shift the (even) coordinate $a^1$ by (the real amount) $x$ to $c=a^1+x$, this is exactly the action of the Heisenberg algebra (with operators $s\mapsto \partial_cs$ and $s\mapsto cs$). 
Computing the generators of the $\ah^3$ and $\alphah^5$ action, we find the maps
$$
s\mapsto -i\ell_3\,s
\qquad,\qquad
s\mapsto -\fracp s{\alpha^5}
\qquad,\qquad
s\mapsto i\alpha^5\,s 
\mapob.
\formula{additionaltoheisenberginell0=0}
$$
Now recall that a function $s(a^1, \alpha^5)$ (which should be written in terms of the shifted coordinate $c$ instead of $a^1$) can be written as $s(a^1, \alpha^5) = s_0(a^1) +\alpha^5\cdot s_1(a^1)$ with $s_i$ functions in $L^2(\RR)$. In terms of the decomposition $s\cong (s_0,s_1)$, the last two transformations in \recalf{additionaltoheisenberginell0=0} are given by
$$
(s_0,s_1) \mapsto (-s_1, 0 )
\qquad,\qquad
(s_0,s_1) \mapsto (0,i s_0) 
\mapob.
$$
We thus are in exactly the same situation as in the proof of \recalt{ellnulnonzeropmisirreducible} (except that we have an action of the Heisenberg algebra instead of the Heisenberg group). As in the proof of \recalt{ellnulnonzeropmisirreducible} we thus may conclude that there does not exist a non-trivial invariant subspace.
\QED\enddemo

To finish our search for the decomposition into irreducible components, it thus remains to analyse the representations $V_{(0,\ell_2,\ell_3,0,\lambda_4),(\lambda_6)}$, which are given in terms of the maps $I_4$ and $I_6$ by
$$
V_{(0,\ell_2,\ell_3,0,\lambda_4),(\lambda_6)}
=
\{\eexp^{\lambda_4I_4+\lambda_6I_6} f \mid f\in Y_{(\ell_2,\ell_3)} \}
\mapob.
$$ 
See this way, these spaces represent a $2$-dimensional odd family decomposition of the space $R_{(\ell_2,\ell_3)}$ defined as
$$
R_{(\ell_2,\ell_3)}
=
Y_{(\ell_2,\ell_3)}\oplus I_4(Y_{(\ell_2,\ell_3)}) \oplus I_6(Y_{(\ell_2,\ell_3)}) \oplus I_6(I_4(Y_{(\ell_2,\ell_3)}))
\mapob.
$$
There are two ways to see that these spaces decompose into $1$-dimensional invariant spaces. The fast way is to introduce the subspace $S_{(\ell_2,\ell_3)} \subset Y_{(\ell_2,\ell_3)}$ of functions that are independent of $\alpha^5$, meaning that the elements of $S_{(\ell_2,\ell_3)}$ are just ($L^2$) functions of a single even variable $a^1$. It then is easy to show that $(S_{(\ell_2,\ell_3)}, I_4,I_5,I_6)$ is a $3$-dimensional invariant odd family decomposition of $R_{(\ell_2,\ell_3)}$. If we finally introduce the ($1$-dimensional) spaces $C_{(\ell_1,\ell_2,\ell_3)}$ consisting of functions (on $G$) of the form
$$
f(a^i, \alpha^j, b,\beta) = c\cdot \eexp^{i(\ell_1a^1+\ell_2a^2+\ell_3a^3)}
\mapob,
$$
then $S_{(\ell_2,\ell_3)}$ is a direct integral of these spaces:
$$
S_{(\ell_2,\ell_3)}
=
\int C_{(\ell_1,\ell_2,\ell_3)} \ \extder\ell_1
\mapob.
$$
Moreover, this is an invariant decomposition which is compatible with the odd family decomposition $(S_{(\ell_2,\ell_3)}, I_4,I_5,I_6)$. More precisely, each subspace $C_{(\ell_1,\ell_2,\ell_3)}$ of $S_{(\ell_2,\ell_3)}$ (actually a generalized subspace in the sense of direct integral as opposed to direct sum) not only is invariant under all maps $\Phi_g$, but also under all commutators $\[\[\Phi_g,(I)]]_P$ with $P\subset \{4,5,6\}$.

If one feels uneasy about distributing direct integrals over the odd-family decomposition, one can start with the subspace $R_{(\ell_2,\ell_3)}$, which is the space of functions on $G$ of the form
$$
f(a^i, \alpha^j, b,\beta)
=
s(a^1,\alpha^4, \alpha^5, \alpha^6) 
\cdot\eexp^{i\ell_2a^2}
\cdot\eexp^{i\ell_3a^3}
\mapob.
$$
The $G$-action on functions of this form, in terms of the function $s$, is given by
$$
(\Phi_\gh s)(a^1,\alpha^4, \alpha^5, \alpha^6) 
=
s(a^1 - \ah^1,\alpha^4 - \alphah^4, \alpha^5 - \alphah^5, \alpha^6 - \alphah^6)  \cdot\eexp^{-i\ell_2\ah^2}
\cdot\eexp^{-i\ell_3\ah^3}
\mapob.
$$
We thus can perform a Fourier transform by introducing the spaces $R_{(\ell_1,\ell_2,\ell_3)}$ of functions of $G$ of the form
$$
f(a^i, \alpha^j, b,\beta)
=
t(\alpha^4, \alpha^5, \alpha^6) 
\cdot\eexp^{i(\ell_1a^1+\ell_2a^2+\ell_3a^3)}
\mapob,
$$
which gives us the invariant direct integral decomposition
$$
R_{(\ell_2,\ell_3)}
=
\int R_{(\ell_1,\ell_2,\ell_3)}\ \extder \ell_1
\mapob.
$$
The space $C_{(\ell_1,\ell_2,\ell_3)}$ introduced above is (also) the subspace of $R_{(\ell_1,\ell_2,\ell_3)}$ of functions independent of the odd variables $\alpha^4, \alpha^5, \alpha^6$. And it is not hard to show that $(C_{(\ell_1,\ell_2,\ell_3)}, I_4, I_5, I_6)$ is an invariant $3$-dimensional odd-family decomposition of $R_{(\ell_1,\ell_2,\ell_3)}$. The difference between this second approach and the first approach is that here we start with the direct integral decomposition and then introduce the odd-family decomposition, whereas in the first approach we start with the odd-family decomposition and then perform the direct integral decomposition, which we then have to ``distribute'' over the odd-family decomposition.

The final remark is that we have the equality
$$
V_{(0,\ell_2,\ell_3,0,\lambda_4),(\ell_1,\lambda_5,\lambda_6)}
=
\{\,\eexp^{\lambda_4I_4 + \lambda_5I_5 + \lambda_6I_6}f \mid f\in C_{(\ell_1,\ell_2,\ell_3)} \,\}
\mapob.
$$

\definition{Summary}
In order to find the irreducible components of $V$ in terms of invariant odd family decompositions, we started by applying a $3$-dimensional Fourier transform $V = \int W_{(\ell_0, \ell_2, \ell_3)}\ \extder\ell_0\,\extder\ell_2\,\extder\ell_3$. Elements of $W_{(\ell_0, \ell_2, \ell_3)}$ are functions which depend upon $4$ odd variables. Inside the various $W_{(\ell_0, \ell_2, \ell_3)}$ we introduced subspaces of functions depending on less and less odd variables as 
$$
\matrix
& &
X_{(\ell_0, \ell_2, \ell_3)}^\pm &\subset &X_{(\ell_0, \ell_2, \ell_3)} &\subset &W_{(\ell_0, \ell_2, \ell_3)}
\\
S_{(\ell_2,\ell_3)} &\subset &
Y_{(\ell_2,\ell_3)} &\subset &
X_{(0,\ell_2,\ell_3)} 
\\
\oversetalign{$0$ odd variables}\to{\ }&&\oversetalign{$1$ odd variable}\to{\ }
&&
\oversetalign{$2$ odd variables}\to{\ }
&&\oversetalign{$4$ odd variables}\to{\ }
\endmatrix
$$
As a last decomposition we applied a Fourier transform $S_{(\ell_2,\ell_3)} = \int C_{(\ell_1, \ell_2, \ell_3)} \ \extder\ell_1$. With these spaces, the link with the decomposition given in \recals{naiveregularrep} is given by the following table.
$$
\matrix
V_{(\ell_0, \ell_2, \ell_3, \lambda_0, \lambda_4)}
&
\rlap{\hss$\leftrightarrow$}
&
\hfill\eexp^{\lambda_0I_0+\lambda_4I_4} 
&\kern-0.9em  X_{(\ell_0, \ell_2, \ell_3)} \hfill
&
\text{irreducible for $\ell_0\neq0\neq\lambda_0$.}\hfill
\\
V_{(\ell_0, \ell_2, \ell_3, 0, \lambda_4)}^\pm
&
\rlap{\hss$\leftrightarrow$}
&
\hfill\eexp^{\lambda_4I_4} 
&\kern-0.9em  X_{(\ell_0, \ell_2, \ell_3)}^\pm \hfill
&
\text{irreducible for $\ell_0\neq0$.}\hfill
\\
V_{(0, \ell_2, \ell_3, \lambda_0, \lambda_4),(\lambda_6)}
&
\rlap{\hss$\leftrightarrow$}
&
\hfill\eexp^{\lambda_0I_0+\lambda_4I_4+\lambda_6I_6} 
&\kern-0.9em  Y_{( \ell_2, \ell_3)}\hfill
&
\text{irreducible for $\lambda_0\neq0$.}\hfill
\\
V_{(0, \ell_2, \ell_3, 0, \lambda_4),(\ell_1, \lambda_5, \lambda_6)}
&
\rlap{\hss$\leftrightarrow$}
&
\hfill\eexp^{\lambda_4I_4+\lambda_5I_5+\lambda_6I_6} 
&\kern-0.9em C_{(\ell_1, \ell_2, \ell_3)}\hfill
&
\text{irreducible.}\hfill
\endmatrix
$$

\enddefinition

\head{ \sectionnum. Determination of the coadjoint orbits }\endhead

In this section we determine the coadjoint orbits of our supergroup. It turns out that there are four families of orbits: a family of $0\vert0$ dimensional orbits with the trivial symplectic form, a family of $2\vert2$ dimensional orbits with an even symplectic form, a family of $2\vert2$ dimensional orbits with an odd symplectic form and a fourth family of $3\vert3$ dimensional orbits with a mixed symplectic form. The existence of orbits with a mixed symplectic form was already estabilshed in \cite{Tu10}.

A basis for the left-invariant $1$-forms on $G$ is given by $\extder a^i$, $\extder \alpha^j$ and
$$
\align
&\extder b - \tfrac12 a^2\extder a^1 +\tfrac12 a^1\extder a^2 - \tfrac12 \extder \alpha^5 \cdot \alpha^5 + \tfrac12 \extder \alpha^6 \cdot \alpha^6
\\
&\extder \beta^1 - \tfrac12 \alpha^4 \extder a^1 - \tfrac12 \alpha^5 \extder a^3 + \tfrac12 a^1 \extder \alpha^4 + \tfrac12 a^3 \extder \alpha^5
\ .
\endalign
$$
The corresponding basis of left-invariant vector fields is given by $\partial_{b}$, $\partial_{\beta}$ and
$$
\alignat3
&\partial_{a^1} + \tfrac12 a^2\, \partial_{b} + \tfrac12 \alpha^4\, \partial_{\beta}
&&\quad,\quad
\partial_{a^2} - \tfrac12 a^1 \partial_{b}
&&\quad,\quad
\partial_{a^3} + \tfrac12 \alpha^5\,\partial_{\beta}
\\
&\partial_{\alpha^4} - \tfrac12 a^1 \,\partial_{\beta}
&&\quad,\quad
\partial_{\alpha^5} + \tfrac12 \alpha^5 \,\partial_{b} - \tfrac12 a^3 \,\partial_{\beta}
&&\quad,\quad
\partial_{\alpha^6} - \tfrac12 \alpha^6\,\partial_{b} 
\mapob.
\endalignat
$$
These vector fields can be identified with the given basis of $E$ in the order $k_0$, $k_1$, $e_1$, \dots, $e_6$. For the Lie algebra this means that the only non-zero commutators (among these basis vectors) are given by
$$
\gather
[e_2,e_1] = - [e_1,e_2] =[e_5,e_5] = -[e_6,e_6] = k_0
\\
[e_4,e_1] = - [e_1,e_4] = [e_5,e_3] = - [e_3,e_5] = k_1
\rlap{\mapob.}
\endgather
$$
A direct computation shows that the triple product $g\gh g\mo$ is given by
$$
\multline
g\gh g\mo = (\ah^1, \ah^2, \ah^3, \bh + a^2 \ah^1 - a^1 \ah^2 + \alphah^5 \alpha^5 - \alphah^6 \alpha^6, 
\\
\alphah^4, \alphah^5, \alphah^6, \betah + \ah^1 \alpha^4 - a^1 \alphah^4 + \ah^3 \alpha^5 - a^3 \alphah^5)
\mapob.
\endmultline
$$
For the adjoint action we get the formul{\ae} $\Ad(g)k_0 = k_0$, $\Ad(g)k_1 = k_1$ and
$$
\gather
\Ad(g)e_1 = e_1 + a^2 k_0 + \alpha^4 k_1
\quad,\quad
\Ad(g)e_2 = e_2 - a^1 k_0
\quad,\quad
\Ad(g)e_3 = e_3 + \alpha^5 k_1
\\
\Ad(g)e_4 = e_4 - a^1 k_1
\quad,\quad
\Ad(g)e_5 = e_5 + \alpha^5 k_0 - a^3 k_1
\quad,\quad
\Ad(g)e_6 = e_6 - \alpha^6 k_0
\mapob.
\endgather
$$

The (left-)dual Lie algebra $\Liealg g^*$ is (isomorphic to) the graded vector space $E$. Seen as a graded manifold it has dimension $8\vert8$ with even coordinates $x_1$, $x_2$, $x_3$, $\xb_4$, $\xb_5$, $\xb_6$, $y_0$, $\yb_1$ and odd coordinates $\xib_1$, $\xib_2$, $\xib_3$, $\xi_4$, $\xi_5$, $\xi_6$, $\etab_0$, $\eta_1$. These are the even and odd parts of the coordinates of a point $\mu\in \Liealg g^*$ with respect to the (left-dual) basis $e_1^*$, $e_2^*$, $e_3^*$, $e_4^*$, $e_5^*$, $e_6^*$, $k_0^*$, $k_1^*$ of $\Liealg g^*$ determined by the formula
$$
\mu = 
\sum_{i=1}^3 
(x_i+\xib_i)\cdot e_i^* + 
\sum_{i=4}^6 (\xb_i+\xi_i)\cdot e_i^* + 
(y_0+\etab_0)\cdot k_0^* + (\yb_1+\eta_1)\cdot k_1^*
\mapob.
$$
The names are chosen such that the unbarred coordinates represent the even part $\mu_0$ of $\mu\in \Liealg g^*$ and the barred coordinates represent the odd part $\mu_1$, while latin characters represent even coordinates and greek characters represent odd coordinates.

For the coadjoint action of $g=(a,b)$ we obtain (from the adjoint action) that the basis vectors $e_i^*$ are left unchanged and that we have
$$
\align
\Coad(g)k_0^* 
&
= k_0^* - e_1^* a^2 + e_2^* a^1 -e_5^* \alpha^5 + e_6^* \alpha^6
\\
\Coad(g)k_1^* 
&
= k_1^* - e_1^* \alpha^4 - e_3^* \alpha^5 + e_4^* a^1 + e_5^* a^3
\mapob,
\endalign
$$
which gives us in terms of coordinates
$$
\gather
x_1\mapsto x_1 - y_0 \cdot a^2 - \eta_1 \cdot \alpha^4
\quad ,\qquad
x_2 \mapsto x_2 + y_0 \cdot a^1
\quad ,\qquad
x_3\mapsto x_3 - \eta_1 \cdot \alpha^5
\\
\xi_4\mapsto \xi_4 + \eta_1\cdot a^1
\quad ,\qquad
\xi_5 \mapsto \xi_5 + y_0 \cdot \alpha^5 + \eta_1 \cdot a^3
\quad ,\qquad
\xi_6 \mapsto \xi_6 - y_0 \cdot \alpha^6 
\endgather
$$
and 
$$
\gather
\xib_1\mapsto \xib_1 - \etab_0 \cdot a^2 - \yb_1 \cdot \alpha^4
\quad ,\qquad
\xib_2 \mapsto \xib_2 + \etab_0 \cdot a^1
\quad ,\qquad
\xib_3\mapsto \xib_3 - \yb_1 \cdot \alpha^5
\\
\xb_4\mapsto \xb_4 + \yb_1\cdot a^1
\quad ,\qquad
\xb_5 \mapsto \xb_5 + \etab_0 \cdot \alpha^5 + \yb_1 \cdot a^3
\quad ,\qquad
\xb_6 \mapsto \xb_6 - \etab_0 \cdot \alpha^6 
\mapob,
\endgather
$$
while the coordinates $y_0$, $\yb_1$, $\etab_0$ and $\eta_1$ remain unchanged.

For the fundamental vector fields  $(v,z)^{\Liealg g^*}$ associated to the element $(v,z) = \sum_{i=1}^6 v^i\cdot e_i + \sum_{i=0}^1 z^i k_i \in \Liealg g\,$  we obtain
$$
\multline
(v,z)^{\Liealg g^*} =  
(v^2y_0 - v^4\eta_1)\fracp{}{x_1} -v^1y_0\fracp{}{x_2} - v^5\eta_1\fracp{}{x_3}
-v^1\eta_1\fracp{}{\xi_4} 
\\
- (v^5y_0+v^3\eta_1)\fracp{}{\xi_5} + v^6y_0\fracp{}{\xi_6}
+(v^2\etab_0 + v^4\yb_1)\fracp{}{\xib_1} - v^1\etab_0\fracp{}{\xib_2} 
\\
+ v^5\yb_1\fracp{}{\xib_3} - v^1\yb_1\fracp{}{\xb_4} + (v^5\etab_0 
- v^3\yb_1)\fracp{}{\xb_5} - v^6\etab_0\fracp{}{\xb_6}
\mapob.
\endmultline
$$

We now look at the orbit $\Orbit_{\mu_o}$ through a fixed point $\mu_o\in \Liealg g^*$ with real coordinates (which we need if we want to be sure that the orbit is a bona fide graded manifold embedded in $\Liealg g^*$). Since the coordinates of $\mu_o$ are real, we have in particular that the odd coordinates of $\mu_o$ are zero: $\xi_i(\mu_o) = \xib_j(\mu_o) = \etab_0(\mu_o) = \eta_1(\mu_o)=0$. Since the coordinates $y_0$ and $\yb_1$ are constant on the orbit, they remain the same real constant. However, the other even coordinates are in general not constant on $\Orbit_{\mu_o}$, so we will denote the remaining even coordinates of $\mu_o$ by $\xnul_1, \xnul_2, \xnul_3, \xnulb_4, \xnulb_5, \xnulb_6\in \RR$. The odd coordinates $\etab_0 = \eta_1 = 0$ are also invariant under the coadjoint action, so they remain constant equal zero on the orbit. Hence the formul{\ae} for the coadjoint action reduce to:
$$
\gather
x_1 \mapsto x_1 - y_0 \cdot a^2
\ ,\ \ 
x_2 \mapsto x_2 + y_0 \cdot a^1
\ ,\ \ 
\xi_5 \mapsto \xi_5 + y_0 \cdot \alpha^5 
\ ,\ \ 
\xi_6 \mapsto \xi_6 - y_0 \cdot \alpha^6 
\\
\xib_1 \mapsto \xib_1 - \yb_1 \cdot \alpha^4
\ ,\ \ 
\xib_3 \mapsto \xib_3 - \yb_1 \cdot \alpha^5
\ ,\ \ 
\xb_4 \mapsto \xb_4 + \yb_1 \cdot a^1 
\ ,\ \ 
\xb_5 \mapsto \xb_5 + \yb_1 \cdot a^3
\rlap{\mapob,}
\endgather
$$
while
all other coordinates remain unchanged. For the fundamental vector fields the formula reduces to
$$
\multline
(v,z)^{\Liealg g^*} =  y_0 \cdot \Bigl(
v^2 \cdot \fracp{}{x_1} -
v^1 \cdot \fracp{}{x_2} -
v^5 \cdot \fracp{}{\xi_5} +
v^6 \cdot \fracp{}{\xi_6} \Bigr)
\\
+\yb_1 \cdot \Bigl(
v^4 \cdot \fracp{}{\xib_1} +
v^5 \cdot \fracp{}{\xib_3} -
v^1 \cdot \fracp{}{\xb_4} -
v^3 \cdot \fracp{}{\xb_5} 
\Bigr)
\mapob.
\endmultline
$$
We now distinguish four cases: (i) $y_0=\yb_1=0$, (ii) $y_0 \neq 0$ (but real!) and $\yb_1 = 0$,
(iii) $\yb_1 \neq 0$ and $y_0 = 0$, and (iv) $y_0 \cdot \yb_1 \neq 0$.

\subhead{Case (i)}\endsubhead
For $y_0=\yb_1=0$, the base point $\mu_o$ of the orbit $\Orbit_{\mu_o}$ is determined by the $6$ real values $\xnul_1, \xnul_2, \xnul_3, \xnulb_4, \xnulb_5, \xnulb_6$, while all other coordinates are zero. Since the coadjoint action reduces to the trivial action on such a point, $\Orbit_{\mu_o}=\{\mu_o\}$ is a point and its symplectic form is the $2$-form which is identically zero (which is indeed symplectic as there are no non-zero tangent vectors). This gives us a $6$-dimensional family of $0$-dimensional orbits labeled by $\xnul_1, \xnul_2, \xnul_3, \xnulb_4, \xnulb_5, \xnulb_6$.

\subhead{Case (ii)}\endsubhead
For $\yb_1=0$ and $y_0\neq0$, the base point $\mu_o$ is determined by the $7$ real values $\xnul_1$, $\xnul_2$, $\xnul_3$, $\xnulb_4$, $\xnulb_5$, $\xnulb_6$, $y_0$, while all other coordinates are zero. The orbit has dimension $2|2$ with even coordinates $x_1, x_2$ and odd coordinates $\xi_5, \xi_6$ (all other coordinates on $\Liealg g^*$ remain unchanged for such an orbit). Substituting the fundamental vector fields associated to basis elements $e_i$ in the formula for the symplectic form gives us the following
identities
$$
\contrf{\fracp{}{x_1}, \fracp{}{x_2}}{\omega} = \frac{1}{y_0}
\quad,\quad
\contrf{\fracp{}{\xi_5}, \fracp{}{\xi_5}}{\omega} = -\frac{1}{y_0}
\quad,\quad
\contrf{\fracp{}{\xi_6}, \fracp{}{\xi_6}}{\omega} = \frac{1}{y_0}
\mapob,
$$
all others being either zero or determined by graded skew-symmetry.
From these identities we deduce that the (even) symplectic form is given as
$$
\omega = (y_0)\mo \cdot (\ \extder x_2 \wedge \extder x_1 - \tfrac12 \,\extder\xi_5 \wedge \extder\xi_5 +
\tfrac12 \,\extder\xi_6 \wedge \extder\xi_6\ )
\mapob.
$$
Since $x_1$ and $x_2$ are coordinates on such an orbit, the different orbits are labeled by $\xnul_3$, $\xnulb_4$, $\xnulb_5$, $\xnulb_6$, $y_0$, thus giving a $5$-dimensional family of $2\vert2$-dimensional orbits with an even symplectic form.

\subhead{Case (iii)}\endsubhead
For $y_0=0$ and $\yb_1\neq0$, the base point $\mu_o$ is again determined by $7$ real values, but now  $\xnul_1$, $\xnul_2$, $\xnul_3$, $\xnulb_4$, $\xnulb_5$, $\xnulb_6$, $\yb_1$, while all other coordinates are zero.  The orbit dimension is still $2|2$, but now with even
coordinates $\xb_4,\xb_5$ and odd coordinates $\xib_1,\xib_3$ (as before, the other coordinates on $\Liealg g^*$ remain unchanged on this orbit). Here we
obtain for the symplectic form the identities
$$
\contrf{\fracp{}{\xib_1}, \fracp{}{\xb_4}, }{\omega} = \frac{1}{\yb_1}
\quad,\quad
\contrf{\fracp{}{\xib_3}, \fracp{}{\xb_5}, }{\omega} = \frac{1}{\yb_1}
\mapob,
$$
from which we deduce that the (odd) symplectic form is given as
$$
\omega = (\yb_1)\mo \cdot (\ \extder\xb_4 \wedge \extder \xib_1 + \extder\xb_5 \wedge \extder\xib_3\ )
\mapob.
$$
Here the $\xb_4$ and $\xb_5$ are coordinates on the orbit, so the different orbits are labeled by $\xnul_1$, $\xnul_2$, $\xnul_3$, $\xnulb_6$, $\yb_1$, giving a $5$-dimensional family of $2\vert2$-dimensional orbits with an odd symplectic form.

\subhead{Case (iv)}\endsubhead
In the last case $y_0\yb_1\neq0$ we have to be slightly more careful. The base point is determined by the $8$ real values $\xnul_1$, $\xnul_2$, $\xnul_3$, $\xnulb_4$, $\xnulb_5$, $\xnulb_6$, $y_0$, $\yb_1$, while all other coordinates are zero. But the combinations $s = y_0\xb_4 - \yb_1x_2$ and $\sigma = y_0\xib_3+\yb_1\xi_5$ are invariant under the group action. Since the base point of the orbit has real coordinates, $\sigma$ is zero at this point, and thus zero on the whole orbit. We introduce the change of coordinates on $\Liealg g^*$ involving only $x_2, \xb_4, \xib_3, \xi_5$ which change into $\xh_2, s, \sigma, \xih_5$ with $\xh_2 = x_2$ and $\xih_5 = \xi_5$. We then can use the coordinates $x_1, \xh_2, \xb_5, \xib_1, \xih_5, \xi_6$ as independent coordinates on the orbit, which thus has dimension $3\vert3$. In terms of these new coordinates (on $\Liealg g^*$) the fundamental
vector field is given as
$$
\multline
(v,z)^{\Liealg g^*} =  y_0 \cdot \Bigl(
v^2 \cdot \fracp{}{x_1} -
v^1 \cdot \fracp{}{\hat x_2}  -
v^5 \cdot \fracp{}{\hat \xi_5}   +
v^6 \cdot \fracp{}{\xi_6} \Bigr)
\\
+\yb_1 \cdot \Bigl(
v^4 \cdot \fracp{}{\xib_1}  -
v^3 \cdot \fracp{}{\xb_5} 
\Bigr)
\mapob.
\endmultline
$$
As before, substituting suitable basis vectors for $v$ in the
formula  for the symplectic form gives us the following identities
$$
\gather
\contrf{\fracp{}{x_1}, \fracp{}{\hat x_2} }{\omega} = \frac{1}{y_0}
\quad,\quad
\contrf{\fracp{}{\xib_1}, \fracp{}{\hat x_2} }{\omega} = \frac{1}{y_0}
\quad,\quad
\contrf{\fracp{}{\xb_5}, \fracp{}{\xih_5} }{\omega} = \frac{1}{y_0}
\\
\noalign{\vskip2\jot}
\contrf{\fracp{}{\xih_5}, \fracp{}{\xih_5}}{\omega} =
\frac{-1}{y_0}
\quad,\quad
\contrf{\fracp{}{\xi_6}, \fracp{}{\xi_6}}{\omega} = \frac{1}{y_0}
\mapob.
\endgather
$$
This results in the (non-homogeneous) symplectic form
$$
\omega = (y_0)\mo (\,\extder\xh_2 \wedge \extder x_1 -\tfrac12\, \extder\xih_5 \wedge \extder\xih_5 + \tfrac12 \,\extder\xi_6 \wedge \extder \xi_6 + \extder\xh_2\wedge \extder\xib_1  + \extder\xih_5 \wedge \extder\xb_5)
\mapob.
$$
In this case the invariants of the orbit are the components $\xnul_3$, $\xnulb_6$, $y_0$, $\yb_1$ and the combination $s^o = y_0\xnul_2 -\yb_1\xnulb_4$, giving a $5$-dimensional family of $3\vert3$-dimensional orbits with a mixed symplectic form.

\head{ \memorize\sectionnum={GQgeneral}. Geometric quantization of the coadjoint orbits }\endhead

In this section we discuss the general theory of geometric quantization of a symplectic supermanifold in the context of the coadjoint orbits of our supergroup $G$. 
As explained in \cite{Tu10} a prequantum bundle is a principal fiber bundle $Y\to \Orbit_{\mu_o}$ with a connection $1$-form $\Gamma$ whose curvature is the (Kostant-Kirillov-Souriau) symplectic form $\omega$. Since the symplectic form can be non-homogeneous, the structure group of $Y$ will be $(\CA_0/d\ZZ)\times \CA_1$ of dimension $1\vert1$ with $d\in (0,\infty)$. On $(\CA_0/d\ZZ)\times \CA_1$ we will use coordinates $(t,\tau)$ with $t$ the canonical even coordinate on $\CA_0$ modulo $d$ and $\tau$ the canonical coordinate on $\CA_1$. The connection $1$-form $\Gamma$ on $Y$, an even Lie algebra valued $1$-form on $Y$, then takes the form
$$
\Gamma = \Gamma_0 \otimes \partial_t + \Gamma_1 \otimes \partial_\tau
$$
with $\Gamma_i$ a $1$-form on $Y$ of parity $i$. It is useful to identify $\Gamma$ with the mixed $1$-form $\Gamma_0+\Gamma_1$ on $Y$ and in the sequel we will often pretend that we have the equality $\Gamma = \Gamma_0 + \Gamma_1$.

In our case the orbits all have a single global chart of the form $(\CA_0)^p \times (\CA_1)^q$, hence are simply connected and thus the prequantum bundle is unique and of the form $Y=\Orbit_{\mu_o} \times (\CA_0/d\ZZ)\times \CA_1$. Moreover, there exists a global potential $\Theta$ for the symplectic form $\omega$, \ie, $\extder\Theta = \omega$. Splitting $\Theta$ in its homogeneous parts, the connection $1$-form $\Gamma$ then is given as
$$
\Gamma_0 = \extder t + \Theta_0
\qquad,\qquad
\Gamma_1 = \extder \tau + \Theta_1
\mapob.
$$

The presence of the (canonical) momentum map on a coadjoint orbit implies that the infinitesimal action of $G$ on the orbit can be lifted to the prequantum bundle (in a way which is compatible with he momentum map). More precisely, if $(v,z) \in \Liealg g$ is an element of the Lie algebra, the infinitesimal generator $(v,z)^{Y}$ of the lifted action on $Y$ is determined by the equations
$$
\contrf{(v,z)^{Y}\restricted_{(\mu,t,\tau)}}\Gamma = \contrs{(v,z)}{\mu}
\qquad\text{and}\qquad
\Lieder((v,z)^{Y})\Gamma = 0
\mapob.
$$
In our examples, these infinitesimal actions on the prequantum bundles all integrate to an action of $G$, which we will provide.

\definition{Remark}
In the non-super orbit method, the lifting of the coadjoint action on an orbit to a prequantum bundle is equivalent to the existence of a character on the isotropy subgroup of the base point whose infinitesimal form is the given base point.

\enddefinition

As in the son-super case, an invariant polarization on a coadjoint orbit $\Orbit_{\mu_o}$ through $\mu_o\in \Liealg g^*$ (where $\mu_o$ has real coordinates) is a foliation which is maximal isotropic with respect to the symplectic form and which is invariant under the group action. And just as in the non-super case, such objects are in bijection with (graded) subalgebras $\Liealg h\subset \Liealg g$ containing the stabilizer subalgebra $\Liealg g_{\mu_o}$ and which are maximal with respect to the condition $\contrs{[\Liealg h, \Liealg h]}{\mu_o} = 0$. 

Since the vectors $k_0, k_1$ generate the center of $\Liealg g$, they always belong to a stabilizer subalgebra. On the other hand, since the commutator of two elements in $\Liealg g$ always lies in the graded subspace generated by $k_0, k_1$, it follows easily that any graded subspace containing the vectors $k_0, k_1$ is a graded subalgebra. If we have a homogeneous basis for $\Liealg h$, then any commutator between two elements of $\Liealg h$ is either a multiple of $k_0$ or $k_1$, but never both at the same time, just because of a parity argument. 

There are several equivalent ways to describe geometric quantization (without half-densities or half-forms). In the non-super case one can either look at the associated complex line bundle $L\to \Orbit_{\mu_o}$ and determine the sections that are covariantly constant in the direction of the (invariant) polarization, but one can also look directly at functions on the principal circle bundle $Y\to \Orbit_{\mu_o}$ that satisfy two conditions: (i) equivariance with respect to the action of the structure group $\SS^1$ and (ii) constant in the direction of the horizontally lifted (invariant) polarization. The lifted action of $G$ to the principal bundle $Y$ preserves the space of functions determined by the second method (using the induced action on functions $(g\cdot f)(y) = f(g\mo y)$). This is the representation of $G$ obtained by geometric quantization of the orbit.

In the super case we have to be a bit more devious because of the odd dimension of the structure group. We thus introduce two parameters, one even (real) $\hbar$ and one odd $\kappa$. The first is usually introduced in physics and can (without loss of generality) be put equal to $1$. We will not do so here in order to maintain a minimum of equality between the even and odd coordinates on the structure group. On the other hand, we cannot fix a specific value of $\kappa$ except $\kappa=0$ because else we obtain results that are no longer within the category of graded manifolds (see \cite{Tu04} for details). We thus are obliged to keep it as a free parameter. With this in mind, the two conditions imposed on functions $f:Y\to \CA^\CC$ on $Y$ by geometric quantization become \recalf{firstGQcondition} and \recalf{secondGQcondition}:
$$
f(y\cdot (t,\tau)) = \eexp^{i\tau\kappa + it/\hbar}\cdot f(y)
\mapob,
\formula{firstGQcondition}
$$
where $y\cdot(t,\tau)$ denotes the right-action of the structure group on the principal bundle and where $ \eexp^{i\tau\kappa + it/\hbar}\cdot f(y) $ denotes the usual multiplication in $\CA^\CC$.
$$
(v,z)^{h} f = 0
\formula{secondGQcondition}
$$ 
for all $(v,z)\in \Liealg g$, where $(v,z)^{h}$ is the horizontal lift to $Y$ of the fundamental vector field $(v,z)^{\Liealg g^*}$. It (thus) satisfies the condition $\contrf{(v,z)^{h}}{\Gamma} = 0$. In our situation is is necessarily of the global form
$$
(v,z)^{h} = (v,z)^{\Liealg g^*} + p_0 \cdot \partial_t + p_1\cdot \partial_\tau
\mapob.
$$
The coefficients $p_i$ are determined by the equations $\contrf{(v,z)^{h}}{\Gamma_i}=0$.

\definition{Remark}
For the even part $\CA_0/d\ZZ$ of the structure group we already have a parameter $d\in \RR^{>0}$ and the choice of $\hbar$ must be compatible with $d$ in the sense that $d/\hbar$ must be a multiple of $2\pi$ in order that condition (i) doesn't reduce to the condition that $f$ must be identically zero. The parameter $d$ was introduced for the general case, in which the symplectic form is not exact, but has a non-trivial group of periods. It should satisfy the condition that the group of periods is included in $d\ZZ$. In our situation here the group of periods reduces to $\{0\}$, so no restriction is imposed a priori on $d$.

\enddefinition

\definition{Remark}
The parameter $\kappa$ is not discussed in \cite{Tu10}. Its presence here will turn out to be crucial if we want to obtain a match between the irreducible representations found in the regular representation and the representations obtained by geometric quantization of coadjoint orbits.

\enddefinition

\head{ \sectionnum. Orbits of dimension $0\vert0$ }\endhead

Since the orbit is a point, the prequantum bundle is $\{\mu_o\} \times (\CA_0/d\ZZ)\times \CA_1$ together with the connection $1$-form $\Gamma = \extder t + \extder \tau$. Since the canonical momentum map takes the (constant) value $\mu_o$, the lifted vector fields $(v,z)^{Y}$ are given by
$$
(v,z)^{Y}
=
\bigl(v^1 \xnul_1 + v^2 \xnul_2 + v^3 \xnul_3 \bigr) \cdot \fracp{}{t} 
+ \bigl(   v^4 \xnulb_4 +  v^5 \xnulb_5 + v^6 \xnulb_6\bigr) \cdot \fracp{}{\tau}
\mapob,
$$
It follows immediately that the lifted action is given by
$$
g\in G: \pmatrix t\\ \tau\endpmatrix \mapsto 
\pmatrix t  -a^1\xnul_1-a^2\xnul_2-a^3\xnul_3 \\
 \tau  -\alpha^4 \xnulb_4 -\alpha^5 \xnulb_5 -\alpha^6 \xnulb_6\endpmatrix
\mapob.
\formula{liftedactioncasei}
$$

For a point orbit $\mu_o$, the stabilizer subgroup $G_{\mu_o}$ is the whole group. An invariant polarization thus corresponds necessarily to $\Liealg h = \Liealg g$. Since the orbit dimension is $0\vert0$, this corresponds to the trivially zero foliation on the $0$-dimensional manifold.

Smooth functions on $Y=(\CA_0/d\ZZ) \times \CA_1$ satisfying condition \recalf{firstGQcondition} are determined by a complex number $c\in \CC$ as
$$
f(t,\tau) = c\cdot \eexp^{i\tau\kappa + it/\hbar}
\mapob,
$$
whereas the second condition \recalf{secondGQcondition} is void. These functions constitute a $1$-dimen\-sio\-nal graded vector space of dimension $1\vert0$. The induced action of $G$ on such a function is given (using \recalf{liftedactioncasei}) as
$$
\align
(g\cdot f)(t,\tau)
&
=
f(g\mo(t,\tau))
=
f(t  +a^1\xnul_1+a^2\xnul_2+a^3\xnul_3 ,
 \tau  +\alpha^4 \xnulb_4 +\alpha^5 \xnulb_5 +\alpha^6 \xnulb_6)
\\&
=
\eexp^{i(\alpha^4 \xnulb_4 +\alpha^5 \xnulb_5 +\alpha^6 \xnulb_6)\kappa +
i(a^1\xnul_1+a^2\xnul_2+a^3\xnul_3)/\hbar} \cdot f(t,\tau)
\mapob.
\tag{\memorize\theoremnummer={00dimorbitaction}}
\endalign
$$

\head{ \sectionnum. Orbits of dimension $2\vert2$ with an even symplectic form }\endhead

As said in \recals{GQgeneral}, the prequantum bundle is just the direct product $Y=\Orbit_{\mu_o} \times (\CA_0/d\ZZ) \times \CA_1$. Using the explicit expression of the symplectic form, it is not hard to see that a connection form $\Gamma$ is given by
$$
\Gamma_0 
= 
\extder t + {(y_0)}\mo ( -x_1\, \extder x_2 - \tfrac12 \xi_5\, \extder \xi_5 + \tfrac12 \xi_6 \,\extder \xi_6 ) 
\qquad\text{and}\qquad
\Gamma_1 
=  
\extder \tau 
\mapob.
$$
With this choice, the lifted fundamental vector fields $(v,z)^{Y}$ are given by
$$
\align
(v,z)^{Y} 
=  &
\phantom{+} y_0 \cdot \Bigl(
v^2 \cdot \fracp{}{x_1} -
v^1 \cdot \fracp{}{x_2}  -
v^5 \cdot \fracp{}{\xi_5}   +
v^6 \cdot \fracp{}{\xi_6} \Bigr)
\\&
+\bigl( v^2 x_2 + v^3 \xnul_3  - \tfrac12 v^5 \xi_5 - \tfrac12 v^6 \xi_6 + z^0 y_0 \bigr) \cdot \fracp{}{t} 
\\&
+ \bigl(  v^4 \xnulb_4 + v^5 \xnulb_5 + v^6 \xnulb_6  \bigr) \cdot \fracp{}{\tau}
\mapob,
\endalign
$$
which integrate to an action of $G$ on $Y$ by
$$
g\in G:
\pmatrix x_1 \\ x_2  \\ \xi_5 \\ \xi_6 \\ t\\  \tau \endpmatrix
\mapsto
\pmatrix
x_1 - y_0a^2
\\ 
x_2 + y_0 a^1
\\
\xi_5 + y_0 \alpha^5
\\
\xi_6 - y_0 \alpha^6
\\
t- a^2x_2 - a^3 \xnul_3 + \tfrac12 \alpha^5 \xi_5 + \tfrac12 \alpha^6 \xi_6 - b y_0 - \tfrac12 y_0 a^1 a^2 
\\
\tau   - \alpha^4\xnulb_4 - \alpha^5\xnulb_5 -
\alpha^6 \xnulb_6 
\endpmatrix
\formula{liftedactioncaseii}
$$

For all these orbits the stabilizer subgroup $G_{\mu_o}$ is given by 
$$
G_{\mu_o} = \{(a^i, \alpha^j,b, \beta)\mid a^1=a^2=\alpha^5=\alpha^6=0\}
\mapob.
$$
The stabilizer subalgebra $\Liealg g_{\mu_o}$ of dimension $2\vert2$ thus is generated by the vectors $e_3, e_4, k_0, k_1$. If a subalgebra $\Liealg h\supset \Liealg g_{\mu_o}$ has even dimension $4$, then necessarily it contains all $4$ even vectors $e_1, e_2, e_3, k_0$. But then 
$$
\contrs{\[e_1,e_2\]}{\mu_o} = \contrs{-k_0}{\mu_o} = -y_0\neq0
\mapob.
$$
Hence for a polarization the even dimension of $\Liealg h$ is at most $3$. 
Similarly, if $\Liealg h$ has odd dimension $4$, then it necessarily contains all odd vectors $e_4,e_5, e_6, k_1$ and in particular $e_5$. And then we have 
$$
\contrs{\[e_5,e_5\]}{\mu_o} = y_0\neq0
\mapob.
$$
Since the same argument applies to $e_6$, it follows that the odd dimension of $\Liealg h$ for a polarization is at most $3$ and that $\Liealg h$ cannot contain the (isolated) vectors $e_5$ or $e_6$. On the other hand, it is easy to check that the subalgebra $\Liealg h$ of dimension $3\vert 3$ generated by the vectors $ae_1+be_2, e_3, e_4, e_5-\epsilon e_6, k_0, k_1$ with $\vert a\vert^2+ \vert b\vert^2=1$ and $\epsilon=\pm1$ satisfies the condition $\contrs{\[\Liealg h, \Liealg h\]}{\mu_o} = 0$. It thus represents an invariant polarization; with the given degrees of freedom in $a,b,\epsilon$ these are the only ones.

Functions $f$ on $Y$ satisfying condition \recalf{firstGQcondition} are determined by functions $g$ on the orbit according to the formula
$$
f(x_1, x_2, \xi_5, \xi_6,t,\tau) = \eexp^{i\tau\kappa + it/\hbar} \cdot g(x_1, x_2, \xi_5, \xi_6) 
\mapob.
\formula{caseiifunctions}
$$
We now choose the polarization generated by $e_2, e_3, e_4, e_5-\epsilon e_6, k_0, k_1$ (the others will give equivalent representations), which is equivalent to the invariant polarization (foliation) on the orbit generated by the vector fields
$$
\partial_{x_1}
\qquad\text{and}\qquad
\partial_{\xi_5} + \epsilon \partial_{\xi_6}
\mapob.
$$
The horizontal lifts of these vector fields are 
$$
\fracp{}{x_1} 
\qquad\text{and}\qquad
\fracp{}{\xi_5} + \epsilon \fracp{}{\xi_6} - \frac{\xi_5-\epsilon\xi_6}{2y_0} \cdot \fracp{}{t} 
\mapob.
$$
Condition \recalf{secondGQcondition} applied to \recalf{caseiifunctions} then gives us the equations
$$
\fracp{g}{x_1} = 0
\qquad\text{and}\qquad
\fracp{g}{\xi_5} + \epsilon \fracp{g}{\xi_6} - i \cdot \Bigl( \frac{\xi_5-\epsilon\xi_6}{2y_0\hbar}  \Bigr) \cdot g = 0
\mapob.
$$
The solutions to these equations are determined by functions $h$ of one even and one odd variable according to
$$
g(x_1, x_2, \xi_5, \xi_6) = h(x_2, \xi_5 - \epsilon \xi_6) \cdot
\eexp^{\frac{-i\epsilon}{2y_0\hbar}\xi_5\xi_6}
\mapob.
$$
The induced action of $G$ on such a function is given (using \recalf{liftedactioncaseii}) as
$$
\align
(\Phi_g f)(x_1, x_2, 
\xi_5, \xi_6,t,\tau)
&
=
h(x_2 - y_0 a^1, (\xi_5 - \epsilon \xi_6) - y_0 ( \alpha^5 + \epsilon \alpha^6)
)
\\&
\kern2em
\cdot
\eexp^{\frac{-i\epsilon}{2y_0\hbar}\xi_5\xi_6}
\cdot
\eexp^{i\tau\kappa + it/\hbar}
\cdot
\eexp^{i(\alpha^4\xnulb_4 + \alpha^5\xnulb_5 +
\alpha^6 \xnulb_6)\kappa}
\\&
\kern3em
\cdot
\eexp^{i(a^2x_2 + a^3 \xnul_3 -\frac12 (\alpha^5 - \epsilon \alpha^6)(\xi_5 - \epsilon\xi_6))/\hbar} 
\\&
\kern4em
\cdot
\eexp^{iy_0( b  - \frac12 a^1 a^2 + \frac12\epsilon \alpha^5\alpha^6)/\hbar}
\mapob.
\endalign
$$
If we introduce the variables $z_1 = (x_2 - \xnul_2)/y_0$ and $\zeta = (\xi_5-\epsilon\xi_6)/y_0$, then the action of $\gh\in G$ on the functions $h$ in terms of these new variables becomes
$$
\align
(\Phi_\gh h)( z_1, 
\zeta)
&=
h(z_1 -  \ah^1, \zeta -  ( \alphah^5 + \epsilon \alphah^6)
)
\cdot
\eexp^{i\frac{y_0}{\hbar}(\ah^2  z_1  -\frac12 (\alphah^5 - \epsilon \alphah^6)\zeta)}
\\&
\kern2em
\cdot
\eexp^{i(\ah^2 \xnul_2 + \ah^3 \xnul_3 )/\hbar}
\cdot
\eexp^{i\frac{y_0}{\hbar}( \bh  - \frac12  \ah^1 \ah^2 + \frac12\epsilon  \alphah^5\alphah^6)}
\cdot
\eexp^{i(\alphah^4\xnulb_4 + \alphah^5\xnulb_5 +
\alphah^6 \xnulb_6)\kappa}
\mapob.
\tag{\memorize\theoremnummer={22evenaction}}
\endalign
$$
This action depends upon the coordinates $\xnul_2$, $\xnul_3$, $\xnulb_4$, $\xnulb_5$, $\xnulb_6$, $y_0$ of the base point $\mu_o$. The last five are the parameters of the orbit, but the first is not an invariant of the orbit. To see that $\xnul_2$ is a spurious parameter in this action, it suffices to change the variable $z_1$ to $z_1 = x_2/y_0$, in which case the $\ah^2 \xnul_2$ term in the exponential disappears. The ``introduction'' of the parameter $\xnul_2$ thus just gives an equivalent description of the same representation. We introduced it to facilitate future comparisons.

\head{ \sectionnum. Orbits of dimension $2\vert2$ with an odd symplectic form }\endhead

As before, the prequantum bundle is the direct product $Y=\Orbit_{\mu_o} \times (\CA_0/d\ZZ) \times \CA_1$. Here a connection form $\Gamma$ is given by
$$
\Gamma_0 
= \extder t 
\qquad\text{and}\qquad
\Gamma_1 
=  \extder \tau + {(\yb_1)}\mo ( \xb_5 \,\extder \xib_3 - \xib_1\, \extder \xb_4 ) 
\mapob.
$$
The lifted fundamental vector fields $(v,z)^{Y}$ are given by
$$
\align
(v,z)^{Y} 
=  &
\phantom{+} \yb_1 \cdot \Bigl(
v^4 \cdot \fracp{}{\xib_1} +
v^5 \cdot \fracp{}{\xib_3} -
v^1 \cdot \fracp{}{\xb_4} -
v^3 \cdot \fracp{}{\xb_5} 
\Bigr)
\\&
\kern2em
+\bigl( v^1 \xnul_1 + v^2 \xnul_2 + v^3 \xnul_3 
\bigr) \cdot \fracp{}{t} 
\\&
\kern4em
+ \bigl( v^3\xib_3 +  v^4 \xb_4 + v^6 \xnulb_6 + z^1 \yb_1 \bigr) \cdot \fracp{}{\tau}
\mapob,
\endalign
$$
which integrate to an action of $G$ on $Y$ by
$$
g\in G:
\pmatrix \xib_1 \\ \xib_3 \\ \xb_4 \\ \xb_5 \\ t\\  \tau \endpmatrix
\mapsto
\pmatrix
\xib_1 - \yb_1 \alpha^4
\\
\xib_3 - \yb_1 \alpha^5
\\
\xb_4 + \yb_1a^1
\\ 
\xb_5+ \yb_1 a^3
\\
t- a^1\xnul_1 - a^2\xnul_2 - a^3 \xnul_3 
\\
\tau  
- a^3 \xib_3 - \alpha^4\xb_4 -
\alpha^6 \xnulb_6 - \beta \yb_1 - \tfrac12 \yb_1 a^1 \alpha^4 + \tfrac12 \yb_1 a^3 \alpha^5
\endpmatrix
\formula{liftedactioncaseiii}
$$

For all orbits of this type the stabilizer subgroup $G_{\mu_o}$ is given by 
$$
G_{\mu_o} = \{(a^i, \alpha^j,b, \beta)\mid a^1=a^3=\alpha^4=\alpha^5=0\}
\mapob.
$$
The stabilizer subalgebra $\Liealg g_{\mu_o}$ of dimension $2\vert 2$ is generated by the vectors $e_2$, $e_6$, $k_0$, $k_1$. Since $y_0=0$, the subalgebra $\Liealg h$ representing an invariant polarization can have an even dimension $4$, \ie, containing all $4$ even vectors $e_1, e_2, e_3, k_0$. But then it cannot contain any combination of $e_4$ and $e_5$ as we have 
$$
\contrs{\[ae_4+be_5, e_1\]}{\mu_o} = a\yb_1
\qquad\text{and}\qquad
\contrs{\[ae_4+be_5, e_3\]}{\mu_o} = b\yb_1
\mapob.
$$
Since $\yb_1\neq0$, the condition that these values are zero implies that we have $a=b=0$. The subalgebra $\Liealg h$ of dimension $4\vert 2$ generated by the vectors $e_1, e_2, e_3, e_6, k_0, k_1$ thus represents an invariant polarization.

Similarly, the subalgebra $\Liealg h$ representing an invariant polarization can have an odd dimension $4$, \ie, containing all $4$ odd vectors $e_4, e_5,e_6, k_1$. But then it cannot contain any combination of $e_1$ and $e_3$ as we have 
$$
\contrs{\[ae_1+be_3, e_4\]}{\mu_o} = -a\yb_1
\qquad\text{and}\qquad
\contrs{\[ae_1+be_3, e_5\]}{\mu_o} = -b\yb_1
\mapob.
$$
The subalgebra $\Liealg h$ of dimension $2\vert 4$ generated by the vectors $e_2, e_4, e_5, e_6, k_0, k_1$ thus represents an invariant polarization.

If we now look at a subalgebra $\Liealg h\supset \Liealg g_{\mu_o}$ of dimension $3\vert 3$, it must be generated by $ae_1+be_3, e_2, ce_4+de_5, e_6, k_0, k_1$ for some constants $a,b,c,d$. The condition $\contrs{\[\Liealg h, \Liealg h\]}{\mu_o} = 0$ then implies in particular that we must have 
$$
\contrs{\[ae_1+be_3, ce_4+de_5\]}{\mu_o} = -ac-bd = 0
\mapob.
$$
Since the couples $(a,b)$ and $(c,d)$ can be multiplied by a constant without changing $\Liealg h$, it follows easily that the ``only'' solution is given by $c=b$ and $d=-a$ and that we can require $\vert a\vert^2+\vert b\vert^2=1$. Since $\contrs{\[\Liealg h, \Liealg h\]}{\mu_o} = 0$ does not give any other conditions, it follows that the subalgebra $\Liealg h$ of dimension $3\vert 3$ generated by the vectors $ae_1+be_3, e_2, be_4-ae_5, e_6, k_0, k_1$ represents an invariant polarization.

As before, functions $f$ on the prequantum bundle $Y$ satisfying condition \recalf{firstGQcondition} are determined by functions $g$ on the orbit according to
$$
f(\xb_4,\xb_5, \xib_1,\xib_3, t, \tau) = \eexp^{i\tau\kappa + it/\hbar} \cdot 
g(\xb_4,\xb_5, \xib_1,\xib_3)
\mapob.
\formula{caseiiifunctions} 
$$
However, in order to proceed with the quantization, we have to choose an invariant polarization. 

\medskip
\noindent$\bullet$
We start with the invariant polarization represented by the subalgebra of dimension $4\vert2$, which corresponds to the foliation on the orbit generated by the vector fields
$$
\partial_{\xb_4}
\qquad\text{and}\qquad
\partial_{\xb_5}
\mapob.
$$
The corresponding horizontal lifts are given by
$$
\fracp{}{\xb_4} + \frac{\xib_1}{\yb_1} \fracp{}{\tau}
\qquad\text{and}\qquad
\fracp{}{\xb_5}
\mapob,
$$
Condition \recalf{secondGQcondition} applied to functions of the form \recalf{caseiiifunctions} then gives the equations
$$
\fracp{g}{\xb_4} + \frac{i\xib_1\kappa}{\yb_1} g = 0
\qquad\text{and}\qquad
\fracp{g}{\xb_5} = 0
\mapob.
$$
The solutions are determined by functions $h$ of two odd variables according to
$$
g(\xb_4,\xb_5, \xib_1,\xib_3)
=
\eexp^{-i\xb_4\xib_1\kappa/\yb_1}
\cdot h(\xib_1,\xib_3)
\mapob.
$$
The induced action of $G$ on such a function is given (using \recalf{liftedactioncaseiii}) as
$$
\align
(\Phi_g f)(\xb_4,\xb_5, 
\xib_1,\xib_3, t, \tau)
&
=
h(\xib_1 + \yb_1 \alpha^4,\xib_3 + \yb_1 \alpha^5)
\cdot
\eexp^{i\tau\kappa + it/\hbar}
\cdot
\eexp^{-i\xb_4\xib_1\kappa/\yb_1}
\\&
\qquad
\cdot \eexp^{i( a^1\xnul_1 + a^2\xnul_2 + a^3 \xnul_3  )/\hbar}
\\&
\qquad\qquad
\cdot \eexp^{i(a^1\xib_1 + a^3 \xib_3  +
\alpha^6 \xnulb_6 + \beta \yb_1 + \frac12 \yb_1 a^1 \alpha^4 + \frac12 \yb_1 a^3 \alpha^5)\kappa}
\mapob.
\endalign
$$
If we introduce the variables $\zeta_4 = -\xib_1/\yb_1$ and $\zeta_5 = -\xib_3/\yb_1$, then the action of $\gh\in G$ on the functions $h$ in terms of the new variables becomes
$$
\align
(\Phi_\gh h)( 
&
\zeta_4,\zeta_5)
=
h(\zeta_4 - \alphah^4,\zeta_5 - \alphah^5)
\cdot \eexp^{-i(\ah^1\zeta_4 + \ah^3\zeta_5)\yb_1\kappa}
\\&
\qquad\qquad
\cdot \eexp^{i( \ah^1\xnul_1 + \ah^2\xnul_2 + \ah^3 \xnul_3 )/\hbar}
\cdot \eexp^{i\alphah^6 \xnulb_6}
\cdot \eexp^{i( \betah  + \frac12 \ah^1 \alphah^4 + \frac12  \ah^3 \alphah^5)\yb_1\kappa}
\mapob.
\endalign
$$

\medskip\noindent$\bullet$
Next we choose the invariant polarization represented by the subalgebra of dimension $2\vert 4$, which corresponds to the foliation on the orbit generated by the vector fields
$$
\partial_{\xib_1}
\qquad\text{and}\qquad
\partial_{\xib_3}
\mapob.
$$
The corresponding horizontal lifts are given by
$$
\fracp{}{\xib_1}
\qquad\text{and}\qquad
\fracp{}{\xib_3} - \frac{\xb_5}{\yb_1} \cdot \fracp{}{\tau}
\mapob,
$$
which gives us the equations
$$
\fracp{g}{\xib_1} = 0
\qquad\text{and}\qquad
\fracp{g}{\xib_3} - \frac{i\xb_5\kappa}{\yb_1} \cdot g = 0
\mapob,
$$
whose solutions are determined by functions $h$ of two even variables according to
$$
g(\xb_4,\xb_5, \xib_1,\xib_3)
=
\eexp^{i\xib_3\xb_5\kappa/\yb_1}
\cdot h(\xb_4, \xb_5)
\mapob.
$$
The induced action of $G$ on such a function is given (using again \recalf{liftedactioncaseiii}) as
$$
\align
(\Phi_g f)(\xb_4,\xb_5, 
\xib_1,\xib_3, t, \tau)
=\ 
&
h(\xb_4 - \yb_1a^1 ,
\xb_5- \yb_1 a^3)
\cdot
\eexp^{i\tau\kappa + it/\hbar}
\cdot
\eexp^{i\xib_3\xb_5\kappa/\yb_1}
\\&
\qquad
\cdot \eexp^{i(a^1\xnul_1 + a^2\xnul_2 + a^3 \xnul_3  )/\hbar}
\\&
\qquad\qquad
\cdot \eexp^{i( \alpha^4\xb_4 + \alpha^5 \xb_5 + 
\alpha^6 \xnulb_6 + \beta \yb_1 - \frac12 \yb_1 a^1 \alpha^4 - \frac12 \yb_1 a^3 \alpha^5)\kappa}
\mapob.
\endalign
$$
If we introduce the variables $\zb_1 = (\xb_4 - \xnulb_4)/\yb_1$ and $\zb_3 = (\xb_5 - \xnulb_5)/\yb_1$, then the action of $\gh\in G$ on the functions $h$ in terms of the new variables becomes
$$
\align
(\Phi_\gh h)(\zb_1,\zb_3 )
&=
h(\zb_1 - \ah^1 ,
\zb_3-  \ah^3)
\cdot \eexp^{i(\ah^1\xnul_1 + \ah^2\xnul_2 + \ah^3 \xnul_3  )/\hbar}
\\&
\qquad
\cdot \eexp^{i( \alphah^4 z_1 + \alphah^5 z_3)\yb_1\kappa}
\cdot \eexp^{i( \alphah^4\xnulb_4  + \alphah^5 \xnulb_5 + 
\alphah^6 \xnulb_6 )\kappa}
\cdot \eexp^{i(\betah - \frac12 \ah^1 \alphah^4 - \frac12 \ah^3 \alphah^5)\yb_1\kappa}
\mapob.
\endalign
$$
As for the case of the representation associated to an orbit of dimension $2\vert2$ with an even symplectic form, the appearance of the parameters $\xnulb_4$ and $\xnulb_5$ is artificial; they disappear when we use the coordinates $\zb_1 = \xb_4 /\yb_1$ and $\zb_3 = \xb_5 /\yb_1$.

\medskip\noindent$\bullet$
Our last choice is the invariant polarization represented by the subalgebra of dimension $3\vert3$ generated by $e_2, e_3, e_4, e_6, k_0, k_1$ (the other choices for $a,b$ will give equivalent representations), which corresponds to the foliation on the orbit generated by the vector fields
$$
\partial_{\xb_5}
\qquad\text{and}\qquad
\partial_{\xib_1}
\mapob.
$$
The corresponding horizontal lifts are given by the same vectors, so we have to look for functions $g$ satisfying $\partial_{\xb_5}g = \partial_{\xib_1}g = 0$. These are determined by functions $h$ of one even and one odd variable according to
$$
g(\xb_4,\xb_5, \xib_1,\xib_3)
=
h(\xb_4, \xib_3)
\mapob.
$$
The induced action of $G$ on such a function is given (using once again \recalf{liftedactioncaseiii}) as
$$
\align
(\Phi_g f)(\xb_4,\xb_5, 
\xib_1,\xib_3, t, \tau)
=\ 
&
h(\xb_4 - \yb_1a^1,\xib_3 + \yb_1 \alpha^5)
\cdot \eexp^{i\tau\kappa + it/\hbar}
\cdot \eexp^{i( a^1\xnul_1 + a^2\xnul_2 + a^3 \xnul_3 )/\hbar}
\\&
\qquad
\cdot \eexp^{i( a^3 \xib_3 + \alpha^4\xb_4 +
\alpha^6 \xnulb_6 + \beta \yb_1 - \frac12 \yb_1 a^1 \alpha^4 + \frac12 \yb_1 a^3 \alpha^5)\kappa}
\mapob.
\endalign
$$
If we introduce the variables $\zb_1 = (\xb_4 - \xnulb_4) /\yb_1$ and $\zeta_5 = -\xib_3/\yb_1$, then the action of $\gh\in G$ on the functions $h$ in terms of the new variables becomes
$$
\align
(\Phi_\gh h)(\zb_1,
\zeta_5)
&=
h(\zb_1 - \ah^1,\zeta_5 - \alphah^5)
\cdot \eexp^{i(  \alphah^4\zb_1-\ah^3 \zeta_5) \yb_1\kappa}
\\&
\qquad
\cdot \eexp^{i( \ah^1\xnul_1 + \ah^2\xnul_2 + \ah^3 \xnul_3  )/\hbar}
\cdot \eexp^{i( \alphah^4 \xnulb_4  +
\alphah^6 \xnulb_6 )\kappa}
\cdot \eexp^{i( \betah - \frac12 \ah^1 \alphah^4 + \frac12 \ah^3 \alphah^5) \yb_1\kappa}
\mapob.
\tag{\memorize\theoremnummer={22odd33polaction}}
\endalign
$$
And as in previous cases, the appearance of the parameter $\xnulb_4$ is artificial and disappears when we use the coordinate $\zb_1 = \xb_4 /\yb_1$ instead of the shifted version $\zb_1 = (\xb_4 - \xnulb_4) /\yb_1$.

\head{ \sectionnum. Orbits of dimension $3\vert3$ }\endhead

As in the other cases, the prequantum bundle is the direct product $Y=\Orbit_{\mu_o} \times (\CA_0/d\ZZ) \times \CA_1$. Here a connection form $\Gamma$ is given by
$$
\align
\Gamma_0 
&
= \extder t + {(y_0)}\mo ( -x_1\, \extder \xh_2 - \tfrac12 \xih_5\, \extder \xih_5 + \tfrac12 \xi_6 \,\extder \xi_6 ) 
\\
\Gamma_1 
&
=  \extder \tau + {(y_0)}\mo ( -\xb_5 \,\extder \xih_5 - \xib_1\, \extder \xh_2 ) 
\mapob.
\endalign
$$
The lifted fundamental vector fields $(v,z)^{Y}$ are given by
$$
\align
(v,z)^{Y} 
=  &
\phantom{+} y_0 \cdot \Bigl(
v^2 \cdot \fracp{}{x_1} -
v^1 \cdot \fracp{}{\hat x_2}  -
v^5 \cdot \fracp{}{\hat \xi_5}   +
v^6 \cdot \fracp{}{\xi_6} \Bigr)
\\&
+\yb_1 \cdot \Bigl(
v^4 \cdot \fracp{}{\xib_1}  -
v^3 \cdot \fracp{}{\xb_5} 
\Bigr)
\\&
+\bigl( v^2 \xh_2 + v^3 \xnul_3  - \tfrac12 v^5 \xih_5 - \tfrac12 v^6 \xi_6 + z^0 y_0 \bigr) \cdot \fracp{}{t} 
\\&
+ \bigl( v^3\xib_3 +  v^4 \xb_4 + v^6 \xnulb_6 + z^1 \yb_1 \bigr) \cdot \fracp{}{\tau}
\mapob,
\endalign
$$
where the functions $\xb_4$ and $\xib_3$ must be seen as the functions $y_0\mo (s^o + \yb_1 \xh_2) = y_0\mo (y_0\xnulb_4- \yb_1\xnul_2 + \yb_1 \xh_2)$ and $-y_0\mo \yb_1 \xih_5$ respectively. These vector fields integrate to an action of $G$ on $Y$ by
$$
g\in G:
\pmatrix x_1 \\ \xh_2 \\ \xb_5 \\ \xib_1 \\ \xih_5 \\ \xi_6 \\ t\\  \tau \endpmatrix
\mapsto
\pmatrix
x_1 - y_0a^2
\\ 
\xh_2 + y_0 a^1
\\ 
\xb_5+ \yb_1 a^3
\\
\xib_1 - \yb_1 \alpha^4
\\
\xih_5 + y_0 \alpha^5
\\
\xi_6 - y_0 \alpha^6
\\
t- a^2\xh_2 - a^3 \xnul_3 - b y_0 + \tfrac12 \alpha^5 \xih_5 + \tfrac12 \alpha^6 \xi_6 - \tfrac12 y_0 a^1 a^2 
\\
\tau  
- a^3 \xib_3 - \alpha^4\xb_4 -
\alpha^6 \xnulb_6 - \beta \yb_1 - \tfrac12 \yb_1 a^1 \alpha^4 + \tfrac12 \yb_1 a^3 \alpha^5
\endpmatrix
\formula{liftedactioncaseiv}
$$

For all these orbits the stabilizer subgroup is given by 
$$
G_{\mu_o} = \{(a^i, \alpha^j,b, \beta)\mid a^i=\alpha^j=0\}
\mapob.
$$
The stabilizer subalgebra $\Liealg g_{\mu_o}$ of dimension $1\vert1$ thus is generated by $k_0,k_1$. If a subalgebra $\Liealg h$ representing an invariant polarization has even dimension $4$, it contains all $4$ even basis vectors, but then 
$$
\contrs{\[e_1, e_2\]}{\mu_o} = -y_0\neq0
\mapob.
$$
The conclusion is that the even dimension of such $\Liealg h$ is at most $3$. Similarly, if the odd dimension is $4$, it contains in particular the vector $e_5$, but then 
$$
\contrs{\[e_5,e_5\]}{\mu_o} = y_0\neq0
\mapob.
$$
Hence the odd dimension of such $\Liealg h$ is at most $3$. A detailed analysis when the even dimension is $3$ shows that this is possible if and only if the even basis vectors are $ae_1+be_2, e_3, k_0$ (with $\vert a\vert^2+\vert b\vert^2=1$). Similarly, the only way to have three odd basis vectors in $\Liealg h$ is when they are $e_4, e_5+\epsilon e_6, k_1$ with $\epsilon=\pm1$. If $\Liealg h$ has dimension $3\vert 3$, it thus contains the vectors $ae_1+be_2, e_3, e_4, e_5+\epsilon e_6, k_0, k_1$. But then 
$$
\contrs{\[e_3, e_5+\epsilon e_6\]}{\mu_o} = -\yb_1\neq0
\mapob,
$$
which shows that dimension $3\vert 3$ is excluded too. A detailed analysis of dimensions $3\vert 2$ and $2\vert 3$ shows that that the only possibilities are the following: for dimension $3\vert 2$ there is a single $\Liealg h$ possible, generated by $e_2, e_3, e_4, k_0, k_1$ and for dimension $2\vert 3$ there are exactly two possibilities, generated by $e_2, e_4, e_5+\epsilon e_6, k_0, k_1$, $\epsilon=\pm1$.

As in the previous cases, functions $f$ on $Y$ satisfying condition \recalf{firstGQcondition} are determined by functions $g$ on the orbit according to
$$
f(x_1, \xh_2, \xb_5, \xib_1, \xih_5, \xi_6, t,\tau)
=
\eexp^{i\tau\kappa + it/\hbar} \cdot 
g(x_1, \xh_2, \xb_5, \xib_1, \xih_5, \xi_6)
\mapob.
\formula{caseivfunctions} 
$$

\medskip\noindent$\bullet$
The invariant polarization corresponding to the subalgebra of dimension $3\vert2$ corresponds to the foliation on the orbit generated by the vector fields
$$
\partial_{x_1}
\qquad,\qquad
\partial_{\xb_5}
\qquad\text{and}\qquad
\partial_{\xib_1}
\mapob.
$$
The horizonal lifts are given by ``the same'' vector fields, hence we have to look for functions $g$ determined by functions $h$ of one even and two odd variables according to
$$
g(x_1, \xh_2, \xb_5, \xib_1, \xih_5, \xi_6)
=
h(\xh_2, \xih_5, \xi_6)
\mapob.
$$
The induced action of $G$ on such a function is given (using \recalf{liftedactioncaseiv}) as
$$
\align
(\Phi_g f)(x_1, \xh_2, \xb_5, 
\xib_1, \xih_5, \xi_6, t, \tau)
=\ 
&
h(\xh_2 - y_0 a^1 ,
\xih_5 - y_0 \alpha^5 ,
\xi_6 + y_0 \alpha^6 )
\cdot \eexp^{i\tau\kappa + it/\hbar}
\\&
\cdot \eexp^{i(a^2\xh_2 + a^3 \xnul_3 + b y_0 - \frac12 \alpha^5 \xih_5 - \frac12 \alpha^6 \xi_6 - \frac12 y_0 a^1 a^2)/\hbar}
\\&
\cdot \eexp^{i(a^3 \xib_3 + \alpha^4\xb_4 +
\alpha^6 \xnulb_6 + b^1 \yb_1 - \frac12 \yb_1 a^1 \alpha^4 + \frac12 \yb_1 a^3 \alpha^5)\kappa}
\mapob,
\endalign
$$
where as before the dependent functions $\xib_3$ and $\xb_4$ represent the functions $-y_0\mo \yb_1 \xih_5$ and  $y_0\mo (s^o + \yb_1 \xh_2) = y_0\mo (y_0\xnulb_4- \yb_1\xnul_2 + \yb_1 \xh_2)$ respectively.
If we introduce the variables $z_1 = (\xh_2 - \xnul_2)/y_0$, $\zeta_5 = \xih_5/y_0$ and $\zeta_6 = -\xi_6/y_0$, then the action of $\gh\in G$ on the functions $h$ in terms of the new variables becomes
$$
\align
(\Phi_\gh h)( z_1, \zeta_5, \zeta_6)
&=
h(z_1 - \ah^1 ,
\zeta_5 - \alphah^5 ,
\zeta_6 - \alphah^6 )
\cdot \eexp^{i(\ah^2\xnul_2  + \ah^3 \xnul_3 )/\hbar}
\cdot \eexp^{i( \alphah^4\xnulb_4 + \alphah^6 \xnulb_6 )\kappa}
\\&
\kern2em
\cdot \eexp^{i(\ah^2 z_1 - \frac12 \alphah^5 \zeta_5 + \frac12 \alphah^6 \zeta_6 + \bh - \frac12 \ah^1 \ah^2) y_0/\hbar}
\\&
\kern4em
\cdot \eexp^{i(-\ah^3 \zeta_5 + \alphah^4 z_1 +
 \betah - \frac12 \ah^1 \alphah^4 + \frac12 \ah^3 \alphah^5) \yb_1\kappa}
\mapob.
\tag{\memorize\theoremnummer={33mixed32polaction}}
\endalign
$$
This action depends upon the basepoint parameters $\xnul_2$, $\xnul_3$, $\xnulb_4$, $\xnulb_6$, $y_0$, $\yb_1$. As in previous cases, the appearance of the parameters $\xnul_2$ and $\xnulb_4$ is artificial. Using the coordinate $z_1 = \xh_2/y_0$ instead of the shifted version, the term $\ah^2\xnul_2$ in the exponent would have disappeared and the term $\alphah^4\xnulb_4$ would have become $\alphah^4 s^o/y_0$. And then it is obvious that the action only depends upon the orbit parameters $\xnul_3$, $s^o$, $\xnulb_6$, $y_0$, $\yb_1$. Another shift would have made the term $\alphah^4\xnulb_4$ disappear and changed the term $\ah^2\xnul_2$ to $-s^o\ah^2/\yb_1$.

\medskip\noindent$\bullet$
The invariant polarization corresponding to the subalgebra of dimension $2\vert3$ corresponds to the foliation on the orbit generated by the vector fields
$$
\partial_{x_1}
\qquad,\qquad
\partial_{\xib_1}
\qquad\text{and}\qquad
\partial_{\xih_5} - \epsilon \partial_{\xi_6}
\mapob.
$$
The horizonal lifts are given by the vector fields
$$
\fracp{}{x_1}
\qquad,\qquad
\fracp{}{\xib_1}
\qquad\text{and}\qquad
\fracp{}{\xih_5} - \epsilon \fracp{}{\xi_6} - \frac{\xih_5+\epsilon \xi_6}{2 y_0} \cdot \fracp{}{t} + \frac{\xb_5}{y_0}\cdot \fracp{}{\tau}
\mapob,
$$
which gives us the equations for the functions $g$
$$
\fracp{g}{x_1} = 0
\quad,\quad
\fracp{g}{\xib_1} = 0
\qquad\text{and}\qquad
\fracp{g}{\xih_5} - \epsilon \fracp{g}{\xi_6} - i \cdot \frac{\xih_5+\epsilon \xi_6}{2 y_0\hbar} \cdot  g + i\cdot\frac{ \xb_5 \kappa}{y_0} \cdot g = 0
\mapob,
$$
whose solutions are determined by functions $h$ of two even and one odd variable according to
$$
g(x_1, \xh_2, \xb_5, \xib_1, \xih_5, \xi_6)
=
h(\xh_2, \xb_5, \xih_5 + \epsilon \xi_6) \cdot
\eexp^{\frac{i\epsilon}{2y_0\hbar}\xih_5\xi_6}
\cdot\eexp^{-\frac{i}{y_0}\xb_5\xih_5\kappa}
\mapob.
$$
The induced action of $G$ on such a function is given (using again \recalf{liftedactioncaseiv}) as
$$
\align
(\Phi_g f)(x_1, \xh_2, \xb_5, &
\xib_1, \xih_5, \xi_6, t, \tau)
\\
=\ 
&
h(\xh_2 - y_0 a^1 ,
\xb_5 - \yb_1 a^3 ,
(\xih_5 +\epsilon\xi_6)  - y_0 (\alpha^5 -Ê\epsilon \alpha^6 )
)
\\& 
\cdot \eexp^{i\tau\kappa + it/\hbar}
\cdot
\eexp^{\frac{i\epsilon}{2y_0\hbar}\xih_5\xi_6}
\cdot
\eexp^{-\frac i{y_0} \xb_5 \xih_5 \kappa}
\\&
\kern1em
\cdot \eexp^{i( a^2\xh_2 + a^3 \xnul_3 + b y_0 - \frac12 (\alpha^5 + \epsilon\alpha^6) ( \xih_5 + \epsilon \xi_6) - \frac12 y_0 a^1 a^2 - \frac\epsilon2 y_0 \alpha^5\alpha^6)/\hbar}
\\&
\kern2em
\cdot \eexp^{i(  \alpha^4\xb_4 +  \alpha^5\xb_5 +
\alpha^6 \xnulb_6 + \beta \yb_1 - \frac12 \yb_1 a^1 \alpha^4 - \frac12 \yb_1 a^3 \alpha^5)\kappa}
\mapob,
\endalign
$$
where as before the dependent functions $\xib_3$ and $\xb_4$ represent the functions $-y_0\mo \yb_1 \xih_5$ and  $y_0\mo (s^o + \yb_1 \xh_2) = y_0\mo (y_0\xnulb_4- \yb_1\xnul_2 + \yb_1 \xh_2)$ respectively.
If we introduce the variables $z_1 = (\xh_2 - \xnul_2)/y_0$, $\zb_3 = (\xb_5 - \xnulb_5)/\yb_1$ and $\zeta = \xi/y_0$, then the action of $\gh\in G$ on the functions $h$ in terms of the new variables becomes
$$
\align
(\Phi_\gh h)( z_1, \zb_3, \zeta)
&=
h(z_1 - \ah^1 ,
\zb_3 -  \ah^3 ,
\zeta - (\alphah^5 -Ê\epsilon \alphah^6 )
)
\cdot \eexp^{i\frac{y_0}{\hbar}( \ah^2 z_1 - \frac12 (\alphah^5 + \epsilon\alphah^6) \zeta)} 
\\& 
\kern2em
\cdot \eexp^{i(  \alphah^4 z_1 +  \alphah^5 \zb_3) \yb_1\kappa}
\cdot \eexp^{i( \ah^2\xnul_2 + \ah^3 \xnul_3 )/\hbar}
\cdot \eexp^{i(  \alphah^4\xnulb_4 +  \alphah^5\xnulb_5 +
\alphah^6 \xnulb_6 )\kappa}
\\&
\kern4em
\cdot \eexp^{i\frac{y_0}{\hbar}( \bh - \frac12 \ah^1 \ah^2 - \frac\epsilon2 \alphah^5\alphah^6) }
\cdot \eexp^{i( \betah - \frac12 \ah^1 \alphah^4 - \frac12 \ah^3 \alphah^5) \yb_1\kappa}
\mapob.
\endalign
$$
And just as in previous cases, the dependence on the parameters $\xnul_2$, $\xnulb_4$ and $\xnulb_5$ is artificial. If we use the coordinates $z_1 = \xh_2 /y_0$ and $\zb_3 = \xb_5 /\yb_1$ instead of the shifted ones given above, the terms $\ah^2 \xnul_2$ and $\alphah^5 \xnulb_5$ in the exponents disappear and the term $\alphah^4 \xnulb_4$ becomes $\alphah^4 s^o/y_0$. And in that form this representation only depends upon the parameters $\xnul_3$, $\xnulb_6$, $y_0$, $\yb_1$, $s^o$ describing the orbit. Shifting $z_1$ differently would have made the term $\alphah^4 \xnulb_4$ disappear and changed the term $\ah^2 \xnul_2$ to $-\ah^2 s^o/\yb_1$.

\head{\sectionnum. Recapitulation and comparison}\endhead

\noindent
If we gather the results for all orbits, we get the following list: 
\roster\item"(i)"
For each $0\vert0$-dimensional orbit we have a ($1$-dimensional) representation.

\item"(ii)"
For each $2\vert2$-dimensional orbit with an even symplectic form we have two different representations by functions on a $1\vert1$-dimensional manifold.

\item"(iii)"
For each $2\vert2$-dimensional orbit with an odd symplectic form we have three different representations by functions on a manifold of dimension $2\vert0$, $0\vert2$ and $1\vert1$ respectively.

\item"(iv)"
For each $3\vert3$-dimensional orbit (with a mixed symplectic form) we have three different representations, two by functions on a manifold of dimension $2\vert1$ and one on a manifold of dimension $1\vert2$.

\endroster
In order to compare this list with the representations we found in the regular representation in \recals{naiveregularrep}, we start with the spaces $V_{(0, \ell_2, \ell_3, 0, \lambda_4),(\ell_1, \lambda_5, \lambda_6)}$. Comparison of the $G$ action given in \recalf{ell0zeroandlambda0zeroaction} with the action given in the representation associated to a $0\vert0$-dimensional orbit \recalf{00dimorbitaction} tells us that they are the same, provided we make the following identifications
$$
\matrix
\ell_1 &=& -\xnul_1/\hbar
&\quad,\quad&
\ell_2 &=& -\xnul_2/\hbar 
&\quad,\quad&
\ell_3 &=& -\xnul_3/\hbar 
\\ 
\lambda_4 &=& -\xnulb_4\kappa
&\quad,\quad&
\lambda_5 &=& - \xnulb_5\kappa
&\quad,\quad&
\lambda_6 &=& - \xnulb_6\kappa 
\mapob.
\endmatrix
$$
Instead of commenting directly on these identifications, let us first complete our list, starting with $V_{(0, \ell_2, \ell_3, \lambda_0, \lambda_4),( \lambda_6)}$ with $\lambda_0\neq0$. Comparing the action \recalf{ell0zerobutnotlambda0action} with the action given in the representation associated to a $2\vert2$-dimensional orbit with an odd symplectic form and a $1\vert1$-dimensional polarization \recalf{22odd33polaction} tells us that they are the same, provided we make the identifications
$$
\matrix
0 &=& -\xnul_1/\hbar
&\quad,\quad&
\ell_2 &=& -\xnul_2/\hbar 
&\quad,\quad&
\ell_3 &=& -\xnul_3/\hbar 
\\ 
\lambda_0 &=& -\yb_1\kappa
&\quad,\quad&
\lambda_4 &=& - \xnulb_4\kappa
&\quad,\quad&
\lambda_6 &=& - \xnulb_6\kappa 
\mapob.
\endmatrix
$$

We then turn our attention to $V_{(\ell_0, \ell_2, \ell_3, 0, \lambda_4)}^\pm$ with $\ell_0\neq0$. Comparing the action \recalf{ell0not0lambda0is0action} with the action \recalf{22evenaction} given in the representations associated to a $2\vert2$-dimensional orbit with an even symplectic form tells us that they are the same, provided we make the identifications
$$
\matrix
\ell_0 &=& -y_0/\hbar
&\quad,\quad&
\ell_2 &=& -\xnul_2/\hbar 
&\quad,\quad&
\ell_3 &=& -\xnul_3/\hbar 
\\ 
\lambda_4 &=& - \xnulb_4\kappa
&\quad,\quad&
0 &=& - \xnulb_5\kappa
&\quad,\quad&
0 &=& - \xnulb_6\kappa 
\mapob.
\endmatrix
$$

We finally look at $V_{(\ell_0, \ell_2, \ell_3, \lambda_0, \lambda_4)}$ with $\ell_0\neq0\neq\lambda_0$. Comparing the action \recalf{overalltransformation} with the action \recalf{33mixed32polaction} given in the representation associated to a $3\vert3$-dimensional orbit with a $2\vert1$-dimensional polarization tells us that they are the same, provided we make the identifications
$$
\matrix
\ell_0 &=& -y_0/\hbar
&\quad,\quad&
\ell_2 &=& -\xnul_2/\hbar 
&\quad,\quad&
\ell_3 &=& -\xnul_3/\hbar 
\\ 
\lambda_0 &=& - \yb_1\kappa
&\quad,\quad&
\lambda_4 &=& - \xnulb_4\kappa
&\quad,\quad&
0 &=& - \xnulb_6\kappa 
\mapob.
\endmatrix
$$

If we look at the required identifications, we see that they all have the same form. The even parameters $\ell_0$, $\ell_1$, $\ell_2$, $\ell_3$ have to be the same as the coordinates of the basepoint $\mu_o$ of the orbit $y_0$, $\xnul_1$, $\xnul_2$, $\xnul_3$ up to a factor $-\hbar$. And the odd parameters $\lambda_0$, $\lambda_4$, $\lambda_5$, $\lambda_6$ have to be the same as the coordinates (of $\mu_o$) $\yb_1$, $\xnulb_4$, $\xnulb_5$, $\xnulb_6$ up to a factor $-\kappa$. But here the situation is not as simple as for the even parameters. For the even parameters we performed a honest Fourier transform of (smooth) functions of even coordinates. As such the even parameters $\ell_0$, $\ell_1$, $\ell_2$, $\ell_3$ thus are real. And identifying them up to a factor $-\hbar$ with the real coordinates $y_0$, $\xnul_1$, $\xnul_2$, $\xnul_3$ poses no problem. On the other hand, the parameters $\lambda_0$, $\lambda_4$, $\lambda_5$, $\lambda_6$ are odd, so identifying them with the real coordinates $\yb_1$, $\xnulb_4$, $\xnulb_5$, $\xnulb_6$ is not obvious. But as said, it should be up to the odd factor $\kappa$. But then all four odd parameters $\lambda_0$, $\lambda_4$, $\lambda_5$, $\lambda_6$ should be real multiples of the same odd factor $\kappa$. Which would suggest that we cannot choose these four parameters freely. 

We thus reach the unsatisfactory situation that we have a nice identification between the irreducible representations appearing in the regular representation and representations associated to coadjoint orbits, but there is nevertheless something strange about this identification. If we take the odd parameters seriously, then we are embarrased by the fact that the identification makes them all real multiples of the same odd factor. However, the sufficient conditions for irreducibility involve (in)dependence of the odd parameters over $\RR$, which seems to indicate thatthere is a link between the odd parameters and real coefficients.

Another point which we have not addressed here is how to define a scalar product on the representation obtained by geometric quantization. It is fairly easy to see that the natural scalar product consisting of integrating over the coordinates on which the functions depend will yield an invariant non-degenerate sesquilinear form. But when the odd dimension is odd (no pun intended; it is the case for the $2\vert2$-dimensional orbit with an even form, for the $2\vert2$-dimensional orbit with an odd form and the $3\vert3$-dimensional polarization and for the $3\vert3$-dimensional orbit with the $2\vert3$-dimensional polarization), this form will be an odd sesquilinear form. Which means that if we take this approach seriously, then the notion of unitary representations in the super setting should not restrict attention to \stress{even} non-degenerate sesquilinear forms.

A third problem to which I have not (yet) found a satisfactory answer is how to define the notion of equivalent representations for odd family decompositions. In view of my claim in the introduction  that the orbits with a non-homogeneous symplectic form are necessary, this is an important question. And indeed, if for some notion of equivalence the representations associated to the orbits with a non-even symplectic form are all equivalent to representations obtained from an orbit with an even symplectic form, then the orbits with a non-homogeneous symplectic form are not needed. Fortunately for me, for any reasonable definition of equivalence, the representations associated to an orbit with a non-homogeneous symplectic form cannot be equivalent to representations obtained via another type of orbit. To substantiate this claim, it suffices to look at the subgroup $H$ of $G$ in which all but $a^1$, $a^2$ and $b$ are non-zero. This $H$ is the classical Heisenberg group. A direct computation shows that the representations $V_{(\ell_0,\ell_2,\ell_3,\lambda_0,\lambda_4)}$ with $\ell_0\neq0$ and $\lambda_0\neq0$ decompose into four copies of the (infinite dimensional) irreducible representation of this Heisenberg subgroup. On the other hand, the representations $V_{(\ell_0,\ell_2,\ell_3,0,\lambda_4)}^\pm$ decompose as a direct sum of just two copies. And the representations $V_{(\ell_0,\ell_2,\ell_3,\lambda_0,\lambda_4),(\lambda_6)}$ do not contain this irreducible representation of $H$. It follows that they cannot be isomorphic under any reasonable notion of equivalence.

As the reader can see, there remain a lot of questions to be answered. However, I hope to have convinced her\slash him that symplectic supermanifolds with a non-homogeneous symplectic form are interesting objects to study.


\Refs
\widestnumber\key{Tu10}

\ref
\key DW
\by B.~DeWitt
\book Supermanifolds
\publ Cambridge UP
\publaddr Cambridge
\yr1984
\endref

\ref\key GS
\by V.~Guillemin \& S.~Sternberg
\book Supersymmetry and Equivariant de~Rham Theory
\publ Springer
\publaddr Berlin
\yr1999
\endref

\ref
\key Ko
\by B.~Kostant
\paper Graded manifolds, graded Lie theory, and
prequantization
\pages 177--306
\inbook Differential geometric methods in mathematical
physics
\eds K.~Bleuler \& A.~Reetz
\publ Springer-Verlag
\publaddr Berlin
\bookinfo Proceedings Conference, Bonn 1975. LNM 570
\yr1977
\endref

\ref
\key Le
\by D.A.~Leites
\paper Introduction to the theory of supermanifolds
\jour Russian Math.\ Surveys
\vol 35
\yr 1980
\pages 1--64
\endref

\ref
\key Ro
\by A.~Rogers
\book Supermanifolds: Theory and Applications
\publ World Scientific Publishing Co. Pte. Ltd.
\publaddr Singapore
\yr2007
\endref

\ref\key Tu04
\by G.M.~Tuynman
\book Supermanifolds and Supergroups: Basic Theory
\publ Kluwer Academic Publishers (nowadays part of Springer Verlag)
\publaddr Dordrecht
\yr2004
\bookinfo Mathematics and Its Applications Vol 570
\endref

\ref
\key Tu10
\by G.M. Tuynman
\paper
Super symplectic geometry and prequantization
\jour Journal of Geometry and Physics 
\vol60
\yr2010
\pages1919--1939
\endref

\endRefs
\enddocument

\end